%% file: 00-LRSR.tex
\documentclass[a4paper,11pt]{article}

\usepackage{subcaption}
\usepackage{standalone}
\usepackage{pgfplots}
\usepackage{framed}
\pgfplotsset{compat=1.16}

\usepackage{jheppub} %
\usepackage{derivative}
\DeclareDifferential{\Dd}{\mathrm{D}}[style-notation=multiple, style-notation-*=single
]

\usepackage{amssymb}
\usepackage{slashed}

\usepackage{multirow}
\newcommand{\inspirebox}[1]{[\href{#1}{\textsc{Inspire}}]} %

\usepackage{iftex}
\ifLuaTeX
\usepackage{fontspec} %
\else
\usepackage[utf8]{inputenc} %
\DeclareUnicodeCharacter{2212}{-} %
\usepackage[LGR,T1]{fontenc} %
\usepackage{textgreek}
\fi

\usepackage{float}
\usepackage{xcolor}
\usepackage{cancel}
\usepackage{bm}
\usepackage[shortlabels]{enumitem} %
\usepackage{listings} %
\usepackage{mathtools}
\usepackage{cleveref}
\usepackage{ytableau}
\usepackage{comment}
\usepackage{amstext} %
\usepackage{lineno}
\usepackage{csquotes}
\usepackage{nicematrix}
\usepackage{xspace} %

\input{shorthand}

\usepackage{tikz}
\usetikzlibrary{calc, intersections}
\usepackage[compat=1.1.0]{tikz-feynman}
\usepackage{silence} %
\usepackage{array}   %
\newcolumntype{L}{>{$}l<{$}} %
\makeatletter
\edef\savedcodes{\catcode`\noexpand\_=\the\catcode`\_}
\edef\@tempa{\csname opt@newtxmath.sty\endcsname}
\def\@tempb{{subscriptcorrection}}
\expandafter\expandafter\expandafter
  \in@\expandafter\@tempb\expandafter{\@tempa}
\ifin@
  \catcode`\_=12 %
\fi
\makeatother
\usepackage{tensor}
\savedcodes

\title{Local CFTs extremise $F$} %

\author[\Delta]{Ludo Fraser-Taliente}%

\affiliation[\Delta]{Rudolf Peierls Centre for Theoretical Physics, University of Oxford}
\affiliation[\Delta]{Department of Physics, Carnegie Mellon University}

\emailAdd{lfrasert@andrew.cmu.edu}

\abstract{
Many CFTs can be extended to lines of nonlocal CFTs parametrised by the scaling dimension $\Delta$ of the fundamental field appearing in the action. 
$\Delta=\tfrac{d}{2}-\zeta$ is set by the exponent of the kinetic term $(-\partial^2)^{\zeta}$, which is nonlocal for noninteger $\zeta$.
If $\Delta$ is tuned to $\Delta_\mathrm{local}$, the scaling dimension of the fundamental field in the local CFT, \cite{Behan:2017emf} showed that we recover the conformal data of that CFT (plus a decoupled sector).
One natural question is: how is the local point special on this line of nonlocal CFTs?
We prove that these local CFTs lie at the extrema of the (universal part of the) sphere free energy $\tilde{F}(\Delta)=\sin(\tfrac{\pi d}{2}) \log Z_{S^d}$ of the long-range CFTs: $\odv{\tilde{F}}{\Delta}|_{\Delta=\Delta_\mathrm{local}}=0$; and for unitary CFTs they locally maximise it.
The simple proof uses the fact that the derivative of $\tilde{F}$ with respect to $\Delta$ receives contributions only from the nonlocal terms in the action.
The nonlocal terms must be absent in the limit $\Delta \to \Delta_\mathrm{local}$, and hence the derivative is zero.
Demonstrating maximisation then requires a proof of the generalised $F$-theorem in conformal perturbation theory to subleading order.
We check our result with the O$(N)$ $\phi^4$ and cubic CFTs in the $\epsilon$ expansion and the large-$N$ limit.
This result provides a minimal encoding of the scaling dimension of the fundamental field in many CFTs, and also explains the curious derivative structure of the large-$N$ expansion of $\Delta_\phi$ in the O$(N)$ model.
Finally, this nonlocal $F$-extremisation can be viewed as a non-supersymmetric version of the known $c,F,a$-extremisation mechanisms.

}


\newcommand{\WeylOp}{\mathcal{D}}

\newcommand{\id}{\mathbb{I}}
\newcommand{\spinid}{\mathbb{I}_s}
\newcommand{\thalf}{\tfrac{1}{2}}
{%
}%

\newcommand{\EulerBeta}{\mathrm{B}}
\newcommand{\Ft}{\tilde{F}}

\newcommand{\FttextOrPDF}{\texorpdfstring{$\Ft$}{F~}\xspace}
\newcommand{\FtextOrPDF}{\texorpdfstring{$F$}{F}\xspace}
\newcommand{\FhatextOrPDF}{\texorpdfstring{$\hat{F}$}{\^{}F}\xspace}

\newcommand{\NPDF}{\texorpdfstring{$N$}{N}\xspace}

\usepackage{empheq} %

\pgfmathsetmacro\MathAxis{height("$\vcenter{}$")} %

\begin{document}
\maketitle
\flushbottom

\include{01-Extremization}
\include{02-Perturbative}

\acknowledgments
We are grateful to Connor Behan, Igor Klebanov, and Grisha Tarnopolsky for discussions and comments on the draft.
We are particularly grateful to John Wheater for extensive discussion, commentary, and support.
We also thank Kit Fraser-Taliente, Barak Gabai, Chris Herzog, Petr Kravchuk, Marco Meineri, and Bernardo Zan for useful conversations.
This work was supported by the Dalitz Scholarship from the University of Oxford and Wadham College, as well as the Simons
Foundation Grant No. 994316.
For the purpose of open access, the author has applied a CC BY public copyright licence to any Author Accepted Manuscript (AAM) version arising from this submission. %

\appendix

\input{09-Appendix}

\bibliographystyle{JHEP}
{\raggedright  %
\bibliography{references-zot}    %
}              %
\end{document}

%% file: shorthand.tex
\newcommand{\dotwo}{\tfrac{d}{2}}

\newcommand{\ONpdf}{\texorpdfstring{$\gO(N)$}{O(N)}}

\newcommand{\p}{{\rm p}}

\def\Tr{\mathop{\rm Tr}\nolimits}
\def\bbra{{\langle\kern-2.5pt\langle}}
\def\kket{{\rangle\kern-2.5pt\rangle}}
\def\Bbra{{\Big\langle\kern-3.5pt\Big\langle}}
\def\Kket{{\Big\rangle\kern-3.5pt\Big\rangle}}

\DeclarePairedDelimiter\abs{\lvert}{\rvert}

\DeclarePairedDelimiterX\braket[2]{\langle}{\rangle}{#1\,\delimsize\vert\,\mathopen{}#2}
\DeclarePairedDelimiter\expval{\langle}{\rangle}
\DeclarePairedDelimiter\floor{\lfloor}{\rfloor}

\newcommand   \x   {\times}

\newcommand   \half{\frac 1 2}
\newcommand   \lptl{\raise .8ex\hbox{$^\leftarrow$} \hspace{-9pt} \partial}
\newcommand   \lrptl{\raise .8ex\hbox{$^\leftrightarrow$} \hspace{-9pt} \partial}

\newcommand   \SO    {\mathrm{SO}}
\newcommand   \gO    {\mathrm{O}}

\DeclareMathOperator{\vol}{vol}

\newcommand   \cF {\mathcal{F}}

\newcommand   \cM {\mathcal{M}}

\newcommand   \cO {\mathcal{O}}

\newcommand   \cR {\mathcal{R}}

\newcommand{\be}{
  \begin{equation}
  \begin{aligned}
}
\newcommand{\ee}{
  \end{aligned}
  \end{equation}
}

\makeatletter
\DeclareRobustCommand\bea{\@ifnextchar[{\@@bea}{\@bea}}
\def\@@bea[#1]#2\eea{\begin{subequations}\begin{align}#2\end{align}\label{#1}\end{subequations}}
\def\@bea#1\eea{\begin{subequations}\begin{align}#1\end{align}\end{subequations}}
\makeatother

\newcommand\Pochhammer[2]{{\left(#1\right)_{#2}}}

%% file: 01-Extremization.tex
\section{Overview}

A fruitful method of computing conformal data has been to promote seemingly quantized constants like the dimension or the number of fields to continuous parameters.
The canonical example of this is the perturbative solution of the critical Ising model in $d=4-\epsilon$, and its most recognisable result is the expansion in $\epsilon$ of the lowest conformal scaling dimension $\Delta_{\phi,\text{Ising}}$. %
Surprisingly, another \enquote{constant} that can be promoted is exactly that scaling dimension $\Delta_{\phi,\text{Ising}}$. 
To do so, we generalise the action of the Ising CFT to
\begin{equation}\label{eq:LRaction}
    S = \int_x \half \phi(-\partial^2)^{\dotwo -\Delta} \phi + \frac{g_0 }{4!}\phi^4.
\end{equation}
Then, for every $\Delta$, we find a CFT at the RG fixed point ($\beta_g=0$), where the scaling dimension of $\phi$ is forced by the nonlocal kinetic term to be exactly $\Delta$. %
The price we pay is the loss of locality. 
We get a continuous family of CFTs parametrised by $\Delta$, but the fractional derivative makes the theory no longer local, as it no longer has a local energy-momentum tensor of dimension $\Delta_T = d$; we call them nonlocal CFTs or long-range CFTs.

In the limit $\Delta \to \Delta_{\phi,\text{Ising}}$ (i.e. $\frac{1}{8}$ in $d=2$, $0.518\cdots$ in $d=3$, and $\tfrac{d-2}{2}=1$ in $d=4$ -- where it is the free theory), \cite{Behan:2017emf} understood that this long-range CFT becomes the usual local Ising CFT (plus a decoupled generalised free field)\footnote{This is somewhat subtle, as one naive expectation from \eqref{eq:LRaction} would be for the local CFT (also called the short-range CFT) to be found for $\dotwo -\Delta =1$. However, the limit of removing the cutoff does not commute with taking the exponent $\dotwo -\Delta \to 1$ -- this is now well established, and is discussed further below.}.
The same story holds for more complicated long-range CFTs: they become the local CFTs as their long-range parameters $\Delta_i$ are tuned to match the local CFT's scaling dimensions $\Delta_i^{\text{local CFT}}$.

\begin{figure}
\centering
\begin{tikzpicture}[>=Latex, line cap=round, line join=round]
        \def\xmin{1.4}
        \def\xmax{11.8}
        \def\ymin{0.9}
        \def\ymax{6}
        \def\xa{3.20}   %
        \def\xb{4.45}   %
        \def\yb{4.2}    %
        \def\xc{8.15}   %
        \def\xd{10.95}  %
        \coordinate (BluePeak)  at (\xa,4.95);
        \coordinate (BlackPeak) at (\xb,\yb);
        \coordinate (Meet)      at (\xc,2.95);
        \coordinate (BlueLeft)  at (\xmin,3.6);
        \coordinate (BlackLeft) at (\xmin,3.3);
        \coordinate (BlueRight) at (10.65,1.72);

        \draw[gray!55, densely dotted] (\xmin,1.12) -- (\xmax-0.10,1.12);
        \node[gray!65, anchor=west] at (8.65,1.3) {\scriptsize empty CFT (\(\tilde{F}=0\))};

        \draw[->, line width=1.1pt] (\xmin,\ymin) -- (\xmin,\ymax) node[left=3pt] {$\tilde{F}$};
        \draw[->, line width=1.1pt] (\xmin,\ymin) -- (\xmax,\ymin);

        \draw[line width=0.9pt] (\xmin-0.08,4.95) -- (\xmin+0.08,4.95);
        \node[anchor=east, text=blue] at (\xmin-0.16,4.95) {\(\tilde{F}_b\!\left(\tfrac{d-2}{2}\right)\)};
        \draw[line width=0.9pt] (\xmin-0.08,\yb) -- (\xmin+0.08,\yb);
        \node[anchor=east] at (\xmin-0.16,\yb) {\(\tilde{F}_\mathrm{LR}(\Delta_\phi^\mathrm{SR}) = \tilde{F}_{\mathrm{SR}}\)};
        \draw[line width=0.9pt] (\xmin-0.08,2.95) -- (\xmin+0.08,2.95);
        \node[anchor=east] at (\xmin-0.16,2.95) {$\tilde{F}_\mathrm{LR}(\tfrac{d}{4}) = $ \textcolor{blue}{$\tilde{F}_b(\tfrac{d}{4})$}};

        \draw[blue, line width=1pt] (\xa,\ymin-0.08) -- (\xa,\ymin+0.08);
        \node[blue, below=3pt] at (\xa,\ymin-0.0) {$\tfrac{d-2}{2}$};
        \draw[line width=1pt] (\xb,\ymin-0.08) -- (\xb,\ymin+0.08);
        \node[below=3pt] at (\xb,\ymin-0.0) {$\Delta_\phi^{\mathrm{SR}}$};
        \draw[line width=1pt] (\xc,\ymin-0.08) -- (\xc,\ymin+0.08);
        \node[below=3pt] at (\xc,\ymin-0.0) {$\tfrac{d}{4}$};
        \node[below=3pt] at (\xd,\ymin-0.0) {$\Delta_\phi$};

        \draw[line width=1pt] (BlackLeft) .. controls +(1.00,0.78) and +(-1.05,0.00) .. (BlackPeak);
        \draw[line width=1pt] (BlackPeak) .. controls +(1.05,0.00) and +(-1.20,0.45) .. (Meet);
        \draw[line width=1pt] (Meet) .. controls +(0.95,-0.30) and +(-1.20,-0.00) .. (10.7,2.35);

        \draw[blue, line width=1.2pt] (BlueLeft) .. controls +(0.65,1.05) and +(-0.60,0.00) .. (BluePeak);
        \draw[blue, line width=1.2pt] (BluePeak) .. controls +(1.20,0.00) and +(-1.10,0.80) .. (Meet);
        \draw[blue, line width=1.2pt] (Meet) .. controls +(1.05,-0.80) and +(-1.15,0.55) .. (BlueRight);

        \fill[blue] (BluePeak) circle (2.2pt);
        \fill (BlackPeak) circle (2.2pt);

        \node[blue, above=3pt] at (BluePeak) {\scriptsize local scalar};
        \node[below=3pt] at (BlackPeak) {\scriptsize local Ising};

        \coordinate (LegendOrigin) at (6.10,6.0);
        \begin{scope}[shift={(LegendOrigin)}]
                \draw[rounded corners=2pt, fill=white, draw=black!25] (0,0) rectangle (5.15,-0.96);
                \draw[blue, line width=1.2pt] (0.28,-0.28) -- (1.18,-0.28);
                \node[anchor=west] at (1.48,-0.28) {\small Generalised free field};
                \draw[line width=1pt] (0.28,-0.61) -- (1.18,-0.61);
                \node[anchor=west] at (1.48,-0.61) {\small Long-range $\phi^4$};
        \end{scope}

\end{tikzpicture}
\caption[$F$-extremisation]{A picture of $F$-extremisation in the simplest pair of long-range CFTs. 
$\tilde{F}$ is shown as a function of $\Delta_\phi$ for the \textcolor{blue}{generalised free field} (GFF, \textcolor{blue}{blue}) and the \textbf{long-range Ising CFT} (\textbf{black}) for $2<d<4$. %
In $d=3$, these CFTs are unitary:
observe that the local/short-range CFTs, marked by dots, maximise $\tilde{F}$ -- they are the usual free scalar of dimension $\frac{d-2}{2}$ and the standard Wilson-Fisher fixed point (i.e. the $\phi^4$ CFT, with free energy $\tilde{F}_{\mathrm{SR}}$ and $\expval{\phi(x)\phi(0)}\propto x^{-2\Delta_\phi^{\mathrm{SR}}}$) respectively. For analytic expressions, see \eqref{eq:bosonF} and \eqref{eq:FtLR}\footnotemark.}\label{fig:Fextremisation}
\end{figure}
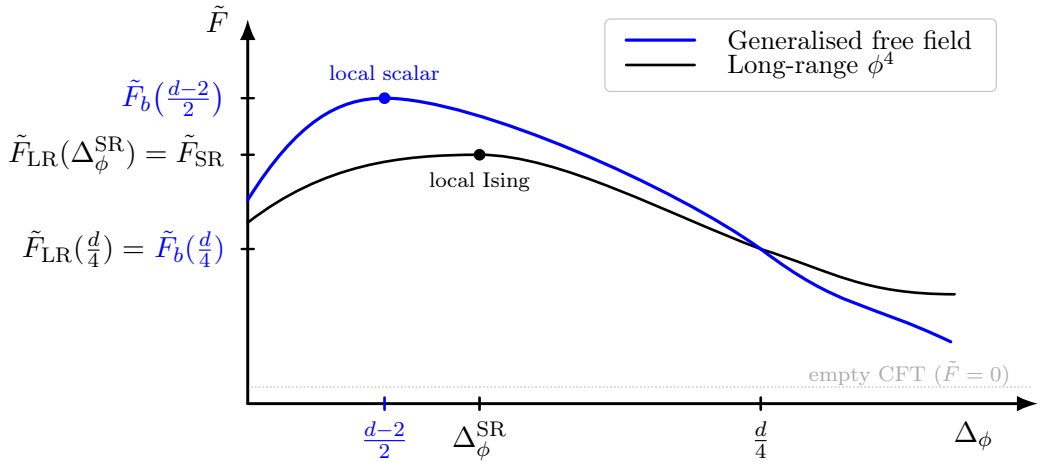

Crucially, \textit{from the long-range perspective there is no obvious signal that the local values of $\Delta$ are special}.
Put another way, \textit{if we did not know about the local Ising CFT already, solving the family described by \eqref{eq:LRaction} would naively not tell us that it exists.}
To remedy this, we prove nonlocal $F$-extremisation:
\begin{framed}
The long-range CFTs become local CFTs at the extrema of the universal part of the sphere free energy $\tilde{F}(\{\Delta_i\})$, i.e.
\begin{equation}
    \pdv{\tilde{F}}{\Delta_i} \Big\rvert_{\Delta_i=\Delta_i^\text{local CFT}} = 0.
\end{equation}
When these theories are unitary, they are local maxima of $\tilde{F}$; this is shown in \cref{fig:Fextremisation}.
\end{framed}
$\tilde{F}$ is perhaps the most important piece of conformal data.
It measures the number of effective degrees of freedom of a CFT: as a candidate weak $C$-function, under unitary RG flows it should only be possible to flow between two CFTs if $\tilde{F}_\text{UV} >\tilde{F}_\text{IR}$. 
As we shall see, this extremisation condition is quite natural -- and resembles known supersymmetric $\tilde{F}$-extremisation results.
In general, $\tilde{F}$ is the coefficient of the universal term in $-\log Z_{S^d}$, which it is natural to expect to be related to the number of degrees of freedom in any theory. 
In $d=2$ it is $\tfrac{\pi c}{6}$; in $d=3$ it is the finite part of $F$; and in $d=4$ it is $\frac{\pi a}{2}$.

\footnotetext{We provide further commentary on \cref{fig:Fextremisation} here.
The two CFTs cross (and are identical) at $d/4$, where $\phi^4$ becomes marginally irrelevant -- then $\phi^4$ becomes irrelevant in the GFF for $\Delta>d/4$.
The generalised $F$-theorem states that it is only possible to flow from higher to lower on this plot.
This makes it manifest why the scaling limit of the lattice long-range Ising gives mean-field behaviour for $s<\dotwo$, i.e. $\Delta>d/4$; it is no longer possible to flow to the WF theory. 
As we take $d \to 4$, the two local CFTs slide to $\Delta_\phi = d/4 \to 1$ and merge, due to the triviality of $\phi^4$ theory in $d=4$.
For numerical values in $d=3$, see \cref{sec:WFnumerics}.}
This long-range construction works for any number $n_f$ of fermions or bosons, giving an $n_f$-dimensional family of nonlocal CFTs parametrised by $\{\Delta_i\}$, which at discrete points become local CFTs. 
We therefore state our result for this maximally general case.
We freely consider nonunitary and complex CFTs, to avoid CFTs apparently popping in and out of existence as we vary $d$ and $\Delta_\phi$, and assume conformal invariance of these LR CFTs.

The proof of $F$-extremisation is remarkably simple, and follows rapidly in \cref{sec:proof} once we have set the problem up correctly: at a conformal fixed point, the derivative of $\tilde{F}$ with respect to $\Delta$ receives contributions only from the nonlocal term in the action, essentially because one-point functions of local operators vanish.
Manifestly, this nonlocal term must be absent in the local limit\footnote{An important caveat: we make the consistent choice to always define $\tilde{F}$ \textbf{without} the contribution of the auxiliary generalised free field $\chi$ that implements nonlocality (see \cref{sec:notincludechi} for a justification).\label{footnote:noChi}
} -- so the derivative vanishes, and we are at an extremum. 

The proof of maximisation is more laborious, and requires second-order conformal perturbation theory about the local CFT.
Despite this, for the unitary theories, we can offer some fuzzy intuition for maximisation, using the fact that $\tilde{F}$ is a weak $C$-function: \textit{if $\tilde{F}$ measures the number of effective degrees of freedom, then the local theory is the one with the largest $\tilde{F}$, as the values of the quantum fields are less dependent on what is happening at separated points, because the action only couples adjacent sites}.

The free energy of these nonlocal theories therefore provides a nice way to identify the local CFTs contained in a given long-range (LR) CFT.
It also simplifies the presentation of the conformal data in the cases that have been computed so far; see, for example, the large-$N$ case presented in \cref{sec:motivation} below. 
It is similar to supersymmetric $\tilde{F}$-maximisation, which is the general-$d$ version of $c,F,a$-maximisation that arises in SCFTs with sufficient supersymmetry (reviewed below).
However, unlike it, we stress that \textit{we still ultimately require perturbative control} to compute the conformal data of the local CFTs -- e.g. a large $N$ or a small $\epsilon$.
The general lesson that we find throughout, though, is that it can be useful to step back and consider broader nonlocal and nonunitary families of CFTs, even when our ultimate goal is understanding the local and unitary CFTs.

The rest of this paper proceeds as follows:
\begin{enumerate}[start=2]
    \item In \cref{sec:motivation}, we give some inspiration for how one might notice this phenomenon, as well as definitions and disambiguations of $\tilde{F}$ and $F$, Weyl covariance, conformal families, locality, and supersymmetric $\tilde{F}$-maximisation/extremisation.
    \item In \cref{sec:creatingLR}, we discuss the two known ways of creating long-range CFTs, which we term GFF$+$local and CFT$+\hat{\phi}\chi$.
    \item In \cref{sec:proof}, we prove that $F$ is extremised if and only if the LR CFT becomes local, by requiring locality of the action, as sketched above. We assume conformal invariance of the LR fixed points, and deliberately remove the contribution of the GFF $\chi$.
    \item In \cref{sec:genFinCPT}, we prove maximisation for unitary theories. 
    We first extend the generalised $F$-theorem in conformal perturbation theory (CPT) to second order. 
    This is used to show that if a given unitary local CFT can be deformed to a line of unitary long-range CFTs, then $\tilde{F}$ is locally maximised by the local CFT on that line (and the Hessian is negative definite assuming IR stability).
    To apply this to the GFF$+$local construction, we must assume the duality of the two LR constructions.
    We check with the free scalar, and also find a superior normalization for $\tilde{F}$, which we call $\hat{F}$. 
    \item In \cref{sec:perturbative}, we provide checks of our result.
    It can be shown in interacting theories where we have perturbative control: this has been done for the large-$N$ expansion of the vector models to order $1/N^2$ \cite{Fraser-Taliente:2025udk}.
    We therefore demonstrate it for the Wilson-Fisher $\epsilon$ expansion in the long-range $\gO(N)$ $\phi^4$ %
and (diagonal) $\sigma\phi^i\phi^i+\sigma^3$ cubic theories.
\item We close by discussing our results in \cref{sec:Discussion}.
\end{enumerate}

The manifold appendices contain supplementary material: \cref{app:conventions} contains our conventions; \cref{app:2PI} contains an alternate proof of a construction used in our proof; \cref{app:multiOperator} and \cref{app:Dsymmetric} contain a proof of the multi-operator generalisation of the generalised $F$-theorem and gradient flow in CPT to second order; %
\cref{app:DiagramValues} %
contains the free energy diagrams for the long-range $\phi^4$ theory, including a simplified form for a complicated double sum; \cref{app:FhatNumerics} shows the improvement of extrapolation between dimensions when using our new normalization $\hat{F}$ for the free energy; \cref{sec:subtlety} contains a review of the spectral continuity of the long-range Ising in the local limit and of the flows between the various long-range CFTs in the Ising model setting.

\section{Motivation (a suspicious fact) and definitions}\label{sec:motivation}

Let us observe the following suspicious fact, from \cite{Fraser-Taliente:2025udk}.
In the $\gO(N)$ $\phi^4$ CFT at large $N$, the scaling dimension of the fundamental field is
\begin{subequations} \label{eq:DeltaPhiSRoN2}
\begin{align}
\Delta^\mathrm{SR}_{\hat{\phi}} &\equiv \frac{d-2}{2} + \frac{\hat{\gamma}_{\phi,1}}{N} + \frac{\hat{\gamma}_{\phi,2}}{N^2}  + O(1/N^3),
\end{align}
where we have the unpleasantly complicated expressions:
\begin{align}%
\hat{\gamma}_{\phi,1} & %
= \frac{-2 \Gamma (d-2)}{\Gamma \left(2-\dotwo\right) \Gamma \left(\dotwo-2\right) \Gamma \left(\dotwo-1\right) \Gamma \left(\dotwo+1\right)}\\
\hat{\gamma}_{\phi,2} &= \frac{\hat{\gamma}_{\phi,1}}{2}  \Big[
    \Gamma _{\text{tet}}(2) \, \left(B(\dotwo -1)-2B(d -2)+B(2)\right) \\
    &\quad \quad+\Gamma_\mathrm{pr}(2) \,\left(B(\dotwo -1)-2B(d -2)+B(d-3)\right)\Big] \notag\\
    &-\hat{\gamma}_{\phi,1}^2 \left(\psi_0(\dotwo -1)-\psi_0\left(\dotwo + 1\right)-1+2 (B(2)-B(d-2))\right). \notag
\end{align}
\end{subequations}
We have used $A(x) \equiv \Gamma(\tfrac{d}{2}-x)/{\Gamma(x)}$ and $B(x) \equiv -\odv{}{x}\log A(x)$, and here
\begin{equation}\begin{aligned}
\Gamma_{\mathrm{tet}}(s) \equiv  -\frac{4}{\Gamma \left(\frac{d}{2}\right)} \frac{A\left(\frac{d-s}{2}\right)^2}{A(d-s)}, \quad \Gamma_{\mathrm{pr}}(s) \equiv \frac{8}{\Gamma \left(\frac{d}{2}\right)} A(s) A\left(d-\tfrac{3 s}{2}\right)  \frac{A\left(\frac{d-s}{2}\right)^3}{A(d-s)^2}.
\end{aligned}\end{equation}

\textit{Vaguely, this \enquote{looks like} it is related to the derivative of something, being made of a bunch of gamma and polygamma functions $\psi_0(x) \equiv \odv{}{x}\log \Gamma(x)$ with the same arguments.}
In \cite[\S 2.6]{Fraser-Taliente:2025udk}, we realized that this $\Delta_\phi$ is the extremal point of $\tilde{F}_\mathrm{LR}(\Delta)$ for the long-range vector model in the large-$N$ expansion.
In dimensional regularization (DREG), $\tilde{F} \equiv \sin(\tfrac{\pi d}{2}) \log Z_{S^d}$, and we find
\begin{align}\label{eq:largeNFt}
    \tilde{F}_{\text{LR}}(\Delta\equiv \tfrac{d-s}{2}) &= N \tilde{F}_b(\Delta) + \tilde{F}_b(s) + \frac{\tilde{F}_b'(s)}{N} (\tfrac{1}{2}\Gamma_\text{tet}(s) + \tfrac{1}{3}\Gamma_\text{pr}(s)) + O(1/N^2),\\
    \Delta  = \Delta^\mathrm{SR}_{\hat{\phi}} & \quad \text{solves} \quad \odv{\tilde{F}_\mathrm{LR}}{\Delta}=0,
\end{align}
where $\tilde{F}_b(\Delta)$ is a standard function giving $\tilde{F}$ for a Gaussian field of dimension $\Delta$, \eqref{eq:bosonF}. 
Hence, \eqref{eq:largeNFt} contains precisely the same information as \eqref{eq:DeltaPhiSRoN2}\footnote{In fact, it contains strictly more, since it is also extremised by the local higher-derivative theories with Lagrangian $\tfrac{Z_\phi}{2} \phi_i (-\partial^2)^k \phi_i + g_0(\phi_i \phi_i)^2$.
 Also, the rest of the conformal data can also be computed as a function of $s$, and the short-range CFT's data is slightly shorter to write down in that form.
For example, the scaling dimension of $\sigma \sim \phi^2$ in the local limit is $\Delta_\sigma^{\mathrm{SR}} = \Delta_\sigma\rvert_{\odv{\tilde{F}}{s}=0} = s + \frac{\Gamma_\text{tet}(s)+\Gamma_\text{pr}(s)}{N} +O(1/N^2)\rvert_{\odv{\tilde{F}}{s}=0}$.
Intuitively, this appears to be because we are removing complexity that only arises in the short-range limit, when we evaluate simple functions of $\Delta$ at the complicated local scaling dimension $\Delta^{\mathrm{SR}}$.
} -- a remarkable simplification.

This built on an earlier observation \cite{Fraser-Taliente:2024hzv} that certain large-$N$ (melonic) CFTs' scaling dimensions $\{\Delta_i\}$ can be identified at leading order by extremising $\tilde{F}$ over a family of CFTs parametrised by $\{\Delta_i\}$ subject to a constraint. 
These two works led to the natural questions:
\begin{enumerate}
    \item \enquote{Does this extremisation extend to more CFTs, to all orders, and non-perturbatively?}
    \item \enquote{What are those families?} 
\end{enumerate}
In this work we answer those two questions with \enquote{yes} and \enquote{they are long-range CFTs}, respectively, under some assumptions which we now explain.

\subsection{A full statement of \FtextOrPDF-extremisation}\label{sec:fullstatement}

Functionally, this means that in this paper we determine how to identify the local CFT from the perspective of the long-range CFTs.
Parametrising the family of long-range CFTs by the conformal scaling dimensions $\{\Delta_i\}$, we can compute (the universal part of) their sphere free energies
\begin{equation}
    \Ft_{\mathrm{LR}}(\{\Delta_i\}).
\end{equation}
We:
\begin{enumerate}
    \item Assume that the long-range theories are constructed either by perturbing scalar and fermionic generalised free fields (GFFs) with local operators (GFF$+$local), or by coupling a local CFT to a GFF (CFT$+\hat{\phi}\chi$).
    \item Assume that the long-range theories are indeed CFTs, not just scale-invariant FTs.
    \item Assume that these constructions give the same CFT (an \enquote{IR duality} \cite{Behan:2017emf}).
    \item Exclude, for consistency, the contribution to $\tilde{F}_\mathrm{LR}$ from the Gaussian fields $\chi$ that are explicitly present only in the CFT$+\hat{\phi}\chi$ construction (see \cref{sec:notincludechi}).
\end{enumerate}
Then, in \cref{sec:proof}, we prove that the local CFTs lie at the extrema,
\begin{equation}\label{eq:Fextred}
    \pdv{}{\Delta_i}\Ft_{\mathrm{LR}}(\{\Delta_j\}) \Big\rvert_{\text{local}} = 0,%
\end{equation}
where we precisely define a CFT as local if it has a spin-$2$ conserved operator of dimension $\Delta_T=d$ in its suitably defined OPE algebra (see \cref{sec:caveats}).
The converse also holds: if $\tilde{F}$ is extremised, then that point is a local CFT.
This extremisation then has the unexpected benefit of explaining the structure visible in the perturbative data above.

For the specific case of local unitary CFTs that can be deformed to unitary long-range CFTs, we find in \cref{sec:specialiseGF} that $\tilde{F}$ is not just extremised but locally\footnote{Local is used here in two different senses, unfortunately!} \textit{maximised}.
To show this, we only need to work perturbatively about the local CFT. 
Thus, we find
\begin{equation}
   \text{unitarity } \implies \,\partial_{\Delta}^2 \tilde{F}(\Delta)\rvert_{\text{local}} <0,
\end{equation}
in the one-field case. 
The specific requirement is the positivity of the two-point function of the long-range deformation and the reality of the action, which are guaranteed by unitarity. 
By the assumption of the IR duality, this holds for both constructions.

We also find a suitable generalisation if there are multiple long-range fields: along any line $\{\Delta_i(t)\}$ of real long-range CFTs that leads to a unitary short-range CFT, $\tilde{F}(\{\Delta_i(t)\})$ is maximised at the local CFT.
We have a stronger result in the more constrained case where we have a real nonlocal family, parametrised by $\Delta_i$, that is stable to the nonlocal deformations in the IR.
That is, if all of the nonlocal operators become irrelevant in the IR near the local CFT, then the Hessian at the local point, $\partial_{\Delta_i} \partial_{\Delta_j}\tilde{F}|_{\mathrm{SR}}$, is negative definite -- and so $\tilde{F}(\{\Delta_i\})$ is at a multidimensional local maximum. 

\subsection{The result should not be surprising}

This observation of $F$-extremisation should not be surprising. 
We have taken a local CFT, generalised it to a line of nonlocal CFTs, and asked how the original CFT is special.
An extremisation principle is natural, and the free energy is the most natural single number used to measure a CFT.
The fact that we find a simpler representation of the scaling dimension of the fundamental fields \textit{is} a little surprising, and very neat.

\subsection{Definitions and disambiguations}\label{sec:caveats}

We end this section with a few definitions and disambiguations, as well as some comments on supersymmetric $F$-extremisation.

\newcommand{\bigslant}[2]{{\raisebox{.2em}{$#1$}\left/\raisebox{-.2em}{$#2$}\right.}}
\textbf{DREG and $\tilde{F}$}. 
We define \enquote{free energy} to mean the universal part of the sphere free energy of the CFT, i.e. schematically
\begin{equation}\label{eq:FmodCT}
    \tilde{F} \; \propto \; {F} \Big/ {\text{ local counterterms}}\; \equiv \; \text{\enquote{$F$ mod local counterterms}}.
\end{equation}
More precisely, let us work in a locally generally-covariant regularization scheme in integer $d$. 
The critical sphere free energy admits the following expansion as the cutoff $\Lambda$ is removed
\begin{equation}
   -\log Z_{S^d}= \underbrace{\sum_{n=0}^{\floor{\tfrac{d}{2}}} b_n (\Lambda R)^{d-2n}}_{\text{UV details}} +\begin{cases} (-1)^{\dotwo} \frac{2}{\pi} \Ft \log(\Lambda  R)  & d \text{ even} \\
	(-1)^{\frac{d+1}{2}}\Ft & d \text{ odd} \end{cases}.
\end{equation}
$\tilde{F}$ is the only part that cannot be freely modified\footnote{If the CFT is supersymmetric the possible counterterms are more constrained, so the other parts of $F$ may become meaningful \cite{Gerchkovitz:2014gta}.
Also, in theories that are not parity invariant (such as 3d Chern-Simons theories, which we do not consider here), $Z_{S^d}$ can have a phase \cite{Jafferis:2010un,Jafferis:2011zi,Klebanov:2011gs}, in which case we must take the real part of the logarithm (equivalently, we define $\tilde{F} \equiv \sin(\pi d/2) \log \abs{Z_{S^d}}$).
}, justifying our schematic \eqref{eq:FmodCT}.
This is because the $b_n$s can be shifted by adding local terms built from the metric (see \cite{Ahmadain:2024hdp} for related discussion in the context of off-shell string theory): schematically, these are $\sim\Lambda^{d-2n-2m}\int \sqrt{g}\nabla^{2m} \cR^n$, which on the sphere scale like $(\Lambda R)^{d-2n-2m}$.
Hence, $\tilde{F}$ is the only part that has a universal meaning, as the $b_n$s depend on UV regularization details. 
We can think of the $b_n$s as encoding the degrees of freedom of that UV regularization, which we do not care about, as they are gapped out in the conformal limit: thus, we say that $\tilde{F}$ counts the number of effective CFT degrees of freedom.

We will therefore refer to $F$, $\tilde{F}$ (and our new normalization $\hat{F}$, which we discuss in \cref{sec:proposedFhat}) as \enquote{the free energy}, moving freely between them and ignoring their overall normalization -- unless we need to know about their signs, in which case we must specialise to $\tilde{F}$ or $\hat{F}$.
In practice, this means that in this paper we compute $\tilde{F}$ by computing the partition function in dimensional regularization\footnote{Further regularization may be required on a model-by-model basis, such as for the large-$N$ theories in \cite{Fraser-Taliente:2025udk,Fraser-Taliente2026ZT} (see also \cite{Vasiliev:1975mq,Ciuchini:1999wy,Tarnopolsky:2016vvd}).} (DREG, nicely described by \cite[\S 4]{Collins:1984xc}), which automatically tunes away those counterterm-dependent power-law divergences $\propto b_n$ in the free energy. 
Thus, we define
\begin{align}
\tilde{F} &\equiv - \sin(\pi d/2) F = \sin(\pi d/2)\log Z_{S^d}. %
\end{align}
In the limit of even $d=2n-\epsilon$ the zero of the prefactor cancels with the $\frac{1}{\epsilon}$ in $-\log Z_{S^d}$ (the analogue of $\log \Lambda$ in DREG, which comes from the Weyl anomaly), ensuring that $\tilde{F}$ is finite. 
As mentioned above, in $d=2$ we find $\tilde{F}=\frac{\pi}{6}c$, and in $d=4$ it is $\tilde{F}=\frac{\pi}{2}a$.
In $d=3$, $\tilde{F}$ is just the finite part of the sphere free energy $F$ \cite{Pufu:2016zxm}. 
It also admits a smooth analytic continuation in $d$, justifying the title of \cite{Giombi:2014xxa}. 
We therefore work in continuous $d$ throughout this paper (see \cite[\S 2.2.5]{Fraser-Taliente:2025qcl} for more discussion of this). %

$\tilde{F}$ is the quantity that is thought to be a weak $C$-function \cite{Barnes:2004jj,Gukov:2015qea,Nishioka:2018khk}, in the sense of counting the effective number of degrees of freedom: in RG flow between unitary CFTs they should always satisfy $\tilde{F}_\text{UV} > \tilde{F}_\text{IR}$ (this has been proven by constructing monotonic quantities that interpolate between these two -- but that are not equal to $\tilde{F}$ away from fixed points -- in $d=2,3,4$ for \cite[$c$]{Zamolodchikov:1986c},\cite[$F$]{Casini:2015woa},\cite[$a$]{Komargodski:2011vj,Komargodski:2011xv,Luty:2012ww} \cite{Cardy:1988cwa}; but not yet in $d=5,6$ \cite{Heckman:2015axa,Fluder:2020pym,Kundu:2019zsl} or continuous $d$ \cite{Giombi:2014xxa,Fei:2015oha}).
As mentioned above, however, throughout this paper we treat all CFTs (real and complex, according to the classification of \cite{Gorbenko:2018ncu}; unitary and nonunitary) on an equal footing.
This allows us to consider the long-range CFTs as existing for all values of $\Delta$, rather than only in the finite window of $\Delta$s for which they are stable and unitary \cite{Behan:2017emf}.

\textbf{Locality (or lack thereof)}. 
We have been fairly blithely discussing locality of a CFT above.
If one has an action, the notion of locality is obvious: a theory is local if its action consists of an integral of a Lagrangian density that depends only on fields and (a finite number of) their derivatives at a single point\footnote{
There also exist apparently nonlocal CFTs which are secretly local, in the sense that they can be made manifestly local by introducing massless degrees of freedom; one such is the action of massless QED after integrating out the massless fermions \cite{Giombi:2015haa,Giombi:2016fct,Fraser-Taliente:2024lea,Fraser-Taliente2026qed}. %
}. 
Additionally, if its action is local, then its variation with respect to the metric should be local as well -- that is, the theory should have a local stress tensor.
We will use this fact to single out the local theories in a family of nonlocal theories, as we know that exactly at those points the action should not contain any nonlocal operators.
This will be crucial for our proof in \cref{sec:proof} that the local CFTs are precisely those which extremise $\tilde{F}$ in these families.

If handed a set of CFT data, it is not immediately obvious whether a CFT is local. 
One must begin with an operator $\phi$ which we define to be local, and a CFT is then $\phi$-local (local with respect to $\phi$) if the closure of the OPE algebra of that operator contains a conserved spin-2 operator $T^{\mu\nu}$ with scaling dimension $\Delta_T =d$, the stress tensor \cite{Fraser-Taliente:2026gdh}.
In our multidimensional long-range theories, it does not matter which $\phi$ we start with, since we assume that none of the fundamental fields are decoupled -- and if any of them are long-range, the stress tensor will not be local.
This is what we call a local stress tensor.

The defining feature of these nonlocal CFTs is that they are nonlocal, and therefore do not have a (1) conserved spin-$2$ operator or (2) the conserved currents for the global symmetries in the closure of the OPE algebra generated by their local operators \cite{Rong:2024vxo} (recall that the presence of a traceless stress-energy tensor is only a sufficient condition for having a CFT \cite[\S 4.1]{behanBootstrappingContinuousFamilies2019}).
That said, they still have some nonlocal stress tensors/global currents, which are conserved but nonlocal in the sense that they do not appear in the OPE. This is obvious given that in known examples \cite{Fraser-Taliente:2026gdh,Fraser-Taliente2026ZT} they contain fractional derivatives.
These nonlocal CFTs are also often called \textbf{conformal theories} (CTs) \cite{behanBootstrappingContinuousFamilies2019} \cite{Schwarz:2015fva,Paulos:2016fapa} \cite[\S 3.2]{Heemskerk:2009pn}, but we will stick to calling them nonlocal CFTs.

\textbf{Not-conformal-manifolds}. Our families of nonlocal CFTs (often called \enquote{conformal theories} (CTs) to distinguish them from CFTs, which in that convention are only those theories possessing a local stress tensor) are not a \enquote{line of CFTs} in the standard sense of a conformal manifold \cite[footnote 1]{Behan:2018hfx} \cite{Komatsu:2025cai}.
This is most clear when we see that the value of $\Ft$ is not constant along the family. %
To be precise, 
\begin{equation}\label{eq:notLocalOperator}
\cO = \phi\odv{(-\partial^2)^{\dotwo-\Delta}}{\Delta}\phi \Big\rvert_{\Delta=\Delta_\phi}
\end{equation}
is not a marginal local operator (it is definitely not local, for one thing). %
Our families are then continuous parametrisations of conformal fixed points in the sense of the $\epsilon$ expansion or the large-$N$ limit, rather than being conformal manifolds\footnote{Thus, it continues to be hard \cite[\S 2]{Perlmutter:2020buo} to find any non-supersymmetric non-defect conformal manifolds \cite{Gaberdiel:2008fn, Bashmakov:2017rko, Behan:2017mwi,Sen:2017gfr,Hollands:2017chb} (however, see \cite{Chaudhuri:2020xxb,Nakayama:2021fgy} for a large-$N$ $d=4$ construction; \cite{Giambrone:2021wsm} and follow-ups for a holographic construction; or \cite{Chen:2026jla} for a construction in QCD).}.

\textbf{Weyl covariance}. When we say \enquote{the CFT}, we refer to the CFT defined on arbitrary manifolds, not just in flat space.
We demand Weyl invariance of the CFT, meaning that we can compute all correlators on the sphere just by Weyl mapping from flat space, and those correlators transform covariantly (see \cite{Stergiou:2022qqj,Parisini:2023nbd,Konechny:2026bqg} for more discussion of this).
As discussed in \cite{Fraser-Taliente:2026gdh}, this is possible for generic (real) $d$, except in even $d$ where it is sometimes possible for local CFTs and never possible for nonlocal CFTs due to Weyl obstructions (not to be confused with Weyl anomalies, which do not prevent Weyl mapping CFTs \cite[\S 2.3.2]{Fraser-Taliente:2026gdh} \cite[\S 4.2]{Farnsworth:2017tbz} \cite{Baume:2014rla}).
The consequence is that $\tilde{F}$ is a number that we can use to characterise the CFT independently of the manifold.

\textbf{The vector SCFT and \FttextOrPDF-extremisation/maximisation}. %
Our story is very similar to that of supersymmetric $\Ft$-extremisation (also called $c,F,a$-maximisation in $d=2,3,4$ \cite{Intriligator:2003jj,Barnes:2004jj,Benini:2012cz,Benini:2013cda,Pufu:2016zxm,Jafferis:2010un}), although admittedly less powerful. 
For comparison, we now briefly review it in the context of the supersymmetric vector model, following \cite{Giombi:2014xxa}. %
In the supersymmetric vector model, there are two superfields, $\Phi_{i=1,\cdots,N}$ and $\Sigma$.
The $R$-charges at the conformal IR fixed point are determined by extremising the free energy over a family of trial QFTs, each of which has free energy
\begin{equation}
\tilde{F}_{\text{SUSY}}(\Delta_\Phi)\equiv N \tilde{F}_S(\Delta_\Phi) +\tilde{F}_S(\Delta_\Sigma=d-1-2\Delta_\Phi), \label{eq:SUSYvecF}
\end{equation}
for standard functions $\tilde{F}_S$. The BPS condition
\begin{equation}
\Delta_X= \frac{d-1}{2} R_X
\end{equation}
then provides the relationship between the trial $R$-charges and the scaling dimensions in \eqref{eq:SUSYvecF}.

The dimension of $\Sigma$ is fixed to be $d-1-2\Delta_\Phi$ by supersymmetry (specifically, by the requirement that the $R$-charge of the superpotential $W=\frac{\lambda}{2} \Sigma \Phi^i \Phi^i$ is $2$); the actual value of $\Delta_\Phi$ is then the one satisfying
\begin{equation}
    \odv{\tilde{F}_{\text{SUSY}}}{\Delta_\Phi} =0.
\end{equation}
The generalisation of this mechanism to more superfields is straightforward. That is, in any SCFT with four supercharges, several chiral superfields (without gauge fields), and some superpotential $W$, we find that the IR R-charges are determined by maximisation of
\begin{equation}\label{eq:SUSYF}
    \tilde{F}_{\text{SUSY}} = \sum_{X \text{chirals}}\tilde{F}_{S}(\Delta_X),
\end{equation}
subject to the constraint that the superpotential has exact $R$-charge $2$. %
Furthermore, in unitary CFTs, this extremum of $\tilde{F}$ is in fact a maximum, which follows from (reflection) positivity of the two-point function \cite[\S 5.2]{Giombi:2014xxa}.

The end result is extremely similar to our mechanism: our long-range fields map to the unconstrained superfields in the supersymmetric mechanism, we extremise $\tilde{F}$ over a family of QFTs, and in the unitary case maximisation follows from the positivity of the two-point function in a unitary theory. 
There are, however, three notable differences.
\begin{itemize}
\item \eqref{eq:SUSYF} is an exact result for any $\Delta_X$, which follows from localisation (in $d=3$) or anomaly matching (in $d=2,4$).
In the long-range theories we typically only have perturbative access to $\tilde{F}$.
\item Not all of the members of the family of trial QFTs over which $\tilde{F}$ is extremised in the SUSY mechanism possess superconformal symmetry -- that is unlike our situation, where all long-range QFTs are conformal.
\item In general, the IR SCFT lies at an extremum because the derivative with respect to the $R$-charge yields an (integrated) one-point function -- which necessarily vanishes in a CFT. 
In nonlocal $F$-extremisation the derivative yields a term proportional to the one-point function of a nonlocal operator in a CFT, which only vanishes in a local CFT.  
\end{itemize}

\section{Creating long-range theories}\label{sec:creatingLR}

To understand $F$-extremisation, we first must understand how to construct long-range CFTs.
They are well-studied, from Fisher, Ma, and Nickel's early work \cite{Fisher:1972zz} through
to the present day \cite{Paulos:2015jfa,Behan:2017dwr,Behan:2017emf,Gubser:2017vgc,Behan:2018hfx,Giombi:2019enr,Basa:2020cyn,Chai:2021arp,Behan:2023ile,Benedetti:2023pbt,Benedetti:2024wgx,Benedetti:2025nzp,Eustachon:2026vjn,Pagni:2025ebi,Ghosh:2026gku,Li:2024uac}. 
There are (at least) two different ways to construct them, which we name
\begin{enumerate}
    \item the GFF$+$local route;
    \item and the CFT$+\hat{\phi}\chi$ route.
\end{enumerate}
Both lead to the same type of action, which is what we are interested in here. %
We stress that though these theories are nonlocal, they are nonlocal in a special way. 
For example, as stressed by \cite{Paulos:2015jfa}, they can be realized as local defects in a local higher-dimensional theory after integrating out the transverse directions.
Throughout this paper we simply assume the conformal invariance of these long-range fixed points. 
We cite the strong argument given in \cite{Paulos:2015jfa} for the conformal invariance of these long-range QFTs to all orders in perturbation theory, assuming it to generalise beyond the long-range Ising model.
That said, their argument, which used the realization of the LR CFT as a defect theory in an auxiliary higher-dimensional space, strictly holds only in perturbation theory around $\Delta=\frac{d-\eta}{4}$, with $\frac{d}{4} >\Delta >\Delta_{\phi,\text{Ising}}$.

We also freely consider nonunitary and complex fixed points, which we know exist for generic values of $d,N$ and the long-range parameter.
It will be convenient to assume a Lagrangian formulation of the original CFT later, but we do not assume that it is at weak coupling or controlled, in the sense of possessing a small number parametrising the size of the perturbation around a solvable theory.

\subsection{The GFF+local route}\label{sec:GFFplusLocal}

In the first route, which is the one with which we began this paper, we take the nonlocal CFT of a generalised free field $\phi$ (GFF, also called a mean field theory, or MFT) of dimension $\Delta_\phi$, and perturb by all of the relevant or marginal local scalar operators $\cO_g$. 
These $\cO_g$s are the relevant/marginal scalars appearing in the closure of the OPE algebra generated by $\phi$ \cite{Fraser-Taliente:2026gdh}.
We therefore call this the GFF$+$local construction.
Making sure to regularize and renormalize, 
\begin{align}\label{eq:Saction}
S_g[\phi] = \int_x \left[\frac{Z_\phi}{2} \phi \WeylOp_{\dotwo -\Delta_\phi} \phi + \sum_g g_0 \cO_g\right],
\end{align}
we find fixed points where all $\beta_{g_\star}^\text{GFF}=0$.
In this, $\WeylOp_{\dotwo-\Delta_\phi}$ is the Weyl-covariant fractional Laplacian, which in flat space is exactly $(-\partial^2)^{\dotwo-\Delta_\phi}$ \cite{Fraser-Taliente:2026gdh}. 
Because we ultimately work on the sphere, we use $\WeylOp$. 
If $\phi$ is a fermionic GFF, then we overload notation such that it is the Weyl-covariant fractional Dirac operator \cite{maalaouiConformalFractionalDirac2025}.

In the Wilsonian RG, divergences occur when two operator insertions approach each other; they therefore generate only local operators.
This means that nonlocal operators are not renormalized, and so the dimension of $\phi$ is protected \cite{Lohmann:2017qyq}.
Thus, the CFT scaling dimension of $\phi$ is $\Delta_\phi$ in the IR\footnote{
The fact that the anomalous dimension of $\phi$ is zero is also derivable from: (1) the construction of the LR CFT as a defect CFT via the Caffarelli-Silvestre trick \cite[footnote 26, \S 4.3]{Behan:2017emf}; and (2) the bulk equations of motion and the bulk-to-defect OPE, assuming conformal invariance.
}.
Of course, the scaling dimensions of all other operators in the theory will be modified from their values in the initial GFF theory.
Defining
\begin{equation}
X(x)\equiv \frac{1}{Z_\phi} \fdv{}{\phi(x)} \int_y \sum_g g_0 \cO_g,
\end{equation}
the nonlocal equation of motion for $\phi$,
\begin{equation}
    \WeylOp_{\dotwo -\Delta_\phi} \phi = X,
\end{equation}
which is nonlocal, means that the scaling dimension $\Delta_X = d-\Delta_\phi$ is also fixed in the IR. 
This is easily checked using the equation of motion on both $\phi$s in the conformal propagator $\expval{\phi(x)\phi(y)}_{g_\star}$.

For $\Delta_\phi < \dotwo -1$ the local kinetic term $\phi \WeylOp_1 \phi$ becomes relevant. 
It therefore must be added to the action and tuned to ensure that we reach a fixed point, as noted in \cite[footnote 3]{Behan:2017emf}, just as we must also typically tune $\phi^2$. 
More generally, if $\Delta_\phi < \dotwo -k$ we require the relevant term $\phi \WeylOp_k \phi$ in the action for any $k\ge0$, as well as any other operators that become relevant, as in \cite{Safari:2017tgs}.
Identical comments apply to fermionic GFFs.

\subsection{CFT\texorpdfstring{$+\hat{\phi}\chi$}{+phihat chi} flow}\label{sec:CFTplusPhihatChi}

Separately, \cite[\S 6]{Behan:2017emf} presented a simple recipe for constructing a long-range CFT from any short-range CFT (for which we do not require a Lagrangian description \cite[(1.1)]{Behan:2017mwi}).
Taking the original theory to be encoded by $S_\mathrm{CFT}$, we:
\begin{enumerate} 
    \item Pick a CFT operator $\hat{\phi}$\footnote{For notational simplicity, we here assume that the operator that we choose is related to the field integrated over in the path integral -- this is not necessary, as we will see in \cref{sec:otherNonlocalInteractions} (see also \cite{Behan:2025ydd}). %
    } of dimension $\Delta^{\mathrm{SR}}$, and couple it to a nonlocal GFF $\chi$ (of matching bosonic/fermionic statistics) with $\Delta_\chi \equiv d-\Delta^{\mathrm{SR}}-\eta$ for some $\eta \ll 1$:
    \begin{align}
    S_\lambda^{\mathrm{LR,\chi}} =  S_{\text{CFT}} + \frac{Z_\chi}{2}\int_x \chi \WeylOp_{\dotwo -\Delta_\chi} \chi + \lambda_0 \int_x \hat{\phi}\chi,
    \end{align}
    for $\lambda_0=Z_\lambda \lambda \mu^\eta$. 
    The equation of motion of $\chi$ is then
    \begin{equation}\label{eq:chiEOM}
        \WeylOp_{\dotwo - \Delta_\chi} \chi = -\frac{\lambda_0}{Z_\chi}\hat{\phi}.
    \end{equation}
    \item Compute the quantum corrections to the beta function, as described in \cite[\S 2]{Behan:2017emf},
    \begin{equation}
    \beta_\lambda =-\eta \lambda + \beta_3 \lambda^3+ O(\lambda^5) \label{eq:CFTflowBeta}.
    \end{equation}
    The coefficient of $\lambda^2$ in the $\beta$ function is zero because it is proportional to $C_{\cO\cO\cO} \rvert_{\lambda=0}=0$ for $\cO=\hat{\phi}\chi$ (by the standard rules of conformal perturbation theory \cite{Komargodski:2016auf} \cite{Gaberdiel:2008fn,Amoretti:2017aze}).
    Since for $\lambda_0=0$ the theory has a $\chi \to -\chi$ symmetry, all even powers of $\lambda$ in the beta function must vanish.
    \item If $\beta_3$ has the correct sign, then we find a controlled IR fixed point at weak coupling $\lambda_\star$\footnote{Otherwise, we find a possibly complex IR/UV fixed point, which we can still consider as existing formally as a CFT that is not physically reachable but nonetheless is a consistent set of CFT data.}.
    We presume that the same conclusions continue to hold for arbitrary $\eta$\footnote{We need to assume $\eta \neq d/2 + k -\Delta^\mathrm{SR}$ for $k \in \mathbb{Z}_{\ge 0}$, in which case $\Delta_\chi=\frac{d}{2}-k$, so $\chi$ is either a local free $\Box^k$ scalar or a non-dynamical Hubbard-Stratonovich field, as noted in \cite[around (1.2)]{Behan:2025ydd}.}.
    \item The IR scaling dimension of $\chi$ is protected by the long-range interaction (though the two-point function will generically be non-conformal along the flow, it returns to being $\Delta_\chi$ in the IR). 
    Thus, applying the equation of motion \eqref{eq:chiEOM} inside $\expval{\chi\chi}$, in the IR we find a new exact scaling dimension for the operator $\hat{\phi}$:
    \begin{equation}\label{eq:chiRelation}
        \Delta_{\hat{\phi}} = d-\Delta_\chi = \Delta^\mathrm{SR}+\eta.
    \end{equation}
\end{enumerate}
Both of these routes trivially generalise to the presence of more nonlocal fields (in the GFF route, we just add another GFF; if using CFT$+\hat{\phi}\chi$, we require one nonlocal $\chi$ for every $\hat{\phi}$ we want to make nonlocal).
However, as mentioned in \cref{footnote:noChi}, we view $\chi$ as just an auxiliary field implementing the nonlocal kinetic term for $\hat{\phi}$, and do not include its contribution to $F$. 
This will be discussed further in \cref{sec:notincludechi}.

\subsection{A duality at fixed points}\label{sec:conjecturalDuality}

It is not immediately obvious that these two routes end up defining the same CFT. 
Indeed, they are not the same quantum field theories away from the fixed point; it is now well established that generically they will lie in the same universality class, in that they will flow to the same IR CFT.
This was the argument of \cite{Behan:2017emf,Behan:2017dwr}; note, however that a full proof has not yet been provided.

We can give a heuristic argument for this in the simple context where the operator $\hat{\phi}$ descends from the fundamental field of a QFT. 
After integrating out the quadratic $\chi$ field in the CFT+$\hat{\phi}\chi$ route we find the \enquote{Sak-style} nonlocal perturbation\footnote{ 
Again, we have pre-emptively normalized our definition of the partition function $Z_\mathrm{LR}$ to automatically cancel the $F$ contribution from integrating out $\chi$ -- something that we will discuss further around \eqref{eq:Fwithchi}.},
\begin{align}
   Z_{\mathrm{LR}} &\equiv e^{F_b(\Delta_\chi)}\int \Dd\chi \Dd{\hat{\phi}} \,e^{-S_\lambda^{\mathrm{LR,\chi}}} = \int \Dd{\hat{\phi}}\, e^{-S_\lambda^{\mathrm{LR,\text{no }\chi}}} \label{eq:ZLRwithAndWithoutChi}\\
   S_\lambda^{\mathrm{LR,\text{no }\chi}} &\equiv S_{\text{CFT}} - \frac{\lambda_0^2}{2 Z_\chi} \int_x \hat{\phi} \WeylOp_{\Delta_\chi-\dotwo} \hat{\phi} \label{eq:SLRnoChi},
\end{align}
which as expected fixes the scaling dimension of $\hat{\phi}$ to exactly $d-\Delta_\chi \equiv \Delta_\phi$ in the IR (assuming that we tune away any dangerous relevant operators like $\hat{\phi}^2$ that would otherwise destabilize the fixed point).

Following the standard rules of quantum field theory, we should add all possible relevant or marginal scalar operators to this action (that is, any given regularization of this theory will not necessarily uniquely pick out the long-range operator).
We can interpret this also as the couplings in the Lagrangian definition of $S_\text{CFT}$ starting to flow.

The $S_\mathrm{CFT}$ is local. For simplicity, let us assume that $\hat{\phi}$ is the fundamental field, so
\begin{align}
S_\mathrm{CFT} \equiv S^\mathrm{SR}_{g_\star}, \quad S^\mathrm{SR}_g \equiv \int_x \sum_g g_0 \hat{\cO}_g,
\end{align}
where the operators $\hat{\cO}_g$ all appear in the closure of the OPE algebra generated by $\hat{\phi}$.
Hence, in the CFT+$\hat{\phi}\chi$ route we seek to find fixed points of 
\begin{equation}\label{eq:SLRnoChiGlam}
   S_{\lambda,g}^{\mathrm{LR,\text{no }\chi}}   \equiv \int_x  - \frac{\lambda_0^2}{2 Z_\chi} \hat{\phi} \WeylOp_{\dotwo-\Delta_\phi} \hat{\phi} +\sum_g g_0 \hat{\cO}_g.
\end{equation}
But we now recognise exactly the recipe that gave us \eqref{eq:Saction}, up to $Z_\phi \equiv -\frac{\lambda_{0}^2}{Z_\chi}$\footnote{As noted in \cite[Footnote 8]{Behan:2017emf} \cite[(9)]{Eustachon:2026vjn}, for real $\lambda$ the generated operator in \eqref{eq:SLRnoChi} and \eqref{eq:SLRnoChiGlam} has the correct ferromagnetic (negative) sign -- this is just like how in the ferromagnetic Ising Hamiltonian we have $H=-J\sum s_i s_j$ with $J>0$.
The negative sign cannot be dangerous here for two reasons: firstly, we could assume that the original theory of a CFT and a decoupled GFF was healthy; secondly, even if it were not healthy, we have permitted ourselves to freely consider complex and nonunitary CFTs. 
}. 
Thus, the beta functions should be the same set of equations. 
Therefore, intuitively but nonrigorously, there should be a one-to-one map between the solutions of these equations, i.e. the fixed points of these theories. %
For this reason, it was called an IR duality by \cite{Behan:2017emf}; though the QFTs will generically differ, the fixed points should be the same.

We explain how this duality works further in \cref{sec:subtlety}. 
The most important takeaway is that regardless of which construction is used, the form of the action is the same (assuming we integrate out $\chi$): either way, we just have \textbf{a long-range kinetic term and then a collection of integrated local operators}, which can be taken to be only primaries.

\subsection{Generalisation: other nonlocal interactions and more fields}\label{sec:otherNonlocalInteractions}

Since we are looking for conformal theories, it makes sense to restrict to actions with power-law-like interactions which will survive the conformal limit.
Though not all such theories take the manifest form GFF+$\text{local}$, it is worth noting that some apparent nonlocal interactions can be made local by coupling the theory to an auxiliary field. 
For example, consider the presence of an interaction
\begin{align}\label{eq:nonlocalPhi2Phi2}
    S_{\text{int}} \supset \int_{x,y} \phi^2(x) \frac{1}{\abs{x-y}^{2\alpha}} \phi^2(y)
\end{align}
which naively might not appear to be in the class of theories that we consider.
However, by adding a long-range auxiliary field $\sigma(x)$ of the correct dimension $\alpha$, this can be made into the GFF$+$local long-range form above:
\begin{equation}\label{eq:SwithSigma}
S_{\text{int}} \supset \int_x \half \sigma \,\WeylOp_{\dotwo-\alpha} \,\sigma + g_0 \sigma\phi^2.
\end{equation}
Integrating $\sigma$ out restores the interaction \eqref{eq:nonlocalPhi2Phi2}.

The generalisation of the argument for the duality in \cref{sec:conjecturalDuality} to multiple fundamental fields is trivial. 
It remains only to consider the case where $\hat{\phi}$ is not a fundamental field. 
This is precisely the CFT$+\hat{\phi}\chi$ side of the same phenomenon that we just described at the start of this subsection.
The resolution is almost identical: we can introduce an auxiliary Hubbard-Stratonovich field to make $\hat{\phi}$ \enquote{act} like a fundamental field. We may need to match it to a different GFF (i.e. one explicitly including more nonlocal fields like \eqref{eq:SwithSigma}), but the same argument as above should then go through. 

\subsection{Rong's approach: finding a conserved \texorpdfstring{$T^{\mu\nu}$}{T\^mn}} %

The same question of identifying the short-range CFT from the long-range perspective (without just assuming knowledge of $\Delta^{\mathrm{SR}}_{\hat{\phi}}$) was considered by Rong in \cite{Rong:2024vxo}.
The approach was to compute the scaling dimension of the lowest spin-2 operator $\tilde{T}^{\mu\nu}$ in the GFF+$\text{local}$ theory. %
In the limit $\Delta \to \Delta^\mathrm{SR}$, it ought to have protected dimension $\Delta_{T^{\mu\nu}}\rvert_{\Delta^\mathrm{SR}} = d$, as it becomes the stress tensor of the local CFT sector of the full theory of a decoupled CFT$+$GFF\footnote{Naturally, the $\chi$ field does not have a local stress tensor. 
This is not an obstruction since the theories are decoupled.}. 
Alternatively, one can compute the scaling dimension of the spin-$1$ operator which at the fixed point becomes the $\gO(N)$ current $J^{[ij]}_\mu = Z_\phi \hat{\phi}^i (\overleftarrow{\partial}_\mu -\overrightarrow{\partial}_\mu) \hat{\phi}^j$ (in the adjoint rep of $\gO(N)$), with conserved dimension $\Delta_{J_\mu}\rvert_{\Delta^\mathrm{SR}} = d-1$. 
Locating a perturbative solution to either of these conditions, %
which requires an expansion in $d=4-\epsilon$, we can recover exactly the short-range $\Delta^{\mathrm{SR}}_{d=4-\epsilon}$.

This suffices for the standard Wilson-Fisher fixed point. 
However, this approach is blind to the other nonunitary CFTs that arise as $\Delta$ is varied. 
To detect the interacting $\Box^k$ theory, one must compute instead the scaling dimension of the operators that are schematically $T^{\mu\nu} \sim \hat{\phi} \partial_\mu \partial_\nu (-\partial^2)^{k-1} \hat{\phi} + g_{\mu\nu} (\cdots)$ and $J^{[ij]}_\mu \sim \hat{\phi}^{[i} \partial_\mu (-\partial^2)^{k-1}\hat{\phi}^{j]}$ (the derivatives must be distributed carefully between the two fields, as shown in \cite{Fraser-Taliente:2026gdh}).
The advantages of our approach are that: (1) we do not need to renormalize any composite operators; (2) 
our expressions for $\hat{F}$ are generally more compact than $\Delta_{T,J}$ \cite[(2.5)]{Rong:2024vxo};
and (3) we treat all the local CFTs that arise from a long-range CFT democratically.

\section{Proof of \FtextOrPDF-extremisation}\label{sec:proof}

Our proof of $F$-extremisation as the long-range CFTs become short-range is straightforward, under the assumptions of \cref{sec:fullstatement}.
First, we observe that the derivative of $\tilde{F}$ of the nonlocal theory with respect to $\Delta_\phi$ gets a contribution only from the explicitly nonlocal term.
Second, if a short-range limit of the long-range CFT exists, the explicitly nonlocal term must be absent.
Hence, that derivative should be zero.
Flipping this round, $\tilde{F}$ can only be extremised by local CFTs, as we will see that all nonlocal terms must vanish to find an extremum.

In slightly more detail, defining $F_\mathrm{LR}(\Delta_\phi)$ as the free energy of the long-range CFT (\textit{without} the contribution from $\chi$ if using the CFT$+\hat{\phi}\chi$ construction), then
\begin{enumerate}
    \item First, %
    we observe that the derivative of the interacting theory with respect to the long-range scaling dimension receives a contribution only from the long-range kinetic term. 
    Hence, it is proportional to the ratio of the two-point function normalizations of the interacting theory and free theory, 
\begin{equation}\label{eq:dFdDelEqCphi}
\pdv{F_{\mathrm{LR}}}{\Delta_\phi} = \frac{C_{\phi}}{C_{\phi,0}} F_\phi'(\Delta_\phi).
\end{equation}
The factor $F_{\phi}'(\Delta_\phi)$ is just the derivative of the free energy of a generalised free field $\phi$ of dimension $\Delta_\phi$.
\item Second, we assume that the CFT becomes local at some point. Then, the locality of the action of the local CFT tells us that as we tune all of the nonlocal scaling dimensions $\{\Delta_i\}$ to their short-range values $\{\Delta^{\mathrm{SR}}_{i}\}$,
\begin{equation}
    \lim_{\{\Delta_i \to \Delta^{\mathrm{SR}}_{i}\}} \frac{C_{\phi}}{C_{\phi,0}} F_\phi'(\Delta_\phi)= 0.
\end{equation}
If we normalize our field such that $\lim_{\{\Delta \to \Delta^{\mathrm{SR}}\}} C_{\phi,0}$ is finite, and work in a theory where $F_\phi'(\Delta^\mathrm{SR}_{\hat{\phi}})\neq 0$ (see \cref{sec:doesTwoPointVanish} for further commentary on this), then this means that the two-point functions of each of the $\phi$s must vanish in the short-range limit, $\lim_{\{\Delta_i \to \Delta^{\mathrm{SR}}_i\}} C_\phi= 0$.  %
\end{enumerate}
Combined, these two observations nonperturbatively prove our main result.

More generally, if there are multiple long-range fields, each labelled by $\phi$, the partial derivative of $F_\mathrm{LR}(\{\Delta_\phi\})$ with respect to each $\Delta_\phi$ yields the associated two-point coefficient
\begin{equation}\label{eq:dFeqTwoPt}
\pdv{F_\mathrm{LR}}{\Delta_\phi}_{\Delta_{\psi\neq\phi}} = F_\phi'(\Delta_\phi) \frac{C_\phi}{C_{\phi,0}}.
\end{equation}

Note that \eqref{eq:dFdDelEqCphi} is already a nice consistency check by itself.
Interestingly, it also makes it clear that it is possible to compute the sphere free energy of these long-range models just by working in flat space, since $C_\phi$ is calculable in flat space. %

\subsection{Proof}

This proof relies on the fact that we know exactly the only term(s) which are nonlocal in the action, as we put them in by hand. 
As discussed in \cref{sec:GFFplusLocal}, no more nonlocal terms can be generated by the RG flow -- this is because in the Wilsonian RG, UV divergences only happen at coincident points, and so only lead to local operators.

We can treat both realizations of the long-range theories simultaneously, by writing 
\begin{align}\label{eq:generalAction}
S_g[\phi] = \int_x \sum_\phi \frac{Z_\phi}{2} \phi \WeylOp_{\dotwo -\Delta_\phi} \phi + \sum_g g_0 \cO_g +S_\mathrm{CFT},
\end{align}
for, e.g., $g_0 \equiv Z_g g \mu^{d-\Delta_\cO} Z_\phi^{m/2}$ in the simple case where $\cO_g \sim \phi^{m}$.
We take $\{\Delta_\phi\}$ to be generic at this point, such that the theory is indeed a long-range CFT.
Observe that $S_\mathrm{CFT}$ does not depend on $\Delta_\phi$ at all, but $g_0$ and $Z_\phi$ do.
$g_0$ refers to both the explicit interaction terms and whatever local counterterms might be induced by the long-range perturbation.
For example, generically to reach IR fixed points we must tune the mass operator $\hat{\phi}^2$, which certainly will appear in the OPE of $\cO \times \cO$ for $\cO =\hat{\phi}\chi$.

The fractional Laplacian on the sphere is defined by\footnote{Note that $\cF_{2(d-\Delta),0} <0$ for $\dotwo-(n+1)<\Delta<\dotwo-n$ and $d+n<\Delta<d+n+1$ for even integer $n\ge 0$; this explains the ferromagnetic sign, in, for example, \cite[(1.2)]{Behan:2017emf}. 
Note also that for $\Delta < \tfrac{d-2}{2}$ the only difference is the presence of more derivative terms. See, for example, \cite[(3.4)]{Gubser:2019uyf}.}
\begin{equation}\label{eq:fractionalLaplacian}
   \WeylOp_{\dotwo -\Delta} \phi(x)\equiv \lim_{r\to 0} \frac{1}{\cF_{2(d-\Delta),0}}\int_{y, \, s(x,y)>r} \frac{\phi(y) -\phi(x)}{s(x,y)^{2(d-\Delta)}} + \frac{1}{R^{d-2\Delta}}\frac{\Gamma(d-\Delta)}{\Gamma(\Delta)} \phi(x),
\end{equation}
for $\dotwo>\Delta>\tfrac{d-2}{2}$, such that in the free theory (with all $g_0=0$ and $S_\mathrm{CFT}=0$) we find the sphere conformal propagator
\begin{equation}\label{eq:freeNormalization}
\expval{\phi(x) \phi(y)}_{0} = \frac{C_{\phi,0}}{s(x,y)^{2\Delta_\phi}}, \quad C_{\phi,0} \equiv \frac{1}{\cF_{2\Delta_\phi,0} Z_\phi}.
\end{equation}
Here, the sphere metric and chordal distance $s(x,y)$ are
\begin{equation}
    g_{\mu\nu}  = \Omega(x)^2 \delta_{\mu\nu}, \quad \Omega(x) = \frac{2R}{1+x^2}, \quad s(x,y) \equiv \sqrt{\Omega(x)\Omega(y)} \abs{x-y},
\end{equation}
and so $\expval{\phi(x) \phi(y)}_{0}$ is evidently just a Weyl map of a flat-space conformal propagator.
We define $\cF_{\lambda,s}$, a standard function associated with Fourier transforms, in \eqref{eq:FlamsDef}.
For simplicity, we work only with scalar fields, though the proof is trivially extended to include fermions in \cref{sec:fermions}.
Let us regularize and renormalize \eqref{eq:generalAction}; then, we tune to the critical point $g\mapsto g_\star$, where due to the long-range nature of the kinetic term we know that
\begin{equation}
\expval{\phi(x) \phi(y)}_{g_\star} = \frac{C_\phi}{s(x,y)^{2\Delta_\phi}}
\end{equation}
exactly.
Without loss of generality we can take the $\cO_g$s to be a basis of conformal primaries, as all descendants are killed by the integral.
Then consider the conformal sphere free energy,
\begin{align}
    Z_{S^d}(\{\Delta_\phi\}) &\equiv \int \Dd\phi \, e^{-S_{g_\star}[\phi]}\\
    \tilde{F}(\{\Delta_\phi\}) &\equiv [-\log Z_{S^d}]_{\text{universal}} = \sin(\tfrac{\pi d}{2}) \log Z_{S^d}\rvert_\text{DREG}.
\end{align}
Clearly the location of the critical point and the values of the counterterms $Z_g, Z_\phi$ depend on the $\Delta_\phi$s. 
The fields $\phi(x)$ and $\cO_g(x)$, on the other hand, do not.
We will find that all contributions from the $\Delta_\phi$-dependence of coefficients in the action (i.e. $g_\cO(\Delta_\phi) \int_x \cO$, where $\cO$ is a local or nonlocal operator) vanish at the fixed point. 

This also means that the local terms $\propto \phi(x) \Box^k \phi(x)$,  required to correctly define the fractional Laplacian for $\Delta_\phi <\dotwo -1$, do not contribute -- and our results remain true even for the associated nonunitary CFT.
The only term which will survive will be the nonlocal kinetic term $\propto \phi(x) \phi(y)$, because it explicitly depends on the long-range scaling dimension. 
So, working in generic $d$,
\begin{align}\label{eq:chainRuleF}
\pdv{F}{\Delta_\Phi} &= \int_x \sum_\phi \frac{Z_\phi}{2}   \expval*{\odv{}{\Delta_\Phi}\left[\phi\WeylOp_{\dotwo -\Delta_\phi}\phi\right]} + \sum_\phi \half \pdv{Z_\phi}{\Delta_\Phi} \expval*{\phi \WeylOp_{\dotwo -\Delta_\phi} \phi} + \sum_g \pdv{g_0}{\Delta_\Phi} \expval*{\cO_g}.
\end{align}
The UV and IR cutoffs provided by DREG and the sphere have been used to allow us to freely commute the derivative, the path integral, and the spatial integral.
Now, we know that all one-point functions of local primaries on the sphere must be either zero or (if they are e.g. $\cO_g=\cR^n \times \mathbb{I}$) contribute only to nonuniversal terms in the free energy -- and hence can be dropped, since we work with an analytic regularization in DREG. 
We can therefore drop all of the $\cO_g$ terms\footnote{If the reader is uncomfortable with this construction, we provide an alternative diagrammatic proof of \eqref{eq:dFdDelEqCphi} in long-range $\phi^4$ theory with the 2PI formalism in \cref{app:2PI}.}. 

Thus, only the operators that explicitly depend on $\Delta_\Phi$ -- the nonlocal kinetic terms -- give a nonzero contribution.
Crucially, contributions from local kinetic terms of the form $\phi \WeylOp_k \phi$ also vanish. 
\begin{align}
\pdv{F}{\Delta_\Phi} &= \int_x \sum_\phi \frac{Z_\phi}{2} \delta_{\phi\Phi}   \expval*{\odv{}{\Delta_\phi}\left[\phi\WeylOp_{\dotwo -\Delta_\phi}\phi\right]} + \sum_\phi \half \pdv{Z_\phi}{\Delta_\Phi} \expval*{\phi \WeylOp_{\dotwo -\Delta_\phi} \phi} \label{eq:onlyKinetics}
\end{align}
We stress that the $\odv{}{\Delta_\phi}$ does not act on the $\phi$s.
Temporarily modifying the fractional Laplacian \eqref{eq:fractionalLaplacian} to have argument $\dotwo - \alpha$, as we need to differentiate with respect to that argument, we see that
\begin{align}
 \int_x \expval{\phi \WeylOp_{\dotwo -\alpha} \phi} &= \lim_{r\to 0} \frac{1}{\cF_{2(d-\alpha),0}}\int_{x,y, \, s(x,y)>r} \frac{\expval{\phi(x)(\phi(y) -\phi(x))}}{s(x,y)^{2(d-\alpha)}} + \frac{1}{R^{d-2\alpha}}\frac{\Gamma(d-\alpha)}{\Gamma(\alpha)} \int_x \expval{\phi(x)\phi(x)}.
\end{align}
The terms $\propto{\expval{\phi(x)\phi(x)}}$ (which can only be a constant) cancel against each other perfectly for arbitrary $\alpha$ after the $y$ integral, and so
\begin{align}
  \int_x \expval{\phi \WeylOp_{\dotwo -\alpha} \phi} &= \frac{C_\phi}{\cF_{2(d-\alpha),0}} \int_{x,y} \frac{1}{s(x,y)^{2(\Delta_\phi + d-\alpha)}} = \frac{C_\phi}{\cF_{2(d-\alpha),0}} I_2(\Delta_\phi + d-\alpha).
\end{align}
To allow us to consider even $d$, we now multiply through by $-\sin(\tfrac{\pi d}{2})$.
Since $-\sin(\tfrac{\pi d}{2})I_2(d)=0$ for any $d$, the contribution from the second term in \eqref{eq:onlyKinetics} vanishes exactly, and we are left with only
\begin{align}
\pdv{\tilde{F}}{\Delta_\phi} &\equiv -\sin(\tfrac{\pi d}{2})\frac{Z_{\phi} C_\phi}{2} \odv{}{\alpha}  \frac{1}{\cF_{2(d-\alpha),0}} I_2(\Delta_\phi + d-\alpha) \Big\rvert_{\alpha=\Delta_\phi}=  \frac{Z_{\phi} C_\phi}{2} \frac{\sin(\tfrac{\pi d}{2})}{\cF_{2(d-\Delta_\phi),0}} I_2'(d)\\
&= \frac{C_\phi}{C_{\phi,0}} \tilde{F}_b'(\Delta_\phi) = \frac{\expval{\phi(\hat{e}_1)\phi(0)}_{g_\star}}{\expval{\phi(\hat{e}_1)\phi(0)}_0} \tilde{F}_b'(\Delta_\phi), \label{eq:dFdDelEqTrGdC}
\end{align}
where, again because $-\sin(\tfrac{\pi d}{2})I_2(d)=0$, the term coming from the derivative of $\cF_{2(d-\Delta_\phi),0}$ does not contribute. 
We have then substituted \eqref{eq:freeNormalization} to recognise the standard derivative of the free energy of a generalised free scalar field \eqref{eq:bosonF}.

Now, this result is almost independent of the normalization of $\phi$.
This is because $C_{\phi}$ and $C_{\phi,0}$ change in the same way under an overall rescaling of $Z_\phi \to Z_\phi'$ (and so also $g_0 \to Z_g g \mu^{d-\Delta_\cO} Z_\phi^{\prime m/2}$).
However, we do need to account for the possibility that $\phi$ with this normalization becomes the zero operator (we will see this later in \cref{sec:doesTwoPointVanish}). 
Thus, we must rescale $\hat{\phi} \equiv \phi/\sqrt{C_\phi}$ such that it always has a unit-normalized two-point function. This does change the coupling constants, but we saw above that the contribution from $\odv{g_0}{\Delta_\phi}$ drops out. 
More generally, this rescaling of $\hat{\phi}$ leads to a modification only of the non-universal parts of the free energy, which we can ignore\footnote{Equivalently, in DREG the Jacobian is trivial: $\Dd{\phi} = \Dd{\hat{\phi}}$ as $\prod_{x \in \mathbb{R}^d} \odv{\phi(x)}{\hat{\phi}(x)} = 1$.}.

But now the final step is simple. 
We just recall that we have tuned the action to the short-range point by tuning $\{\Delta_i\}$ to $\{\Delta^{\mathrm{SR}}_i\}$.
Thus, the CFT should be local (recall we have explicitly integrated out the nonlocal field $\chi$ if it was present, and removed its free energy contribution); that is, the action should no longer contain any nonlocal terms once we have appropriately tuned the couplings.
The only nonlocal term in the new action (with our rescaled $\phi$) is
\begin{equation}
    S_g[\hat{\phi}] \supset \frac{Z_\phi C_\phi}{2} \frac{1}{\cF_{2(d-\Delta_\phi),0}} \int_x\lim_{r\to 0}\int_{y, s(x,y)>r} \frac{\hat{\phi}(x)(\hat{\phi}(y)-\hat{\phi}(x))}{s(x,y)^{2(d-\Delta_\phi)}}.
\end{equation}
Because this term is a nonlocal operator\footnote{Assuming that $\Delta \neq \dotwo -k$ exactly, which is a trivial case.}, its prefactor must vanish at the fixed point, to avoid the CFT being nonlocal. 
Hence:
\begin{equation}
\lim_{\{\Delta_i \to \Delta^{\mathrm{SR}}_i\}} \frac{Z_\phi C_\phi}{\cF_{2(d-\Delta_\phi),0}}= 0.
\end{equation}
Thus, from \eqref{eq:dFdDelEqTrGdC} we find our result
\begin{equation}\label{eq:FextrInProof}
    \lim_{\{\Delta_i \to \Delta^{\mathrm{SR}}_i\}}\pdv{F}{\Delta_\phi}  =0.
\end{equation}
Now, use again the fact that in these nonlocal CFTs we know exactly how nonlocality enters the action, as we put it in by hand.
Since $\sin(\tfrac{\pi d}{2}) I_2'(d)$ in \eqref{eq:dFdDelEqTrGdC} is nonzero for all $d>0$, this means that the converse of \eqref{eq:FextrInProof} also holds: 
\begin{equation}
\text{Extremal } \tilde{F} \implies \quad \frac{Z_\phi C_\phi}{\cF_{2(d-\Delta_\phi),0}} =0 \quad \forall \quad \phi,
\end{equation}
and therefore their actions contain no nonlocal terms. %
Since every member of the long-range family is conformally invariant by assumption, those extrema must be local CFTs!

Strictly, one must differentiate with respect to the scaling dimension of each of the long-range fields separately.
However, as we shall see in \cref{sec:perturbative}, if we \textit{assume} that in the relevant short-range theories the relevant fields have the same scaling dimension, we can cheat and just consider the diagonal long-range theory. 
Assuming that some of the long-range fields have equal scaling dimension would also allow us to consider a set of long-range fields with a non-diagonal two-point coefficient, i.e. $\expval{\phi_i \phi_j}= A_{ij}/\abs{x}^{2\Delta}$ for $A_{ij}$ not proportional to the identity.

In summary: the derivative with respect to $\Delta_\phi$ of the free energy only receives a contribution from the explicit dependence of operators (not their coefficients) on the scaling dimension $\Delta_\phi$. 
But at the short-range points there cannot be any long-range operators in the action, and so the derivative vanishes if and only if we are at a local CFT.

\subsection{Extension to include fermions}\label{sec:fermions}

The treatment is easily extended to nonlocal field theories containing both bosons and fermions.
Assuming all fields have been broken into their real components for simplicity, we can write the critical action
\begin{equation}
    S=\int_{x,y} \half \Phi_i(x) C_{\Delta_\Phi}^{-1}(x,y)_{ij} \Phi_j(y)+ \sum_g g_0 \cO_g + S_\mathrm{CFT}.
\end{equation}
Conformal symmetry and the statistics of $\Phi$ mean that 
\begin{equation}
\expval{\Phi_i(x)\Phi_j(y)} \equiv G_{\Delta_\Phi}(x,y)_{ij} = (-1)^{\mathrm{F}^\Phi} G_{\Delta_\Phi}(y,x)_{ji} = \frac{C_\Phi}{C_{\Phi,0}} C_{\Delta_\Phi}(x,y)_{ij}.
\end{equation}
As before, all one-point functions vanish, so we can use this to see that
\begin{align}
   \pdv{F}{\Delta_\Phi} &= \odv{}{\Delta}\int_{x,y} \half  C^{-1}_{\Delta}(x,y)_{ij} G_{\Delta_\Phi}(x,y)_{ij} \Big\rvert_{\Delta=\Delta_\Phi}=\odv{}{\Delta} (-1)^{\mathrm{F}^\Phi} \half \Tr[C^{-1}_{\Delta} G_{\Delta_\Phi}] \Big\rvert_{\Delta=\Delta_\Phi}\\
   &= \frac{C_\Phi}{C_{\Phi,0}} \odv{}{\Delta} (-1)^{\mathrm{F}^\Phi} \half \Tr[C^{-1}_{\Delta} C_{\Delta_\Phi}] \Big\rvert_{\Delta=\Delta_\Phi}= \frac{C_\Phi}{C_{\Phi,0}} F_\Phi'(\Delta_\Phi), \label{eq:dFmultiEqCmulti}
\end{align}
where in the last line we recognise the equation for $F_\Phi'(\Delta_\Phi)$ for a GFF $\Phi$ with respect to the scaling dimension \cite[\S 4.2.1]{Fraser-Taliente:2025udk}.
As discussed there, if $\Phi$ transforms in a nontrivial symmetry representation $\rho_\Phi$ of $\SO(d+1,1) \times G_\text{internal}$, $F_{\Phi}(\Delta)$ must contain a factor of $\dim \rho_\Phi$, because the trace includes a sum over any representation indices.
For the same reason as in the single-field case, each of the derivatives \eqref{eq:dFmultiEqCmulti} must vanish at the local point, and we are done.

\subsection{Why we do not include \texorpdfstring{$\chi$}{chi}}\label{sec:notincludechi}
 
The structure of this proof makes it clear why the form of the LR fixed point with the $\chi$ field integrated out is preferable. 
Of course, one reason is that it permits a unified treatment of both GFF$+$local and CFT$+\hat{\phi}\chi$ theories.
The second is that considering the contribution from $\chi$ leads to an extra term in the derivative of $F$.
Let us say we are in the CFT+$\hat{\phi}\chi$ route and the $\chi$ field is present, say
\begin{equation}
S_\text{with $\chi$}^{\text{CFT+$\hat{\phi}\chi$}} = \int_x \frac{Z_\chi}{2} \chi \WeylOp_{\Delta_\phi -\frac{d}{2}} \chi + \lambda_0 \hat{\phi}\chi + \frac{Z_\phi}{2} \hat{\phi} \WeylOp_1 \hat{\phi} + \frac{g_0}{4!}\hat{\phi}^4 \Big\rvert_{g=g_\star},
\end{equation}
computing $F=-\log \int \Dd{{}_{\mathrm{CFT}}}\Dd{\chi}\, e^{-S}$, we would find
\begin{equation}
\pdv{F_\text{with $\chi$}^{\text{CFT+$\hat{\phi}\chi$}}}{\Delta_\phi} = -F_b'(d-\Delta_\phi)+ F_\phi'(\Delta_\phi) \frac{C_\phi}{C_{\phi,0}},
\end{equation}
which is not zero in the short-range limit -- unlike \eqref{eq:dFdDelEqCphi}.
Likewise, $\lim_{\Delta_\phi \to \Delta^{\mathrm{SR}}_{\hat{\phi}}} F_\text{with $\chi$}^{\text{CFT+$\hat{\phi}\chi$}} = F_\mathrm{SR} + F_b(d-\Delta^{\mathrm{SR}}_{\hat{\phi}})$.
It therefore makes sense to subtract off the $\chi$ contribution $F_b(d-\Delta_\phi)$, which is trivial to do, as $\chi$ enters only quadratically,
\begin{equation} \label{eq:Fwithchi}
   F_\text{with $\chi$}^{\text{CFT+$\hat{\phi}\chi$}} = F_\text{without $\chi$}^{\text{CFT+$\hat{\phi}\chi$}} +F_b(d-\Delta_\phi).
\end{equation}
Thus, as in \eqref{eq:SLRnoChi}, we define $F_\mathrm{LR} \equiv F_\text{without $\chi$}^{\text{CFT+$\hat{\phi}\chi$}}$; this also is the quantity that matches both $F$ computed in the GFF$+$local route and $F_\mathrm{SR}$ in the short-range limit\footnote{This also explains the reason that it was necessary to not include the $\chi$ contribution when \cite{Giombi:2024zrt} attempted to obtain $\tilde{F}$ using the long-range approach: considering the $\Delta_\phi \to \Delta^\mathrm{SR}$ limit of the CFT+$\hat{\phi}\chi$ picture, it is clear that in general one must remove the free energy of the decoupled $\chi$ field from $F_{\text{with }\chi}$ in order to obtain that of the local CFT in the short-range limit. 
}.
It is important to account for $\chi$ when ensuring continuity of the spectrum of primaries in the short-range limit, as we review in \cref{sec:subtlety}.
However, we treat $\chi$ as just an auxiliary field (in a generalised sense) as (1) it just serves to implement the nonlocal kinetic term for $\hat{\phi}$, and (2) all of its correlators can be obtained from those of $\hat{\phi}$ by using the equations of motion.
Hence, given our goal of counting independent degrees of freedom, it is very reasonable to integrate it out and not consider its free energy contribution. 
To reiterate, in this paper \textbf{we always consider free energies excluding the contribution from $\chi$}.

One might then object to the manual subtraction of $F_b(d-\Delta_\phi)$ only from $F_\text{with $\chi$}^{\text{CFT+$\hat{\phi}\chi$}}$ on the grounds that it seems ad-hoc. 
The resolution is that to make sure that we find the same free energy, regardless of the construction route, we must be consistent with how we introduce nonlocality to create a family of nonlocal CFTs. 
So, \textit{if} we did want to keep the $\chi$ field, then we should also make sure it is there when using the GFF$+$local route -- the GFF should also be built from a normal scalar using a $\chi$ field!

We have two local CFTs, which we want to promote to lines of nonlocal CFTs: the free scalar of dimension $\dotwo-1$, and the SR Ising with field $\hat{\phi}$.
In both cases, we perform the promotion by coupling it to a $\chi$ field:
\begin{align}\label{eq:FreeScalarWithChi}
S_{\text{GFF${}_\phi$, with }\chi} &= \lim_{\lambda \to \lambda_\star}\int_x \half \phi \WeylOp_1 \phi + \lambda_0 \phi \chi_1 + \frac{Z_\chi}{2} \chi_1 \WeylOp_{\Delta_\phi -\frac{d}{2}} \chi_1, \\
S_{\text{LR, with }\chi} &= \lim_{\lambda \to \lambda_\star} S_\text{CFT}  + \int_x \lambda_0 \hat{\phi} \chi_2  + \frac{Z_\chi}{2} \chi_2 \WeylOp_{\Delta_\phi -\frac{d}{2}} \chi_2.
\end{align}
Once we have built the GFF, we can then proceed with the GFF$+$local route by deforming with $\phi^4$ to build the long-range CFT.

The free energies of \eqref{eq:FreeScalarWithChi} (and its $+\phi^4$ deformation), computed by $-\log \int \Dd{\phi} \Dd{\chi_i}\, e^{-S_i}$, will include a contribution $F_b(d-\Delta_\phi)$ (we will see this explicitly in \cref{sec:GenFFreeCheck}).
Because this exact term appears in both, we make the choice to consistently integrate out $\chi$ on both sides, and drop the contribution $F_b(d-\Delta_\phi)$ in both.

In summary, consistently ignoring the contribution of $\chi$ is permissible.
To be explicit, the free energy of a particular long-range CFT can be found using either of 
\begin{align}
&e^{-F_{\mathrm{LR}}(\{\Delta_i\})} \equiv \lim_{g^i \to g_\star^i} \int \Dd{\phi_i} \exp -\left[\half \phi_i \WeylOp_{\dotwo-\Delta_i}\phi_i + \text{local}\right]\\
&\quad\,\overset{\text{IR duality}}{=} \lim_{\lambda^i \to \lambda_\star^i} \int \Dd{{}_\mathrm{CFT}} \exp -\left[S_{\mathrm{CFT}} - \frac{(\lambda_0^i)^2}{2} \hat{\phi}_i \WeylOp_{\dotwo-\Delta_{i}}\hat{\phi}_i\right]. %
\end{align}
This definition has the two advantages that our long-range $\tilde{F}$s at the crossover point (1) agree exactly with the short-range ones and (2) are extremised there, rather than the locality condition being $\tilde{F}_{\text{LR, with }\chi}'(\Delta_\phi) = - \tilde{F}_b(d-\Delta_\phi)$.

\section{The generalised \texorpdfstring{$F$}{F} theorem in conformal perturbation theory and maximisation} \label{sec:genFinCPT}

The obvious next question is the type of extremum that we find in the limit where a long-range theory becomes short-range. 
This is a question that we can answer in conformal perturbation theory around whichever local CFT is realized. 
Of course, this does require use of the IR duality to describe the long-range theory infinitesimally close to the CFT by the $+\hat{\phi}\chi$ route.
Since we require no perturbative description of the local CFT itself, the proof is nonperturbative in the couplings of that CFT.
We will find that $\tilde{F}$ is maximised for unitary CFTs.
This section is self-contained and somewhat technical, and is not required for the $F$-extremisation result.

We prove a slightly stronger result: that the generalised $F$-theorem of \cite{Fei:2015kta} extends in CPT to the case where the leading quantum term in the beta function vanishes.
\begin{quote}
   Consider the flow from any unitary CFT$_\mathrm{UV}$ to a CFT$_\mathrm{IR}$ due to deformation by a relevant operator $\cO$ of dimension $d-\eta$, in the special case where the OPE coefficient $C_{\cO\cO\cO}=0$.
   If the deformation generates a real perturbative IR fixed point, then the generalised $F$-theorem holds to leading order in conformal perturbation theory:
    \begin{equation}
        \tilde{F}_\mathrm{UV} > \tilde{F}_\mathrm{IR}.
    \end{equation}
\end{quote}
We will assume only that $\cO$ is a local operator, as the rules of CPT are not clear for nonlocal perturbations \cite[\S 1]{Behan:2017emf}.
When applied to the deformation by $\cO=\hat{\phi}\chi$ of a unitary CFT$+$GFF, maximisation follows immediately, since $C_{\cO\cO\cO}$ vanishes by the $\mathbb{Z}_2$ symmetry of a GFF. 

Our calculation is a modification of \cite[\S 4]{Fei:2015oha}, where the generalised $F$-theorem was proven to leading order in conformal perturbation theory under the assumption that $C_{\cO\cO\cO} \neq 0$.
Specifically, they assumed that the coefficient of the $\lambda^2$ term in the beta function, $\beta_2\propto C_{\cO\cO\cO}$, was nonzero -- this is not the case for the long-range deformation.
In \cref{sec:specialiseGF} we will then specialise the result to our case.
This result also proves for the second time that for the CFT$+\hat{\phi}\chi$ route of construction of the long-range theories, the free energy is necessarily an extremum, since we find that the change in the free energy is $\propto \eta^2$.
However, it does not prove it for the GFF$+$local construction, which is the more interesting case -- as in that construction the local CFTs are less obviously special. 
Happily, our easier proof from \cref{sec:proof} sufficed for that construction.

We now give a sketch of this section. 
In \cref{sec:FasGenerator} we set up the problem of CPT. 
In \cref{sec:countertermComputation} we solve for the counterterms to second order in the coupling $\lambda$, and use that to compute the free energy to order $\lambda^4$ in \cref{sec:CPTfreeEnergy}. 
These computations are specialised to the case where the OPE coefficient vanishes in \cref{sec:COOOzero}, which is the one relevant in the long-range setting. 
This allows us to show $F$-maximisation for unitary long-range theories in \cref{sec:specialiseGF}.
There is one important detour that we take on the way: this computation leads to a natural proposal for a new normalization for $\tilde{F}$, which we name $\hat{F}$ in \cref{sec:proposedFhat}.

\subsection{CPT setup: the free energy as a generating function}\label{sec:FasGenerator}

Let us begin by defining the generating function of connected correlators of $\cO$, which is just $F$, but where $\lambda$ is now a function of spacetime $\lambda(x)$ \cite{Gerchkovitz:2014gta,Gomis:2015yaa}.
That is, defining $S=S_\text{CFT} + \int_x \lambda_b(x) \cO(x)$ and $\cO(x) = Z_\cO \cO_\text{ren}(x)$, we can compute that
\begin{equation}
 \fdv{F}{\lambda(x)} \Big\rvert_{\lambda =\text{const}} = \mu^\eta \expval*{\pdv{\mu^{-\eta} \lambda_b}{\lambda} Z_\cO \cO(x) }_\lambda.
\end{equation}
Thus, if we want $\fdv{F}{\lambda(x)}\rvert_{\lambda=\text{const}}=\mu^\eta \expval{\cO_\text{ren}(x)}_\lambda$, we must have \cite[(4.8)]{Fei:2015oha} %
\begin{equation}
    Z_\cO^{-1} = \pdv{\mu^{-\eta} \lambda_b}{\lambda}.
\end{equation}
The factors of $\mu^\eta$ here just serve to keep both sides dimensionless.
Recall that $\beta_\lambda = -\eta/\pdv{\log \mu^{-\eta} \lambda_b}{\lambda}$, so $\beta_\lambda = -\eta Z_\cO \mu^{-\eta}\lambda_b$.

This means that
\begin{align}
\expval{\cO_\text{ren}(x_1) \cdots \cO_\text{ren}(x_n)}_{\lambda,c} = \mu^{-n\eta} (-1)^{n-1} \fdv{}{\lambda(x_1)}\cdots \fdv{}{\lambda(x_n)} F \Big\rvert_{\lambda=\text{const}},
\end{align}
where evidently the $\lambda^{\ge 2}$ terms in $\lambda_b$ give the contact terms $\propto \delta^{(d)}(x_i - x_j)$ in the correlator.
Thus, renormalizing the theory -- determining the correct counterterms -- just corresponds to choosing a function $\lambda_b(\partial_x,\lambda)$ which makes all observables finite. 
The fact that the theory is renormalizable tells us that we only need a single function to do this (i.e. a single coupling -- we are, as usual, ignoring any other relevant operators that are generated by the OPE and considering only the near-marginal ones).

We will need to use the OPE of our original CFT, 
\begin{equation}\label{eq:OOope}
    \cO(x) \cO(y) \sim \frac{C_{\cO\cO \id} \id}{\abs{x-y}^{2\Delta}} + C_{\cO\cO\cO} \frac{\cO(x)}{\abs{x-y}^{\Delta}} + \text{irrelevant},
\end{equation}
where we ignore the contribution of any more relevant operators like $:\cO^2:$ here as we work with an analytic regularization\footnote{One can also think of this as meaning that we are constantly tuning the coefficients of those more relevant operators in the action to zero.}. 

\subsection{Computing the counterterms}\label{sec:countertermComputation}
Let us work in a minimal subtraction scheme. 
This means that $\lambda_b(x)$, a local functional of $\lambda(x)$, contains only poles in $\eta$:
\begin{equation}
   \lambda_b = \mu^\eta \left( \lambda + \frac{\beta_2}{\eta} \lambda^2 + \lambda^3\left(\frac{\beta_2^2}{\eta^2} +  \frac{\beta_3}{2\eta}\right) + O(\lambda^4)\right),
\end{equation}
with $\beta_2$ and $\beta_3$ independent of $\eta$ such that
\begin{equation}\label{eq:CPTbeta}
    \beta_\lambda = -\eta \lambda + \beta_2 \lambda^2 + \beta_3\lambda^3 + O(\lambda^4).
\end{equation}
The fact that the coefficient of $\lambda^3/\eta^2$ is exactly $\beta_2^2$ is a check of the consistency of our calculations.
The free energy can be expanded as\footnote{
Assuming $\beta_2=0$, as in our case, the beta function \eqref{eq:CPTbeta} tells us that the leading divergence at each order in $\lambda$ in $\lambda_0$ is
 \begin{equation}
\lambda_0 \supset \mu^{\eta}\lambda \sum_{n=0}^\infty \binom{2n}{n} \left(\frac{\beta_3 \lambda^2}{4\eta}\right)^n = \frac{\mu^{\eta}{\lambda}}{\sqrt{1-\beta_3 \lambda^2/\eta}}.
\end{equation}
With this analytic regularization, the critical value of $\lambda_0$ is $\infty$, as we might expect. 
Of course, this is the signal that we should never perturbatively expand in $\lambda_0$.
} 
\begin{equation}
\begin{aligned}
    \label{eq:etaFlamexpansion}
    \delta F_\lambda &=  -\frac{\mu^{2\eta}}{2!} \expval*{\left[\int_x  \left(\lambda+  \frac{\beta_2}{\eta} \lambda^2 + \lambda^3\left(\frac{\beta_2^2}{\eta^2} +  \frac{\beta_3}{2\eta}\right)\right)\cO\right]^2}_{c}\\
    & + \frac{\mu^{3\eta}}{3!} \expval*{\left[\int_x  \left(\lambda+  \frac{\beta_2}{\eta} \lambda^2\right) \cO\right]^3}_{c}- \frac{\mu^{4\eta}}{4!} \expval*{\left[\int_x  \lambda \cO\right]^4}_{c} + O(\lambda^5),
\end{aligned}
\end{equation}
where $\expval{\cdot}_c$ denotes a connected correlator in the original CFT, and we have dropped the term $\expval{\int_x\lambda_b \cO}_{\mathrm{CFT}}=0$.

In the correctly renormalized theory, all observables should be finite.
We work on the sphere, but because we are only dealing with UV divergences the computations are identical.
The first instinct is to use the one-point function $\expval{\cO_\text{ren}}$ to identify $\beta_3$. 
However, on the sphere, it is not divergent at this order because $I_2(d-\eta)\propto \eta$ vanishes as $\eta \to 0$ (though it does diverge at the next order).  %
It is therefore necessary to find a correlator that does have divergences. Once we have fixed the counterterms by making that correlator finite, the fact that the theory is renormalizable tells us that all correlators will now be finite.

The next correlator is the renormalized connected two-point function.
Let us consider this $\expval{\cO_\text{ren}(x)\cO_\text{ren}(y)}_{\lambda,c}$ term by term,
\begin{align}
    & -\mu^{-2\eta}\fdv{}{\lambda(x)} \fdv{F}{\lambda(y)} \Big\rvert_{\lambda =\text{const}} = \left(\pdv{\lambda_b}{\lambda}\right)^2 \Bigg[\expval{\cO(x)\cO(y)}_c - \lambda_b \mu^\eta \int_{w}  \expval{\cO(w) \cO(x)\cO(y)}_c\notag \\
    &\qquad\qquad + \frac{\lambda^2}{2!} \mu^{2\eta}\int_{w,z}  \expval{\cO(w) \cO(x)\cO(y) \cO(z)}_c+ O(\lambda^5)\Bigg]\\
    &=  \left(1+\tfrac{3 \beta _3 \lambda ^2+4 \beta _2 \lambda }{\eta }+\tfrac{10 \beta _2^2 \lambda ^2}{\eta ^2}\right) \expval{\cO(x)\cO(y)}_c   -\left(\lambda+\tfrac{5 \beta _2 \lambda ^2}{\eta }\right) \mu ^{\eta }\int_w \expval{\cO(w)\cO(x)\cO(y)}_c \notag\\
    &\qquad\qquad + \tfrac{\lambda ^2 \mu ^{2 \eta }}{2} \int_{w,z} \expval{\cO(w) \cO(x)\cO(y) \cO(z)}_c. \label{eq:OOtoOrd4}
\end{align}
We drop all terms proportional to $\pdv{\lambda(x)}{\lambda(y)}=\delta(x-y)$, such as the contact term $\propto \lambda_b''(\lambda) \lambda_b(\lambda) \delta(x-y)$, assuming that we consider $\cO$ correlators at separated points only.
We then need to ensure that \eqref{eq:OOtoOrd4} is finite order by order in $\lambda$.

\subsubsection{Order \texorpdfstring{$\lambda$}{lambda}}

Finiteness of \eqref{eq:OOtoOrd4} at order $\lambda$ means that
\begin{equation}\label{eq:OOfiniteLam1}
   \frac{4 \beta _2 }{\eta }  \expval{\cO(x)\cO(y)}_c  - \mu ^{\eta }\int_w \expval{\cO(w)\cO(x)\cO(y)}_c = O(\eta^0).%
\end{equation}
The $1/\eta$ pole can be easily evaluated using the OPE. 
The divergences only come as $w$ approaches either $x$ or $y$. 
Writing $\cR_{\not{x}}$ and $\cR_{\not{y}}$ for the region of size $R$ where $x$ and $y$ respectively can be considered as well separated from the other points, we see that
\begin{align}
    \int_w \expval{\cO(w)\cO(x)\cO(y)}_c \sim \int_{w \in \cR_{\not{x}}} \expval{\cO(w)\cO(x)\cO(y)}_c+\int_{w \in \cR_{\not{y}}} \expval{\cO(w)\cO(x)\cO(y)}_c
\end{align}
where $\sim$ indicates that they have the same divergence structure. 

Applying the OPE \eqref{eq:OOope} and using $\expval{\id \cO} = \expval{\cO} = 0$, we see that the divergence comes from the region of small $w-x$. 
Defining $r\equiv \abs{w-x}$, this becomes
\begin{align}
    \sim 2 C_{\cO\cO\cO}\int_{w \in \cR_{\not{y}}} \frac{1}{\abs{w-x}^\Delta}\expval{\cO(x)\cO(y)}_c \sim 2 S_{d-1} C_{\cO\cO\cO}\int_0^{\sim R} \frac{\odif{r}}{r} r^{d-\Delta} \expval{\cO(x)\cO(y)}_c,
\end{align}
where $S_{d-1} \equiv \vol S^{d-1} =  2\pi^{d/2}/\Gamma(d/2)$, the volume of the unit $(d-1)$-sphere.
We have been deliberately vague about the upper bound $\sim R$ of the $r$ integral, since it simply comes from the characteristic size $R$ of the region $\cR_{\not{y}}$, and does not contribute to the pole:
\begin{equation}\label{eq:Rnotcontribute}
   \int_0^R \frac{\odif{r}}{r} r^{\nu} = \frac{R^\nu}{\nu} = \frac{1}{\nu} + O(\eta^0).
\end{equation}
Hence, \eqref{eq:OOfiniteLam1} becomes
\begin{align}
 \frac{4 \beta _2 }{\eta } = 2\mu ^{\eta } S_{d-1} C_{\cO\cO\cO}\int_0^{\sim R} \frac{\odif{r}}{r} r^{d-\Delta} + \text{finite} = 2S_{d-1} C_{\cO\cO\cO} \frac{1}{\eta} + \text{finite}\\
\end{align}
i.e. we have
\begin{equation}\label{eq:beta2val}
     \beta_2 = \half S_{d-1} C_{\cO\cO\cO},
\end{equation}
where because we work in minimal subtraction this must be the value of the OPE coefficient at $\eta=0$ (see \cite[\S 3.1, (3.31)]{Komargodski:2016auf} for more comments on this).

\subsubsection{Order \texorpdfstring{$\lambda^2$}{lambda\^{}2}} \label{sec:orderlam2TwoPoint}
Substituting \eqref{eq:beta2val} into \eqref{eq:OOtoOrd4}, we see that the first divergence is now order $\lambda^2$, and
\begin{equation}\label{eq:OOfiniteLam2}
    \begin{aligned}
    \expval{\cO_\text{ren}(x) \cO_\text{ren}(y)} &\sim \lambda^2\Big[\frac{3 \beta _3 \eta +10 \beta_2^2}{\eta ^2}  \expval{\cO(x)\cO(y)}_c  - \frac{5 \beta_2}{\eta}\int_w\expval{\cO(w)\cO(x)\cO(y)}_c\\
    & \qquad \qquad+ \frac{\mu ^{2 \eta }}{2}  \int_{w,z} \expval{\cO(w)\cO(x)\cO(y)\cO(z)}_c\Big].
    \end{aligned}
\end{equation}
Ensuring this is $O(\eta^0)$ will determine $\beta_{3}$.

At order $1/\eta^2$, after substituting the value of $\int_w \expval{\cO(w)\cO(x)\cO(y)}_c$, we see that
\begin{align}
     \beta_2^2 = +\frac{1}{20} \left[\mu^{2\eta}\frac{\int_{w,z} \expval{\cO(w) \cO(x)\cO(y) \cO(z)}_c}{\expval{\cO(x)\cO(y)}_c}\right]_{\eta^{-2}},\label{eq:beta2sqdef}
\end{align}
where $[\cdot]_{\eta^{-a}}$ means that we extract the coefficient of the $1/\eta^a$ pole. 
The pairwise divergences $\propto C_{\cO\cO\id}^2$ do not contribute, since we are considering the connected correlator; and the contributions from relevant operators can be ignored, since we are analytically regulating via $\eta$.
It only remains to compute this a little more carefully.

Consider the ways we can obtain a double divergence:
\begin{equation}\label{eq:doubleDiv}
     \int_{w,z} \expval{\cO(w) \cO(x)\cO(y) \cO(z)}_c = J_{1} + 2\times J_2.
\end{equation}
\begin{enumerate}
\item $J_1$: if $w$ approaches $x$ and $z$ approaches $y$ (or vice versa). 
Using the OPE \eqref{eq:OOope}, we find
\begin{align}
    &J_1 \sim 2 \int_{w \text{ near x},z \text{ near } y} \expval{\cO(w) \cO(x)\cO(y) \cO(z)}_c\\
    \sim&  2 C_{\cO\cO\cO}^2 \int_{a,b \text{ near 0}} \frac{\expval{\cO(x) \cO(y)}}{\abs{a}^\Delta \abs{b}^\Delta} = 2 \left(C_{\cO\cO\cO} S_{d-1} \int_0^R \frac{\odif{r}}{r} r^{d-\Delta}\right)^2 \expval{\cO(x) \cO(y)}\\
    =& 8 \frac{\beta_2^2}{\eta^2} \expval{\cO(x) \cO(y)} \label{eq:J1eval}
\end{align}
\item $J_2$: if $w$ and $z$ approach $x$ (or $y$, giving a second identical copy, so $2\times J_2$). Then, defining $r_1 = |w - x|$, $r_2 = |z - x|$, $\rho = |w - z|$, we have three regions:
\begin{enumerate}[i]
\item $r_1 < r_2$ and $r_1 < \rho$ — the pair $(w, x)$ is closest
\item $r_2 < r_1$ and $r_2 < \rho$ — the pair $(z, x)$ is closest
\item $\rho < r_1$ and $\rho < r_2$ — the pair $(w, z)$ is closest
\end{enumerate}
The first two are the same under the exchange of $w$ and $z$, so
\begin{equation}
    J_2 = 2J_{2i} + J_{2iii}.
\end{equation}
Let us consider first $J_{2i}$. Applying the OPE to $w$ and $x$, we find
\begin{align}
    J_{2i} &\sim \int_{\text{w near x}} \frac{C_{\cO\cO\cO}}{\abs{w-x}^\Delta}\int_{\text{z near x}} \expval{\cO(z)\cO(x)\cO(y)}_c\\
    & \sim \int_{\text{w near x}} \frac{C_{\cO\cO\cO}}{\abs{w-x}^\Delta}\int_{\text{z near x}} \frac{C_{\cO\cO\cO}}{\abs{z-x}^\Delta} \expval{\cO(x)\cO(y)}
\end{align}
For the pole, only the hierarchical region $r_1 \ll r_2 \ll R$ matters (where $\rho \approx r_2$, so the sector constraint is automatically satisfied). The angular integrals just produce factors of $S_{d-1}$:
\begin{equation}
J_{2i} = C_{\mathcal{OOO}}^2\,S_{d-1}^2 \int_0^R \frac{dr_2}{r_2}\,r_2^{\eta} \int_0^{r_2} \frac{dr_1}{r_1}\,r_1^{\eta}\;\langle \cO(x)\cO(y)\rangle.
\end{equation}
The inner integral gives $r_2^\eta/\eta$. The outer integral then gives:
\begin{equation}
\frac{1}{\eta}\int_0^R \frac{dr_2}{r_2}\,r_2^{2\eta} = \frac{1}{2\eta^2} + O(1/\eta).
\end{equation}
Hence, $J_{2i} = \frac{2\beta_2^2}{\eta^2}\,\expval{\cO(x)\cO(y)}_c$.

Now we turn to $J_{2iii}$, where  $w$ and $z$ are closest to each other.
The leading singularity is the OPE $\cO(w)\cO(z) \sim C_{\mathcal{OOO}}\,\cO(z)/\rho^\Delta$. After this OPE, what remains is $\int_z \langle \cO(z)\cO(x)\cO(y)\rangle_c$ where the $z \to x$ singularity of the three-point function produces the second pole.

Parametrise by $\rho = |w - z|$ and $s = |z - x|$. In the hierarchical regime $\rho \ll s \ll R$ (where $r_1 \approx s$, so the sector constraint $\rho < r_1$ is satisfied):
\begin{equation}
J_{2iii} = C_{\mathcal{OOO}}^2\,S_{d-1}^2 \int_0^R \frac{ds}{s}\,s^{\eta} \int_0^{s} \frac{d\rho}{\rho}\,\rho^{\eta}\;\langle \cO(x)\cO(y)\rangle = \frac{2\beta_2^2}{\eta^2}\,\langle \cO(x)\cO(y)\rangle.
\end{equation}
Thus, $i$, $ii$, and $iii$ contribute identically, and 
\begin{equation}\label{eq:J2eval}
J_2 \sim 3  \times \frac{2\beta_2^2}{\eta^2}\,\langle \cO(x)\cO(y)\rangle.
\end{equation}
\end{enumerate}
Slotting \eqref{eq:J1eval} and \eqref{eq:J2eval} into \eqref{eq:doubleDiv}, $8+2(6)=20$ means that \eqref{eq:beta2sqdef} is satisfied exactly, and the renormalization procedure is consistent.
Thus, 
\begin{align}
    -3\beta_3 & = \lim_{\eta\to 0} \eta\left[\frac{\frac{\mu ^{2 \eta }}{2}  \int_{w,z} \expval{\cO(w)\cO(x)\cO(y)\cO(z)}_c- \frac{5 \beta_2}{\eta}\int_w\expval{\cO(w)\cO(x)\cO(y)}_c}{\expval{\cO(x)\cO(y)}_c} -10 \frac{\beta_2^2}{\eta^2} \right],\label{eq:beta3def}
\end{align}
again making it clear that $\beta_3$ does not depend on $\eta$ in this MS scheme.

\subsubsection{Rewriting \texorpdfstring{$\beta_3$}{beta\_3}}\label{sec:Beta3Expression}

To find a more useful expression for $\beta_3$ from \eqref{eq:beta3def}, let us work for the moment in flat space with $C_{\cO\cO\cO}=0$, and consider the divergences arising when three of the points coincide.
Since now $\cO(w)$ is separated, and $\cO$ is a local operator by assumption, we can use the OPE
\begin{equation}\label{eq:OPE}
   \cO(x)\cO(y) \cO(z) \sim f(y-x,z-x) \cO(x) + \text{less singular}.
\end{equation}
Here, $f$ is a normalized four-point function, where one of the operators has been sent to infinity $\cO(\infty) =\lim_{x \to \infty} \abs{x}^{2\Delta} \cO(x)$:
\begin{equation}\label{eq:fDef}
    f(y,z) =\frac{1}{ C_\cO}\expval{\cO(0) \cO(y) \cO(z) \cO(\infty)}_c,
\end{equation}
an identity which is easily checked by performing \eqref{eq:OPE} on the RHS and using $\expval{\cO(0)\cO(\infty)} =C_\cO$.
Since this part of the calculation depends only on the local divergences, we could do it in flat space.

Again, the divergences occur in the same regions $\cR_{\not{x}}$ and $\cR_{\not{y}}$, where $x$ and $y$ are well separated from the other points respectively -- which we take to be of size $R$.
Performing \eqref{eq:OPE} on \eqref{eq:beta3def}, we can separate the two divergences and perform the OPE in each region
\begin{align}
    &\int_{w,z} \expval{\cO(w) \cO(x)\cO(y) \cO(z)}_c \sim \int_{\cR_{\not{x}}} \expval{\cO(w) \cO(x)\cO(y) \cO(z)}_c + \int_{\cR_{\not{y}}} \expval{\cO(w) \cO(x)\cO(y) \cO(z)}_c\notag \\
&\sim  \int_{\cR_{\not{x}}} f(w-x,z-x) \expval{\cO(x) \cO(y)}_c + \int_{\cR_{\not{y}}}  f(w-y,z-y)\expval{\cO(x) \cO(y)}_c.
\end{align}
These two integrals are identical up to $x\leftrightarrow y$, so shifting $w \to w+x, \quad z \to z+x$ in the first, we can send $x$ and $y$ to infinity to obtain
\begin{align}
    \frac{\int_{w,z} \expval{\cO(w) \cO(x)\cO(y) \cO(z)}_c}{\expval{\cO(x)\cO(y)}_c}  \sim 2 \int_{w,z \in \cR_{\not{\infty}}} f(w,z).
\end{align}
We now proceed exactly as for \cite[(3.28-30)]{Komargodski:2016auf}. 
We break the integral over $\cR_{\not{\infty}}$ into three subregions, which are those in which $w \to z$, $w \to 0$, and $z \to 0$. 
By symmetry, these contribute identically, and so setting $z=r \hat{e}_1$ by rotational invariance we rewrite \eqref{eq:beta3def} as
\begin{align}\label{eq:beta3simThis}
    \frac{\beta_3}{\eta} & \sim -\frac{2}{6}\int_{w,z \in \cR_{\not{\infty}}} f(w,z) \sim - \int_{\substack{w,z:\, \abs{w}<\abs{z}, \, \abs{w} <\abs{w-z},\\ w,z \in \cR_{\not{\infty}}}}  f(w,z)  \\
    =&-\frac{S_{d-1}}{C_\cO} \int_{\substack{\abs{w}<r \text{ and }\abs{w}<\abs{r\hat{e}_1-w}\\ w,r\hat{e}_1 \in \cR_{\not{\infty}}}} \odif[d]{w} \,r^{d-1} \odif{r} \,\expval{\cO(0) \cO(w)\cO(r \hat{e}_1) \cO(\infty)}_c\notag\\
     =&-\frac{S_{d-1}}{C_\cO} \int_0^{\sim R}\frac{\odif{r}}{r} r^{2\eta} \int_\cR \odif[d]{x} \,\expval{\cO(0) \cO(x)\cO(\hat{e}_1) \cO(\infty)}_c
\end{align}
for $\cR = \{x:\, \abs{x} <1, \abs{x} < \abs{\hat{e}_1 - x}\}$.
In the last line we replaced $w=rx$ and then used the known behaviour of a four-point function under a constant rescaling to factor out $r$.
Thus, we find a finite integral expression for $\beta_3$:
\begin{equation}\label{eq:beta3val}
    \beta_3 =-\frac{S_{d-1}}{2 C_\cO} \int_\cR \odif[d]{x} \,\expval{\cO(0) \cO(x)\cO(\hat{e}_1) \cO(\infty)}_c
\end{equation}
which agrees with \cite[(3.28-30)]{Komargodski:2016auf} after identifying $\beta_3=\beta_{2,\text{there}}$ and setting $C_\cO=1$ and $C_{\cO\cO\cO}=0$, despite our slightly different choice of renormalization scheme\footnote{
    The discussion in that paper corresponds to setting $\lambda(x) = \lambda\theta(R-\abs{x})$ for some large $R$ in flat space after taking the functional derivatives of $F$.
Their renormalization condition is then the finiteness of $\expval{\cO_\text{ren}(x)} \rvert_{\abs{x}>R}$.
This simplifies the counting of divergences, as the operator at infinity is never approached by the insertions. 
This means that the infinity is not renormalized, which agrees with the fact that $Z_\cO\rvert_{\abs{x}>R} =1$; hence this yields the same result as in our approach. 
Since we work on the sphere, we prefer our approach.
}. 
As expected, $\beta_3/C_\cO$ is independent of the normalization of $\cO$.

If we actually needed to numerically calculate the value of $\beta_3$, we would also need to perform further subtractions to remove the contributions of relevant operators, as discussed in \cite{Komargodski:2016auf} and \cite{Behan:2025ydd}.
Because we are working in DREG and do not actually need to compute \eqref{eq:beta3val}, we have no such problem here.
Once such subtractions are performed, positivity of $\beta_3$ is not manifest, as discussed in \cite[\S 2.3]{Behan:2017emf} and \cite[\S 3.1]{Behan:2017emf}.

Now that we have determined the value of $\beta_3$, we can compute the free energy.

\subsection{The free energy in CPT}\label{sec:CPTfreeEnergy}

The critical coupling is either going to be order $\eta$ or order $\sqrt{\eta}$, so to compute \eqref{eq:etaFlamexpansion} we need to perform the three sphere integrals
\begin{equation}
    \int_{x,y} \expval{\cO(x)\cO(y)}, \quad
    \int_{w,x,y} \expval{\cO(w)\cO(x)\cO(y)}, \quad
    \int_{w,x,y,z} \expval{\cO(w)\cO(x)\cO(y)\cO(z)}
\end{equation}
to at least orders $\eta^2$, $\eta$, and $\eta^0$ respectively.
The first two are easy, as we know the two and three-point sphere correlators of a scalar operator of dimension $\Delta= d-\eta$ exactly:
\begin{align}
\int_{x,y} \expval{\cO(x)\cO(y)} &= C_\cO I_2(\Delta)\\
    \int_{w,x,y}\expval{\cO(w)\cO(x)\cO(y)}&=\int_{w,x,y}\frac{C_\cO C_{\cO\cO\cO}}{s(w,x)^\Delta s(x,y)^{\Delta} s(y,w)^\Delta} = C_\cO C_{\cO\cO\cO} I_3(\Delta).
\end{align}

To compute the free energy on the sphere, we cannot just integrate \eqref{eq:beta3def} over $x$ and $y$, as it does not account for the singularities where $x$ and $y$ approach each other.
However, the divergence counting is similar to the one we performed in \cref{sec:orderlam2TwoPoint}.
For clarity, we will walk it through explicitly.

We treat the double divergences first, which can be derived by performing two OPEs.
After they have been performed, we are left with an integral over the two remaining operators, which must take the form
\begin{equation}
    \frac{\beta_2^2}{\eta^2}\int_{x,y} \expval{\cO(x)\cO(y)} s(x,y)^{2\eta} = \frac{\beta_2^2}{\eta^2}C_\cO I_2(d-2\eta),
\end{equation}
where the $s(x,y)^{2\eta}$ is required because there are no other scales in the problem.
When this is expanded, the first term is $\propto 1/\eta$ -- but we must also keep the $\eta^0$ term, since we are trying to compute the free energy.

Now let us consider the triple-OPE divergences, ignoring the double divergences which we have now accounted for.
We should start by dividing up the region of integration by which pair of operators is the closest together.
For the triple-OPE divergences, there are six choices of pairs of coordinates, and each of the six sectors contributes equally, so we find:
\begin{equation}
    I \equiv \int_{w,x,y,z} \expval{\cO(w)\cO(x)\cO(y)\cO(z)}_c \sim 6 \int_{\substack{w,x \text{ closest}\\
    y,z}} \expval{\cO(w)\cO(x)\cO(y)\cO(z)}_c.
\end{equation}
Now, at any given point in the $w,y$ integrations that leads to a triple-OPE divergence, the pair $w$ and $y$ might be closer to either $x$ or $z$.
Let us say it is $x$ (if it is $z$ instead, we just do the following with $z$ and relabel $z \leftrightarrow x$). 
Let us perform the OPE \eqref{eq:OPE} $\cO(x)\cO(w) \cO(y) \sim f(w-x,y-x) \cO(x)$,
\begin{align}
    I &\sim 6 \int_{x,z} \int_{w,y \in \cR_{\not{z}}} f(w-x,y-x) \expval{\cO(x) \cO(z)} \\
     &\sim 6  \int_{w,y \in\cR_{\not{\infty} }} f(w,y) \int_{x,z}\expval{\cO(x) \cO(z)},
\end{align}
where we have used the fact that $x$ and $z$ are well separated and shifted $w$ and $y$.
The UV divergence comes only from integrating the shifted $w,y$ in a small region around $0$, and so we can substitute \eqref{eq:beta3simThis} to find:
\begin{align}
    I &\sim  \frac{6(-3\beta_3)}{\eta} \int_{x,z} \expval{\cO(x) \cO(z)},
\end{align}

Performing the final integral, we find 
\begin{align}\label{eq:finalIntegral}
\frac{1}{4!} \int_{w,x,y,z}  \expval{\cO(w) \cO(x)\cO(y) \cO(z)}_c = \frac{5}{4} \frac{\beta_2^2}{\eta^2} C_\cO I_2(d-2\eta) +C_\cO \frac{3}{4} I_2'(d) \beta_3 +O(\eta).
\end{align}
The counting of these divergences is in principle straightforward, but we strongly recommend cross-checking against simple perturbative examples; we do so later in \cref{sec:GenFFreeCheck} with the generalised free field, and our results in \cref{sec:perturbative} for the $\phi^4$ long-range theory are also consistent. 

Hence, \eqref{eq:etaFlamexpansion} becomes
\begin{align}
   \delta F_\lambda &= -\frac{\lambda_b^2}{2} C_\cO I_2(d-\eta)  +\frac{\lambda_b^3}{3!} C_\cO C_{\cO\cO\cO} I_3(d-\eta)  - \lambda^4 C_\cO \left(\frac{3}{4} I_2'(d) \beta_3 + O(\eta) \right)+O(\lambda^6).
\end{align}
Then, we find that the only divergence, which comes at order $\lambda^4$ from the $1/\eta$ pole in \eqref{eq:finalIntegral}, cancels perfectly against the $1/\eta$ pole from the $\lambda_b^2$ term.

We want to compute $\tilde{F}$ for a critical point with $\lambda \sim \eta$.
Hence, we must keep $\tilde{F}$ to a consistent order in $\lambda^{n-m}\eta^{m}$ $\forall m \in \mathbb{N}_0$, i.e.
\begin{equation}
\begin{aligned}
\delta\tilde{F}_\lambda &= \frac{2\pi^{d+1}}{\Gamma(d+1)} C_\cO \Bigg[ -\frac{\eta }{2} \lambda^2 (1+\eta  L)+ \frac{S_{d-1}}{2} C_{\cO\cO\cO} \lambda^3 \left(\frac{1}{3} + \eta  L\right)\\
&\qquad\qquad+\lambda^4 \Big(\frac{\beta_3}{4}-\left(\frac{S_{d-1}}{2}\right)^2\frac{C_{\cO\cO\cO}^2 L}{2}\Big)+O(\lambda^{5-m}\eta^m)\Bigg],
\end{aligned}
\end{equation}
where we choose the convenient combination $L\equiv \log(2\mu R)+\frac{\psi_0(-d/2)+\gamma}{2}$.

\subsection{The case \texorpdfstring{$C_{\cO\cO\cO}=0$}{C\_OOO=0}}\label{sec:COOOzero}

If $\beta_2 \propto C_{\cO\cO\cO}=0$, as in our long-range case, then the perturbative zero of the beta function \eqref{eq:CPTbeta} exists at $\lambda \sim \sqrt{\eta}$.
Keeping $\tilde{F}$ to a consistent order in $\lambda^{n-2m}\eta^{m}$ $\forall m \in \mathbb{N}_0$, we find that the factors of $L$ drop out:
\begin{align}
  \delta \tilde{F}_\lambda &= \frac{2\pi^{d+1}}{\Gamma(d+1)} C_\cO \left( -\eta \frac{\lambda^2}{2} + \beta_3\frac{\lambda^4}{4}  + O(\lambda^{6-2m}\eta^{m})\right).
\end{align}
Therefore, at the critical value of $\lambda_\star= \sqrt{\eta/\beta_3}(1+O(\eta))$ we find
\begin{align}\label{eq:deltaFtlam}
    \delta \tilde{F}_{\lambda_\star} &= -\frac{\pi^{d+1}}{\Gamma(d+1)} \frac{C_\cO}{\beta_3}\frac{\eta^2}{2}+ O(\eta^3).
\end{align}
The ratio $\beta_3/C_\cO$ can be seen via \eqref{eq:fDef} to be manifestly independent of the normalization of $\cO$, as it should be.
Assuming the original CFT is unitary, we know that $C_\cO>0$. 
If the CFT is real we know that $\beta_3/C_\cO$ must be real, but as mentioned above its positivity properties are not manifest.

As expected, the renormalized operator at the IR fixed point is irrelevant,
\begin{equation}\label{eq:IRoperator}
    \Delta_\cO\rvert_{\lambda_\star} = d+ \gamma_\cO= d+ \lim_{\lambda \to \lambda_\star} \beta_\lambda \pdv{}{\lambda} \log Z_\cO \rvert_{\lambda_\star} = d+ \odv{\beta_\lambda}{\lambda_\star} = d+2\eta + O(\eta^2),
\end{equation}
where we substituted $\beta_\lambda = -\eta/\pdv{\log \mu^{-\eta} \lambda_b}{\lambda}$ and used $\beta_{\lambda_\star}=0$. 

Recall that $\eta>0$, as we have assumed that we are flowing towards the IR. 
This means that if $\beta_3<0$ then $\lambda_\star$ would be imaginary, and so we would not flow to the CFT on perturbing with $\mathbb{R} \cO$, as the fixed point does not lie in the space of real coupling constants.
Hence, if there is an IR CFT that can be reached (i.e. for real coupling), then we are free to assume $\beta_3>0$ when considering the CPT flow between CFTs. %
We also demonstrate this, as well as gradient flow of the beta functions, in the case of a perturbation by multiple operators in \cref{app:multiOperator}.
We have therefore proven the generalised $F$-theorem to leading order in conformal perturbation theory, for the case where the quadratic term in the beta function vanishes.

In noninteger dimensions, the existence of evanescent operators means that all theories are nonunitary\footnote{More properly, the $d$-continued CFT is always nonunitary, and for integer $d$ there is a subset of operators which are closed under the OPE that defines the unitary theory \cite[\S 2.5]{Henriksson:2025kws} \cite{Zan:2026oyb}.}. 
However, if the $g_\star^i$s are real and the scalars with which we deform the action have positive two-point function, we also find maximisation for $g\propto$ both $\sqrt{\eta}$ and $\eta$, even in noninteger $d$\footnote{Compare a similar comment \cite[footnote 4]{Fei:2015oha}, made in the context of the applicability of the generalised $F$-theorem to noninteger $d$.}.
Of course, in some ranges of $d$ the CFT may become complex, and then maximisation may no longer apply, as with the $\gO(N)$ model in $4<d<6$ \cite{Giombi:2019upv}.

\subsection{A proposal for \texorpdfstring{$\hat{F}$}{\^ F}} \label{sec:proposedFhat}

We now take a slight detour to introduce a new normalization for the free energy -- this is not strictly necessary, but it will simplify our formulae considerably.
When compared to \eqref{eq:FtLR} and the other known computations of $\Ft$, our result \eqref{eq:deltaFtlam} further suggests that the most natural factor to remove from $F$ for a \enquote{nice} result is 
\begin{align}\label{eq:proposedFhat}
    &F= \frac{1}{-\sin(\pi d/2)} \tilde{F} \equiv \frac{2\pi}{-\sin(\pi d/2)}\frac{\Gamma(\dotwo)^2}{\Gamma(d+1)} \hat{F}, \\
   &\boxed{ \hat{F} = \frac{\Gamma(d)}{\Gamma(-\dotwo) \Gamma(\dotwo)^3} F , } %
\end{align}
which defines our proposed quantity $\hat{F}$.
This further factor $2\pi\Gamma(\dotwo)^2/\Gamma(d+1)$ is both monotonic in $d$ and always positive for $d>0$, and so does not modify the conjectural generalised $F$-theorem \cite[footnote 2]{Fei:2015oha}, but it is neater, in removing overall factors that do not carry information.
Notably, it makes the leading term in $\hat{F}$, when computed in conformal perturbation theory, precisely the integral of the Zamolodchikov metric times the beta function, as we show in \cref{app:multiOperator}\footnote{
Accounting for the nontrivial two-point normalization of $\cO=\phi^4$, $C_{\cO}=24 C_\phi^2$, our $\hat{F}$ now matches $A = (\pi/288)^{-1} \tilde{F}$ defined in \cite[(4.5)]{Pannell:2025ixz} in the context of gradient flow in the RG.
}.

$\hat{F}$ proves most useful when dealing with the $\epsilon$ expansion: the $\pi^2 \epsilon^5$ term in the forthcoming result \eqref{eq:Ft4meps} for the free energy of the long-range $\phi^4$ theory is eliminated when we give $\hat{F}$ instead in \eqref{eq:Fhat4meps}, along with the overall factors of $\pi$ and the large denominators.
In the cubic theory, the general-$d$ result is not so clean, but it is still simpler than $\tilde{F}$.
The known large-$N$ results are essentially unchanged, as in \eqref{eq:largeNFt} and \cite{Fraser-Taliente:2025udk} they are written with an overall prefactor $\propto \tilde{F}_{b/f}'(s)$, which we just replace with $\hat{F}_{b/f}'(s)$.

\begin{figure}[H]
\centering
\includegraphics[width=0.72\linewidth]{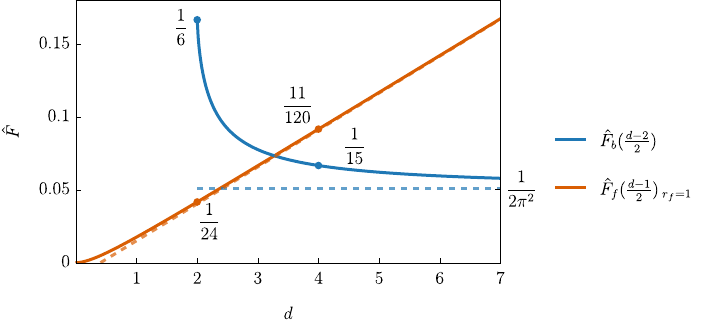}
\caption{The free energy of a free scalar $\hat{F}_b(\tfrac{d-2}{2})$ and a single \textit{real} fermion component $\hat{F}_f(\tfrac{d-1}{2})\rvert_{r_f=1}$ as functions of the spacetime dimension $d$.
We show the exact rational values at $d=2,4$; for bosons we must stop at $d=2$.
We take $r_f=1$, which conveniently ensures that they are on approximately the same scale.
The large-$d$ asymptotic limits from \eqref{eq:largeD} are also shown dashed: they are constant and affine respectively.}
\label{fig:FfreePlot}
\end{figure}

One pleasant improvement is that we now find for the free scalar and one component of a free real fermion ($r_f=1$), the new versions of $\tilde{F}_b'(\Delta)$ \eqref{eq:bosonF} and $\tilde{F}_f'(\Delta)$ \eqref{eq:fermionF} are
\begin{align}\label{eq:FhatsFB}
    \hat{F}_b'(\Delta) &= \half \frac{1}{\EulerBeta(\dotwo, \dotwo-\Delta)\EulerBeta(\dotwo, \Delta-\dotwo)}= \half \frac{1}{\EulerBeta(\dotwo, \zeta)\EulerBeta(\dotwo,-\zeta)}\\
     \hat{F}_{f}'(\Delta) &= -\frac{r_f}{2} \frac{1}{\EulerBeta(\frac{d}{2},\frac{d}{2}-\Delta +\frac{1}{2}) \EulerBeta(\frac{d}{2},\Delta-\frac{d}{2}+\frac{1}{2})} = -\frac{r_f}{2} \frac{1}{\EulerBeta(\frac{d}{2},\frac{1}{2}+\zeta) \EulerBeta(\frac{d}{2},\frac{1}{2}-\zeta)},
\end{align}
for $\Delta=\dotwo -\zeta$; the formulae are extremely similar, and compact in terms of Euler beta functions. %
We find the large-$d$ limits of the free scalar and fermion to be
\begin{equation}
\lim_{d\to \infty}\hat{F}_b(\tfrac{d-2}{2}) = \frac{1}{2\pi^2} +O(1/d), \quad \lim_{d\to \infty}\hat{F}_f(\tfrac{d-1}{2}) = \frac{r_f}{\pi^2}\left(\frac{d}{4} - \frac{1}{\pi^2} + O(1/d)\right), \label{eq:largeD}
\end{equation}
where it is notable that the contribution of a single free scalar is asymptotically constant, and that of a free fermion component is asymptotically affine.
Additionally, as we can see in \cref{fig:FfreePlot}, the next-order correction to $\hat{F}_f$ is very small, $\frac{r_f}{d}(12-\pi^2)/\pi^6 \simeq 0.0022/d$, and so $\hat{F}_f$ as given in \eqref{eq:largeD} is wrong by no more than 4\% for all $d>2$.
The next-order correction to $\hat{F}_b$ is $(\pi^2 - 6)/(\pi^4 d)$, which is again quite similar.

We note that $\hat{F}_{2d}=\frac{c_{2d}}{6}=\frac{a_{2d}}{2}$, $\hat{F}_{3d}=\frac{12}{\pi^2} F$, and $\hat{F}_{4d}=6 a_{4d}$; in general even dimension, $\hat{F}$ is related to the conventional Weyl anomaly by $\hat{F}_{d=2n}=\frac{n^2}{4} \binom{2n}{n} a$, and 
\begin{align}
    \hat{F}_b(\dotwo-1)\rvert_{d=2n} &= \half \int_0^1 \odif{x}\, x^2 \prod _{k=1}^{n-1} \left(1-\frac{x^2}{k^2}\right)\\
    \hat{F}_f(\tfrac{d-1}{2})\rvert_{d=2n} &= \frac{r_f}{2} \int_0^{\thalf} \odif{x}\, x(n-x) \prod _{k=1}^{n-1} \left(1-\frac{x^2}{k^2}\right).
\end{align}

\subsection{\FhatextOrPDF-maximisation for long-range theories}\label{sec:specialiseGF}

We can now specialise our results to the long-range case, where the CFT that we are perturbing about is a combination of some short-range CFT and a GFF $\chi$.
That is, we find an operator $\hat{\phi}$ in the short-range CFT of dimension $\Delta^\mathrm{SR}$, and then define
\begin{equation}\label{eq:CPTaction}
    S_\lambda \equiv S_\text{SR CFT} + \int_x \frac{Z_\chi}{2}\chi \WeylOp_{\tfrac{d}{2}-\Delta_\chi} \chi + \lambda_b \hat{\phi} \chi,
\end{equation}
where $\Delta_\chi = d-\Delta^\mathrm{SR}-\eta$. 
In this theory, by symmetry we know that all $\lambda^{2n+1}$ corrections to the free energy and all $\lambda^{2n}$ corrections to the beta function vanish.
In the theory with $\lambda=\lambda_b=0$ we have
\begin{equation}
    \expval{\hat{\phi}(x)\hat{\phi}(y)}_{\lambda=0} = \frac{\hat{C}_\phi}{s(x,y)^{2\Delta^\mathrm{SR}}},
    \quad \expval{\chi(x)\chi(y)} = \frac{\hat{C}_\chi}{s(x,y)^{2(d-\Delta^\mathrm{SR}-\eta)}},
\end{equation}
where $\hat{C}_\chi = 1/(\cF_{2\Delta_\chi,0} Z_\chi)$\footnote{
$\hat{C}_\chi$ should be positive in a unitary theory, since in radial quantization it is the norm of the state created by the operator; for e.g. $\dotwo <\Delta_\chi<\dotwo +1$ this means that $Z_\chi$ should be negative. 
}. 
Because $\chi$ is decoupled, $\cO=\hat{\phi}\chi$ is an operator of dimension $d-\eta$:
\begin{equation}
   \expval{\cO(x)\cO(y)}_{\lambda=0} = \frac{C_\cO}{s(x,y)^{2(d-\eta)}}.
\end{equation}
If the original CFT is unitary, then we should have $C_\cO=\hat{C}_\phi \hat{C}_\chi>0$. 
The $\mathbb{Z}_2$ symmetry of \eqref{eq:CPTaction}, $\chi \to -\chi$, indicates that $C_{\cO\cO\cO}=0$, so we need our result above to compute the free energy.

The free energy of the unperturbed theory is
\begin{equation}
    \hat{F}_{\lambda=0} = \hat{F}_\text{CFT} + \hat{F}_b(\Delta_\chi).
\end{equation}
Our goal is to compute the change in the free energy of \eqref{eq:CPTaction} at the IR fixed point, i.e. $\delta \hat{F}_{\lambda_\star}$.
Additionally, evidently if $\beta_3 <0$ then $\lambda_\star$ must be complex.
So, assuming that our long-range CFTs are not complex, we know that $\beta_3 >0$. 
Thus, along a line of real long-range fixed points that are found by perturbing around a unitary short-range CFT, we see that $\delta \hat{F}_{\lambda_\star}<0$.

Note that the maximisation applies to the change in free energy. 
However, $ \hat{F}_{\lambda_\star}$ also contains the Gaussian contribution $\hat{F}_b(\Delta_\chi)$, the auxiliary field implementing the nonlocal kinetic term for $\hat{\phi}$.
Thus, as discussed in \cref{sec:notincludechi}, we consistently choose to subtract off that contribution
\begin{equation}
    \hat{F}_\mathrm{LR}\equiv \hat{F}_{\lambda_\star}-\hat{F}_b(d-\Delta^\mathrm{SR}-\eta) = \hat{F}_\mathrm{CFT} + \delta \hat{F}_{\lambda_\star}.
\end{equation}
Taking $\Delta_\phi \equiv \Delta^\mathrm{SR}+\eta$, we can use \eqref{eq:deltaFtlam} to find that if $\beta_3/C_\cO >0$, then
\begin{align}
    \hat{F}_\mathrm{LR} &=\hat{F}_{\mathrm{CFT}} -\frac{\pi^{d}}{\Gamma(\dotwo)^2}\frac{ C_\cO}{\beta_3} \frac{\eta^2}{4} + O(\eta^3) < 0,\\
    \implies \,\odv[2]{\hat{F}_\mathrm{LR}}{\Delta_\phi}\Big\rvert_{\eta=0} &= - \frac{\pi^{d}}{\Gamma(\dotwo)^2}\frac{C_{\cO}}{2\beta_3} < 0, \label{eq:FhatOneField} %
\end{align}
i.e. we find $\hat{F}$-maximisation if the unitary CFT can be deformed to a real IR fixed point (i.e. $\beta_3/C_\cO>0$)\footnote{The extra factor of $\pi^d/\Gamma(d/2)^2$ here is only present because we are not yet using the natural normalization for the coupling $\lambda$, cf. \eqref{eq:multiLRDeformation}. \label{footnote:naturalNorm}}.
This is indeed maximisation: $\hat{F}$ decreases for both signs of $\eta$, even though for $\eta<0$ we find that $\lambda_\star$ is complex.
The only condition we needed is that there exists a real IR fixed point in one of the directions (i.e. $\beta_3 >0$).

\subsubsection{Multiple long-range fields}

This can be generalised to a multi-operator deformation of a unitary theory (i.e. a multidimensional family of long-range theories) using our generalisation of the above CPT computations to multiple operators in \cref{app:multiOperator}.
Take the original CFT to have operators $\hat{\phi}_i$, which without loss of generality can be taken to have correlators
\begin{equation}
    \expval{\hat{\phi}_i (x) \hat{\phi}_j(y)} = \frac{\delta_{ij}}{s(x,y)^{2\Delta^{\mathrm{SR}}_i}}.
\end{equation}
We set each $\chi_i$ to have scaling dimension $\Delta_{\chi_i} = d- \Delta^{\mathrm{SR}}_i- \kappa^i \eta$.
Our operators are then
\begin{equation}
    \cO_i =\hat{\phi}_i \chi_i,
\end{equation}
which we define to have two-point coefficient $C_{ij}$ (which is automatically diagonal from our assumption of non-equal $\kappa_i$s). Since each $\chi_i$ has its own $\mathbb{Z}_2$, we also know that all $C_{ijk}=0$.
Then, we deform the combination of the unitary CFT and $\sum_i \chi_i$ by
\begin{equation}\label{eq:multiLRDeformation}
S_\mathrm{int} = \frac{\Gamma(\dotwo)}{\pi^{d/2}} \sum_i \int_x \lambda_{0}^i \cO_i,
\end{equation}
where we have used the convenient normalization defined in the appendices.
We then tune to the long-range fixed point at $\lambda^i=\lambda_\star^i$ (the value of which depends on the $\kappa^i$s), where we find that the long-range fields have scaling dimensions
\begin{equation}
    \Delta_{i} = \Delta^{\mathrm{SR}}_i + \kappa^i \eta.
\end{equation}

From \eqref{eq:FmaxMultiDirections}, (as usual, we do not consider the contribution from $\sum_i\hat{F}_b(\Delta_{\chi_i})$) we find that the change in $\hat{F}$ is
\begin{equation}\label{eq:FmaxForMultis}
    \hat{F}_{\mathrm{LR}} - \hat{F}_{\mathrm{CFT,SR}} = - \frac{C_{ij}^{(\kappa)}}{4} \eta \lambda^i_\star \lambda^j_\star +O(\eta^3).
\end{equation}
The fixed points $\lambda_\star^i= \sqrt{\eta} (h^i(\kappa) + O(\eta))$ solve the beta function \eqref{eq:nonEqualDimsBeta} to leading order, so
\begin{equation}
-\kappa^i h^i + D\indices{^i_{jkl}}h^j h^k h^l=0 \quad \text{(no sum on $i$)}.
\end{equation}
Thus, assuming that $h^i(\kappa)$ is real (i.e. there exists a real IR long-range CFT in the direction given by $\kappa^i$), we find that \eqref{eq:FmaxForMultis} becomes
\begin{equation}\label{eq:FhatVarKappaLR}
    \hat{F}_{\mathrm{LR}} - \hat{F}_{\mathrm{CFT,SR}} = - \frac{C_{ij}^{(\kappa)}}{4} \eta^2 h^i h^j +O(\eta^3) < 0.
\end{equation}
This follows because $C_{ij}$ is positive definite, so $C_{ij}^{(\kappa)} = \kappa^i C_{ij} = \kappa^j C_{ij}$ is also; hence the change in $\hat{F}$ when moving to the long-range theory must be negative if the $\lambda_\star^i$s are real. 
Hence, in every allowed direction given by $\kappa^i$ we find 
\begin{equation}\label{eq:FmaxAlongLine}
    \odv[2]{\hat{F}_{\mathrm{LR}}}{\eta}\Big\rvert_{\eta=0}= -\half C_{ij}^{(\kappa)} h^i(\kappa) h^j(\kappa) <0.
\end{equation}  %
Thus, along any line $\{\Delta_{i}(t)\}$ of real long-range CFTs parametrised by $t$, leading to a unitary local CFT, we find that $\hat{F}_\mathrm{LR}(\{\Delta_{i}(t)\})$ is maximised by the local CFT. %

\subsubsection{The Hessian of the long-range \texorpdfstring{$\hat{F}$}{\^{}F}}\label{app:Hessian}

Having computed the multidimensional long-range $\hat{F}_\mathrm{LR}(\{\Delta_i\})$, we might want to investigate its Hessian.
At the moment, we have only shown that it is maximised along any line of real long-range CFTs that leads to the unitary CFT \eqref{eq:FmaxAlongLine}.

We find that for $n_c$-dimensional long-range families where all of the $\hat{\phi}_i \chi_i$ couplings become irrelevant in the IR, the Hessian with respect to the $n_c$ scaling dimensions $\Delta_i$ is negative definite.
Hence, if the CFT is stable to the nonlocal deformations, then $\hat{F}_\mathrm{LR}(\{\Delta_i\})$ is maximised.

However, a full Hessian only makes sense if nearby LR fixed points form a smooth family.
Let us work at the fixed point, and for convenience we take the UV Zamolodchikov metric $C_{ij} = \delta_{ij}$; we therefore henceforth ignore whether an index is raised or lowered.
Throughout the following the index $i$ will never be summed.
We will need to use the multi-operator CPT machinery from \cref{app:multiOperator}.

Obstructions to the existence of the Hessian arise if we are on a non-generic branch of solutions; that is, if we are approaching the short-range CFT in a family of long-range solutions where one of the couplings $\lambda^k=0$.
However, by changing which $\hat{\phi}$s we include in our set we can guarantee that all $\lambda^i$s are nonzero. 
This is completely natural, as that is exactly the requirement that the $\Delta_i$s which we are differentiating $\hat{F}$ with respect to are indeed the scaling dimension of fundamental fields. 
We have implicitly assumed this throughout this paper, but we make this assumption explicit in this section for the avoidance of confusion.
We also define $\nu^i = \kappa^i \eta$ such that $\Delta_i = \Delta^{\mathrm{SR}}_{i}+\nu^i$.

Consider the long-range CFTs that exist around some short-range CFT. 
Select one long-range CFT where $\nu^i=\nu^i_{(0)}$ and $\lambda^i=\lambda^i_{(0)}(\nu_{(0)})$ solve the beta functions $\beta_i(\nu_{(0)},\lambda_{(0)})=0$.

\textbf{The CPT beta functions simplify in the LR}.
Now, we have two facts: (1) we have an independent $\mathbb{Z}_2$ symmetry for each $\lambda^i$, and (2) if we do not turn on a long-range deformation, it will not be spontaneously generated by other deformations.
These mean that $\beta^i$ is $\lambda^{m\neq i}$-even and $\lambda^i$-odd, and is overall proportional to $\lambda^i$, i.e.
\begin{equation}
    \beta^i = \lambda^i G_i(\nu, x), \quad x^i \equiv (\lambda^i)^2,
\end{equation} 
for a function $G_i$ that is smooth at $x=0$.
Specifically, this means that
\begin{equation}\label{eq:Gi}
    G_i = -\nu^i + \sum_j A_{ij} x^j +O(x^2), \quad A_{ii} = D_{iiii}, \quad A_{ij} = 3D_{iijj}\quad (i \neq j),
\end{equation}
where $A$ is evidently symmetric, and depends only on the data $D_{ijkl}$ of the local CFT. %
Differentiating $G_i=0$ with respect to $\nu$, and assuming that the family of LR theories is smooth, we see that  
\begin{equation}\label{eq:Ainvertible}
    A_{ij} \pdv{x_j}{\nu^k}\rvert_{\nu=0} = \delta_{ik},
\end{equation}
and so $A$ is invertible.

\textbf{The anomalous dimension matrix, and a related quantity $A$}.
Assuming that in this family the long-range CFT has no marginal operators, we know that the anomalous dimension matrix at the fixed point $\lambda_{(0)}$,
\begin{equation}
\Gamma_{ij} \equiv \pdv{\beta_i}{\lambda^j}\Big\rvert_{(0)} = 2 \lambda_{(0)}^i \lambda_{(0)}^j A_{ij} + O(\nu^2),
\end{equation}
must be symmetric, with all eigenvalues nonzero, and therefore is invertible (the equality here follows from $\beta_i=0$). 
We call $\Gamma_{ij}$ the anomalous dimension matrix because its eigenvalues $\Gamma_i$ appear in the anomalous dimensions $\Delta_i = d+ \Gamma_i +O(\nu^2)$ of the near-marginal operators at the long-range fixed point.

Defining $D_\lambda \equiv \mathrm{diag}(\lambda^1, \lambda^2, \cdots)$, we see that $\Gamma = 2 D_{\lambda_{(0)}} A D_{\lambda_{(0)}} + O(\nu^2)$. 
Since all $\lambda_{(0)}$s are nonzero, $D_{\lambda_{(0)}}$ is invertible, confirming that $A+O(\nu)$ must be invertible. 
Further, we see that if $\Gamma$ is positive definite, i.e. $0 < v^T \Gamma v = 2( D_{\lambda_{(0)}}  v)^T A ( D_{\lambda_{(0)}}  v) +O(\nu^2)$ for all vectors $v \neq 0$, then so is $A+O(\nu)$, because $\lambda_{(0)}\neq 0$.
Taking $\nu \to 0$, this tells us that $A$ is nonnegative definite.
However, since \eqref{eq:Ainvertible} tells us that $A$ is also invertible, we finally conclude that $A$ must be positive definite if $\Gamma$ is. 

\textbf{Computing the Hessian (it's just $-\half A$)}.
Since $A$ is invertible, %
we have a unique smooth solution to $G_i=0$, namely
\begin{equation}
    x^i(\nu) = (A^{-1})_{ij} \nu^j + O(\nu^2),
\end{equation}
which we note is not guaranteed to give real values of the couplings.
Then, using \eqref{eq:FmaxMultiDirections}, the LR free energy is 
\begin{equation}
\hat{F}_{\mathrm{LR}} - \hat{F}_{\mathrm{CFT,SR}} = -\frac{1}{4} \nu^j x^j(\nu) +O(\nu^3) = -\frac{1}{4} \nu^j (A^{-1})_{jk} \nu^k +O(\nu^3),
\end{equation} 
and so the Hessian of the long-range free energy around the local point in a full-dimension family (i.e. all $\lambda^i \neq 0$) is\footnote{This agrees with \eqref{eq:FmaxAlongLine} in the one-dimensional case (taking $C_{11}=1$, $D_{1111}=D$, $\nu^1=\eta$, and $\lambda^1 = \sqrt{\eta/D}$), since we know from \eqref{eq:IRoperator} that $\Gamma\indices{_{11}}=2\eta =2(\lambda^1)^2 A_{11}$ and so $(A^{-1})_{11} = D^{-1}$.}
\begin{align}
\pdv{\hat{F}_{\mathrm{LR}} }{\Delta_i,\Delta_j} \Big\rvert_{\nu=0}= -\frac{1}{2} (A^{-1})_{ij},
\end{align}
for $\Delta_i = \Delta^{\mathrm{SR}}_i+\nu^i$ -- this evidently depends only on short-range data. 

Recall that this matrix $A$ is positive definite if the anomalous dimension matrix $\Gamma$ of a generic long-range CFT near the short-range point is as well -- that is, if none of the operators become slightly relevant (i.e. $\Gamma$ has no eigenvalue $\Gamma_i <0$).

Thus, our complete statement is the following.
Deforming any CFT by $n_c$ couplings $\lambda^i$ to GFFs, we find a manifold of long-range CFTs. 
Assume that all of those couplings are nonzero, i.e. we approach the local CFT from a generic direction in that ${n_c}$-dimensional real manifold parametrised by the $\Delta_i$ (and, critically, we are not on a lower-dimensional branch). 
We know that the local CFT extremises $\hat{F}(\{\Delta_i\})$. 
We have now shown that it also locally maximises $\hat{F}(\{\Delta_i\})$ if and only if $A$ is positive definite. 
This is true if and only if none of the GFF couplings are slightly relevant in the long-range theories near the fixed point.

It would be nice to prove the positivity of $A$ directly from the form of $D$, but that may be difficult, as mentioned regarding the positivity of $\beta_3$ at the end of \cref{sec:Beta3Expression}.

\subsection{Cross-check: perturbing the free scalar}\label{sec:GenFFreeCheck}

When perturbing around the local free scalar, we know everything, and can use it to check the computations in this section for the case where $C_{\cO\cO\cO}=0$.
Consider for the moment the generalised free field, which has action and two-point function
\begin{equation}\label{eq:GFFGaussian}
    S=  \int_x\half \phi \WeylOp_{\dotwo-\Delta} \phi, \quad \expval{\phi(x)\phi(y)} = \frac{C_\phi}{s(x,y)^{2\Delta}}.
\end{equation}
The computation of $\hat{F}$ is standard, \cite[\S A.3]{Fraser-Taliente:2025udk}, and it is easy to demonstrate $F$-extremisation. 
The derivative of $\hat{F}$ is
\begin{equation}\label{eq:FhatDerivGaussian}
    \hat{F}_{b}'(\Delta)= \half \frac{\Gamma (\Delta ) \Gamma (d-\Delta )}{\Gamma (\frac{d}{2})^2 \Gamma (\frac{d}{2}-\Delta ) \Gamma (\Delta -\frac{d}{2})},
\end{equation}
which evidently vanishes at the local free scalar points (or $\Box^k$ CFTs) $\Delta^{\mathrm{SR}}=\frac{d}{2}-k$ with flat-space action\footnote{This holds not just for $k\in\mathbb{N}$ but also for $k \in \mathbb{Z}$: %
the negative-$k$ regime, $\phi(-\partial^2)^{-m}\phi$, where $k=-m$, can be implemented by integrating out an auxiliary local $\Box^m$ scalar $\psi$ from the action $S=\int_x \half \psi (-\partial^2)^m \psi +  \thalf \phi (-\partial^2)^{K>0} \phi+ g_0\psi \phi$. That is, they admit a local completion (see e.g. \cite[\S 7]{Gracey:2017erc}). Hence, despite appearances, even without the explicit completion, they will have a local stress tensor, and extremise $\hat{F}_b$. 
Extremising \eqref{eq:largeNFt} shows that large-$N$ fixed points with $\Delta_\phi = \dotwo + m +O(1/\sqrt{N})$ and $\Delta_\phi = \dotwo + m +O(N^{-2/5})$ appear to exist (though strangely, we do find solutions $\Delta_\phi = \dotwo + m + \half +O(1/N)$).  %
We will not see them in \cref{sec:perturbative}, as there we study only the theories that are weakly coupled around $d=4k>0$. %
\label{footnote:boxNegKAlso}} $S=\int_x \phi (-\partial^2)^k\phi$. 
We can also see that $\hat{F}_{b}(\Delta)$ is only continuous in the range $0<\Delta<d$ to the poles in the derivative \eqref{eq:FhatDerivGaussian} at $\Delta=0,d$.

For $k=1$, the second derivative is $\hat{F}_b''(\dotwo -1)$, which is less than zero for all $d>2$, and so we have a maximum -- agreeing with the unitarity of the free scalar in arbitrary integer $d > 2$.
For general $k$, we find that $\mathrm{sgn}(\hat{F}_b''(\dotwo -k))=(-1)^k$ for all $d>2k$, providing trivial examples of minimisation. 
The same applies for the generalised free fermion \eqref{eq:FhatsFB}: it vanishes for $\dotwo -\thalf - k_{\ge 0}$, and the canonical free fermion satisfies $\hat{F}_f''(\tfrac{d-1}{2})<0$ for all $d>0$.
For general $k$, we then have $\mathrm{sgn}(\hat{F}_f''(\tfrac{d-1}{2} -k))=(-1)^{k+1}$ for all $d>2k$.

We can also try to replicate this result diagrammatically, using that CFT+$\hat{\phi}\chi$ construction of the same GFF theory.
This is a useful cross-check on the divergence counting of the previous section.

\subsubsection{Exact computation}

Thus, consider the long-range theory obtained by deforming the higher-derivative free scalar CFT with $\hat{\phi}\chi$:
\begin{equation}
    S_{\mathrm{CFT}+\hat{\phi}\chi} \equiv \int_x \half \hat{\phi} \WeylOp_k \hat{\phi} +  \frac{1}{2}\chi \WeylOp_{\eta-k} \chi +\lambda_0 \cO, \quad \cO=\hat{\phi}\chi.
\end{equation} 
Setting $\lambda_0=0$ and computing the free energy $F_\lambda=-\log \int \Dd{\hat{\phi}}\Dd{\chi} \, e^{- S_{\mathrm{CFT}+\hat{\phi}\chi} }$, we find
\begin{equation}\label{eq:GaussianFhatLambdaIsZero}
    \hat{F}_{\lambda=0} \equiv \hat{F}_b(\dotwo - k) + \hat{F}_b(\dotwo+k-\eta),
\end{equation}
which we note includes the contribution from $\chi$.

\Cref{sec:CFTplusPhihatChi} then tells us that at the IR fixed point of $\lambda$, we find a long-range scalar of dimension $\Delta=\dotwo-k + \eta$ in the IR, i.e. 
\begin{equation}\label{eq:GaussianFhatLambdaStar}
    \hat{F}_{\lambda=\lambda_\star} \equiv \hat{F}_b(\dotwo - k+\eta) + \hat{F}_b(\dotwo+k-\eta)=0.
\end{equation}
This happens to be exactly zero in this simple Gaussian case.
Although perhaps surprising, this is consistent.
If we attempt to follow \cref{sec:proof} with this action, differentiating $F_{\lambda_\star}$ with respect to $\eta$ we find $\odv{}{\eta}F_{\lambda_\star}\propto\expval{\chi(x)\chi(y)}_{\lambda_\star}=0$ for all values of $\eta$\footnote{Note that we do not modify the normalization of the $\chi$ field with a $Z_\chi$ term here, which is the reason that this correlator vanishes. This can be seen exactly by inverting the $2\times 2$ matrix of $\phi,\chi$ kinetic terms and sending $\lambda_0 \to \infty$.}.

We can derive this nonperturbatively by integrating out $\chi$.
That is,
\begin{align}
    F_{\lambda} &= F_b(\dotwo+k-\eta)-\log\int \Dd{\hat{\phi}} \exp(-\int_x \half \hat{\phi} \left[\WeylOp_k -\lambda_0^2\WeylOp_{k-\eta} \right]\hat{\phi})\\
    &= F_b(\dotwo+k-\eta)+\half \Tr \log\left[\WeylOp_k -\lambda_0^2\WeylOp_{k-\eta} \right]
\end{align}
Using the procedure described in \cite[\S A.3]{Fraser-Taliente:2025udk} \cite{Benedetti:2021wzt}, this second term can be evaluated directly.
It interpolates between $F_b(\dotwo-k)$ when $\lambda_0=0$ and $F_b(\dotwo-k+\eta)$ as the bare coupling $\lambda_0 \to \infty$ (i.e. $\lambda \to \lambda_\star$)\footnote{We use the fact that in DREG $\Tr \log(\lambda_0^2 \delta(x-y))=0$.}, yielding \eqref{eq:GaussianFhatLambdaIsZero} and \eqref{eq:GaussianFhatLambdaStar}.

However, as discussed in \cref{sec:notincludechi}, we do not include the $\chi$ contribution in our definition of $\hat{F}_\mathrm{LR}\equiv  \hat{F}_{\lambda}-\hat{F}_b(\dotwo+k-\eta)$, and so define in general
\begin{equation}
   F_{\mathrm{LR},\lambda} = \half \Tr \log\left[\WeylOp_k -\lambda_0^2\WeylOp_{k-\eta} \right] \quad \implies \quad \hat{F}_{\mathrm{LR},\lambda_\star} = \hat{F}_b(\dotwo -k + \eta).
\end{equation}
$\hat{F}_{\mathrm{LR},\lambda_\star}$ is the quantity that matches \eqref{eq:GFFGaussian}, and hence is extremised by the local theories.

\subsubsection{Comparison to CPT}

For small $\eta$, the change in free energy should be 
\begin{equation}\label{eq:freeScalarChange}
    \delta \hat{F}_{\lambda_\star} = \hat{F}_{\lambda_\star}-\hat{F}_{\lambda=0} = \hat{F}(\dotwo-k+\eta)-\hat{F}(\dotwo-k) = \hat{F}''(\dotwo-k) \frac{\eta^2}{2} +O(\eta^3).
\end{equation}
Let us now reproduce this perturbatively.

Considering the $\lambda^2$ contribution to the two-point function of $\cO$ in flat space, we find six different Wick contractions, which yield three diagrams:
\begin{align}
&\frac{1}{2!}\int_{w,z}\expval{\cO(w)\cO(x)\cO(y)\cO(z)}_c = C_\cO^2\int_{w,z} P_{xy} C_{yz} P_{zw} C_{wx} + P_{xz} C_{zy} P_{yw} C_{wx} + P_{xz} C_{zw} P_{wy} C_{yx}\notag\\
&=\frac{C_\cO^2}{\abs{x-y} ^{2(d-2\eta)}}
\Big[U(\tfrac{d}{2}-\eta ,d-\eta -\Delta^{\mathrm{SR}},2 \eta +\Delta^{\mathrm{SR}}-\tfrac{d}{2}) U(\eta ,d-\eta -\Delta^{\mathrm{SR}},\Delta^{\mathrm{SR}})\\
&\quad + U(\eta ,d-\eta -\Delta^{\mathrm{SR}},\Delta^{\mathrm{SR}})^2 + U(\tfrac{d}{2}-\eta ,\tfrac{d}{2}+\eta -\Delta^{\mathrm{SR}},\Delta^{\mathrm{SR}}) U(\eta ,d-\eta -\Delta^{\mathrm{SR}},\Delta^{\mathrm{SR}})\Big]\notag\\
& = \frac{C_\cO^2}{\abs{x-y} ^{2(d-2\eta)}} \frac{1}{\eta}\frac{3 \pi ^d (-1)^k}{k \Gamma (\frac{d}{2}-k) \Gamma (\frac{d}{2}+k)} + O(\eta^0),
\end{align}
where $P_{xy}=1/\abs{x-y}^{2\Delta^{\mathrm{SR}}=d-2k}$ and $C_{xy}=1/\abs{x-y}^{2\Delta_\chi=2(d-\Delta^{\mathrm{SR}}-\eta)}$.
We can perform the same computation on the sphere with Mellin-Barnes (MB) integrals, exactly as described in \cite[\S B]{Fei:2015kta}; unsurprisingly, we find the same result, as guaranteed by the Weyl invariance of the CFT.
Thus, \eqref{eq:beta3def} gives
\begin{align}
\frac{\beta_3}{C_\cO} = \frac{-1}{\hat{F}_b''(\dotwo -k)} \frac{\pi^d}{\Gamma(\dotwo)^2} =  \frac{2(-1)^{k+1} }{k} \frac{\pi^d}{\Gamma(\dotwo-k) \Gamma(\dotwo+k)}. \label{eq:beta3} 
\end{align}
Plugging this into the free energy \eqref{eq:deltaFtlam}, we indeed recover \eqref{eq:freeScalarChange}.

As a further cross-check, we can compute the sphere free energy diagrams directly using MB integrals, and indeed recover the result predicted by \eqref{eq:finalIntegral},
\begin{align}
    \notag
    &\frac{1}{4!} \int_{w,x,y,z} \expval{\cO(w)\cO(x)\cO(y)\cO(z)}_c = \frac{3!}{4!} \int_{w,x,y,z} P_\phi(w,x) P_\chi(x,y) P_\phi(y,z) P_\chi(z,w)\\
    &\qquad\qquad = \int \frac{\sqrt{g_w}\sqrt{g_x} \sqrt{g_y}\sqrt{g_z}}{s(w,x)^{2\Delta_\phi}s(x,y)^{2\Delta_\chi}s(y,z)^{2\Delta_\phi}s(z,w)^{2\Delta_\chi}}\\
    &\qquad\qquad= \frac{3C_\cO}{4} I_2'(d) \beta_3 +O(\eta)=\frac{C_\cO}{4} \frac{6}{\sin(\pi d/2)} \frac{\pi^{d+1}}{\Gamma(d+1)} \beta_3 +O(\eta).\notag
\end{align}

%% file: 02-Perturbative.tex
\section{Perturbative checks of \FtextOrPDF{}-extremisation}\label{sec:perturbative}

In this section, we will perform some perturbative checks of long-range $F$-extremisation.
We will first consider the long-range $\gO(N)$ $\phi^4$ CFT, and explicitly find the free energy for arbitrary $N$. 
This will also demonstrate a slight subtlety: the critical two-point function does not have to vanish.
If a symmetry-breaking quadratic interaction is chosen, the same procedure is possible -- but of course an explicit choice of solution $\{g_\star\}$ is required, so we cannot explicitly find a single closed form for the free energy.
To demonstrate when this choice is required, we will take the cubic theory that was recently considered in \cite[\S 4.3]{Giombi:2024zrt}.
This section will also provide further justification for our new normalization $\hat{F}$ proposed in \cref{sec:proposedFhat}.
We leave demonstrations of the multi-field extremisation to forthcoming work \cite{Fraser-Taliente2026mmls}.

\subsection{Large-\NPDF comments}

In \cref{sec:GenFFreeCheck}, we showed that $F$ is trivially extremised by local free scalars and fermions, and so we will discuss those no further.
Additionally, we began the paper by demonstrating $F$-extremisation in the large-$N$ $\gO(N)$ $\phi^4$ CFT; it and Gross-Neveu models were treated in detail in \cite{Fraser-Taliente:2025udk}, and shown to satisfy $F$-extremisation to order $1/N$ in the free energy (and so $1/N^2$ in the fundamental field anomalous dimension). 

We shall therefore only make two observations here.
Firstly, we observe that the second derivative with respect to the scaling dimension $\Delta_\phi$ in the long-range $\gO(N)$ model has its sign determined by $N\hat{F}_b''(\dotwo -1)$, which is indeed negative for $d>2$, indicating $\hat{F}$-maximisation indeed holds in the strict large-$N$ limit for $d=3$, where this theory is unitary\footnote{
Of course, $N\hat{F}_b''(\dotwo -1)$ is negative for all higher $d> 2$ as well. 
This just reflects the fact that to leading order in $N$ $\hat{F}_{\gO(N)}$ is identical to $\hat{F}$ for $N$ free scalars. 
The $\Box^k$ theories (for even $k$) then give examples of interacting nonunitary theories where $\hat{F}$ is minimised.}. 
Secondly, we observe that the field $\phi$ with the standard long-range normalization has its two-point function vanish at the short-range point (to the order calculated).
Somewhat nontrivially, in the short-range limit, each of the terms $A_{\phi,i}$ blows up $\propto N$ such that not just the leading $1$, but all lower terms in the series cancel to zero:
\begin{align}
\lim_{\Delta_\phi \to \Delta^\mathrm{SR}} \frac{C_{\phi}}{C_{\phi,0}} &= \lim_{\Delta_\phi \to \Delta^\mathrm{SR}}  1 + \frac{A_{\phi,1}}{N} + \frac{A_{\phi,2}}{N^2} + O(1/N^3) = O(1/N^3).
\end{align}
This is as it should be to ensure the vanishing of the derivative of \eqref{eq:largeNFt},
\begin{equation}\label{eq:Long_rangeONTwoPointFunc}
    \odv{\hat{F}_{\text{LR}}}{\Delta_\phi} = N \hat{F}_b'(\Delta_\phi) \frac{C_{\phi}}{C_{\phi,0}},
\end{equation}
in the short-range limit, since $\lim_{\Delta_\phi \to \Delta^\mathrm{SR}} N\hat{F}_b'(\Delta^\mathrm{SR})$ does not vanish to leading order in $1/N$.
If we unit-normalize $\hat{\phi}\equiv \frac{\phi}{\sqrt{C_{\phi}}\rvert_{\Delta_\phi}}$ for generic $\Delta_\phi$, we find that in the limit $\Delta_\phi \to \Delta^\mathrm{SR}$ the three-point coefficients of this $\hat{\phi}$ match those of the unit-normalized short-range model, as discussed in \cite[\S 6.2]{Chai:2021arp}.
Both observations also hold for the large-$N$ Gross-Neveu model \cite[\S 3.5]{Chai:2021wac}.

\subsection{\ONpdf{} \texorpdfstring{$\phi^4$}{phi\^{}4} theory in the \texorpdfstring{$\eta$}{eta} expansion}

Consider the long-range upgrade of the standard $\phi^4$ theory \cite{Rong:2024vxo},
\begin{equation}
S_\mathrm{LR} = \int_x \half \phi^i_0 \WeylOp_{s/2} \phi^i_0 + \frac{g_0}{4!} (\phi^i_0 \phi^i_0)^2. %
\end{equation}
We take $s\equiv \frac{d+\eta}{2}$, such that $\phi^i_0$ has $\Delta=\frac{d-\eta}{4}$ and so $(\phi^i_0 \phi^i_0)^2$ is a nearly marginal operator of dimension $d-\eta$. 
This allows a perturbative treatment.
Recall that the nonlocal kinetic term is not renormalized, so we do not need to introduce a $Z_\phi$.
Defining the renormalized coupling $g$ using the zero momentum BPHZ subtraction of \cite{Benedetti:2020rrq,Benedetti:2024mqx} %
\begin{align}
    g_0 &= \mu^\eta Z_g g, \quad Z_g \equiv 1+ \frac{\delta_g}{g}, \quad \tilde{g} \equiv \frac{g}{(4\pi)^{d/2} \Gamma(d/2)}
\end{align}
we find the beta function
\begin{align}\label{eq:LRbetaExpansion}
\beta_{\tilde{g}} &= -\eta \tilde{g} + \beta_{\tilde{g},2} \tilde{g}^2 +  \beta_{\tilde{g},3} \tilde{g}^3 +  \beta_{\tilde{g},4} \tilde{g}^4 + O(\tilde{g}^5)\\
\beta_{\tilde{g},2} &= \frac{\alpha_D}{3}(N+8), \quad \beta_{\tilde{g},3} = \frac{2\alpha_S}{9}(5N+22)\\
\beta_{\tilde{g},4} &= \frac{1}{27}\Big[(3N^2+22N+56)(2\alpha_{I_2}+\alpha_T) + 2(N^2+20N+60)(2\alpha_{I_1}+\alpha_U) \notag\\
&\quad \quad\quad+ 3(N+8)(N+2)\alpha_{I_3} + (5N+22)\alpha_{I_4}\Big]
\end{align}
where each of the $\alpha$s admits an expansion in $\eta$ and depends explicitly on $d$, $\alpha_X= \alpha_{X,0} + \alpha_{X,1} \eta + O(\eta^2)$ \cite[(2.3)]{Benedetti:2024mqx}.
$\alpha_{I_4}$ is complicated, but we provide a simplified expression for it as a double sum in \cref{app:alphaI4}.
The coupling is 
\begin{align}
\frac{\tilde{g}_0}{\mu^\eta} 
= \tilde{g}
+ \frac{\beta_{\tilde{g},2}}{\eta}\tilde{g}^2
+ \left(\frac{\beta_{\tilde{g},3}}{2\eta}+\frac{\beta_{\tilde{g},2}^2}{\eta^2}\right)\tilde{g}^3
+ \left(\frac{\beta_{\tilde{g},4}}{3\eta}+\frac{7\,\beta_{\tilde{g},3}\beta_{\tilde{g},2}}{6\eta^2}+\frac{\beta_{\tilde{g},2}^3}{\eta^3}\right)\tilde{g}^4 + O(\tilde{g}^5).
\end{align}
Evidently, this is a non-minimal subtraction scheme\footnote{For treatment of the same theory in a minimal subtraction scheme to one loop order lower, see \cite[(5.4)]{Giombi:2019enr}.}. %
Of course, the value of $\tilde{F}$ at the fixed point is scheme-independent, so it does not matter what scheme we use. 

Notably, unlike in the short-range computation of the free energy \cite[(3.3)]{Fei:2015oha}, we do not need to consider the presence of geometric couplings.
This is because we work in generic $d$: there are therefore no marginal geometric couplings which we would otherwise be forced to add.
This is very natural, because the fact that $\hat{F}_\mathrm{LR}'(\Delta_\phi)\propto C_\phi$ means that we have shown that the long-range free energy can be computed from flat-space data alone. 

Define the integrated connected $n$-point function of $\phi_0^4 =:(\phi^i_0 \phi^i_0)^2:$ in the free theory:
\begin{equation}
    H_n \equiv \int_{x_1, \cdots, x_n} \expval{\phi_0^4(x_1) \cdots \phi_0^4(x_n)}_0^\mathrm{conn}.
\end{equation}
Then $F_{\mathrm{LR}_\eta} =-\log Z_{S^d} = N F_b(\Delta_\phi) + \delta F_{\mathrm{LR}_\eta}$ has the perturbative expansion: %
\begin{equation}
\delta F_{\mathrm{LR}_\eta} =-\frac{ g_0^2}{2! (4!)^2}H_2+\frac{g_0^3}{3! (4!)^3}H_3-\frac{g_0^4}{4! (4!)^4} H_4+\frac{g_0^5}{5! (4!)^5} H_5 + O(g_0^6),
\end{equation}
where
\begin{align}
H_2&=8 N (N+2) C_\phi^4 I_2(4 \Delta_\phi),\\
H_3&=64 N (N+2) (N+8) C_\phi^6 I_3(4 \Delta_\phi)\\ %
H_4 &=32 (4!) N (N+2) C_\phi^8 \left(\left(N^2+6 N+20\right) H_4^{(1)} +8  (N+2)H_4^{(2)}+4 (5 N+22)H_4^{(3)}\right)\\
H_5 & = 2^{12} \cdot 3 C_\phi^{10} N (N+2)(20(N+2)(N+8) H_5^{(1)} + 10 (3N^2 + 22N +56) H_5^{(2)}\notag \\
&+ (N^3 +8N^2 +24N+48)H_5^{(3)} + 10(N+2)^2 H_5^{(4)} \notag\\
&+20(N^2 + 20N + 60) H_5^{(5)} + 8(5N+22)H_5^{(6)}).
\end{align}
We give the values of $H_{4}^{(i)}$ and $H_{5}^{(i)}$ computed in DREG in \cref{app:DiagramValues}.

Putting this all together, and making sure that we keep the $\eta$ terms to high enough order, we find the arbitrary-$\tilde{g}$ free energy in this scheme begins as
\begin{equation}
    \begin{aligned}
&\frac{\delta\hat{F}_{\mathrm{LR}_\eta}}{N(N+2)}= -\frac{e^{-\gamma  \eta } M^{2 \eta } \Gamma(\tfrac{d-\eta }{4})^4 }{ \Gamma (\eta ) \Gamma(\tfrac{d+\eta }{4})^4} \frac{\Gamma(\eta -\tfrac{d}{2})}{\Gamma(-\tfrac{d}{2})} \frac{\tilde{g}^2}{2(6)^2} \\
&+ \frac{(N+8) M^{2 \eta } \Gamma(\tfrac{d-\eta }{4})^4 \left(\frac{4 \pi ^{3/2} M^{\eta } \Gamma \left(\frac{d-\eta }{4}\right)^2 \Gamma(\tfrac{3 \eta }{2}-\tfrac{d}{2})}{\Gamma(\tfrac{\eta +1}{2})^3}-\frac{3\ 8^{\eta } e^{\frac{\gamma  \eta }{2}} \Gamma \left(\eta -\frac{d}{2}\right) \Gamma \left(\frac{d+\eta }{4}\right)^2}{\Gamma (\eta +1)}\right)}{2^{3 \eta -1} e^{\frac{3}{2} \gamma  \eta}  \Gamma(-\tfrac{d}{2}) \Gamma(\tfrac{d+\eta }{4})^6} \frac{\tilde{g}^3}{3(6)^3}\\
&+ (\# + \# \eta) \tilde{g}^4 + \# \tilde{g}^5 + O(\tilde{g}^{6-m} \eta^m),
    \end{aligned}
\end{equation}
where $M=\mu R e^{\gamma/2}$, and we have switched to $\hat{F}$ as defined in \cref{sec:proposedFhat}.
We do not give the expansion of the higher values, represented by $\#$s, as they are lengthy and can be reconstructed from the diagram values given in \cref{app:DiagramValues}.
As they must, all of the $1/\eta$ poles have cancelled, so our result is finite order by order in $\tilde{g}$. 

Plugging in the fixed point value,
\begin{equation}
\tilde{g}_\star
= \frac{\eta }{\beta_{\tilde{g},2}}
-\frac{\beta_{\tilde{g},3} \eta ^2}{\beta_{\tilde{g},2}^3}
+\frac{\eta ^3 \left(2 \beta_{\tilde{g},3}^2-\beta_{\tilde{g},2}\beta_{\tilde{g},4}\right)}{\beta_{\tilde{g},2}^5}
+\frac{5 \eta ^4 \left(\beta_{\tilde{g},2}\beta_{\tilde{g},3}\beta_{\tilde{g},4}-\beta_{\tilde{g},3}^3\right)}{\beta_{\tilde{g},2}^7}
+O(\eta^5)
\end{equation}
(recalling that the $\tilde{\beta}$s themselves have expansions in $\eta$), we find the free energy\footnote{Note that $\odv{}{\eta}\tilde{F}_b(\frac{d+\eta}{2}) = -\tfrac{\pi}{8} \tfrac{\Gamma \left(\frac{d}{2}\right)^2}{\Gamma (d+1)} \eta^2 + O(\eta^4)$, i.e.  $\tilde{F}_b(\frac{d+\eta}{2}) = -\tfrac{\pi}{8} \tfrac{\Gamma \left(\frac{d}{2}\right)^2}{\Gamma (d+1)} \frac{\eta^3}{3} + O(\eta^5)$, i.e. the leading correction to $\tilde{F}$ is $ \frac{N(N+2)}{(N+8)^2}\tilde{F}_b(\frac{d+\eta}{2})$, i.e.  $\frac{N(N+2)}{(N+8)^2}\tilde{F}_b(\tilde{\Delta}_{\phi^2,\mathrm{UV}})$ (since $\tilde{\Delta}_{\phi^2,\mathrm{UV}} = d-2\Delta_\phi$). Hence, in the $N\mapsto\infty$ limit, we manifestly reproduce the $O(N^0)$ contribution from the $\sigma$ field in \eqref{eq:largeNFt}.
}
\begin{align}\label{eq:FtLR}
    \tilde{F}_{\mathrm{LR}_\eta} &= N \tilde{F}_b(\Delta_\phi)+\tfrac{\pi N (N+2)}{8} \tfrac{\Gamma(\frac{d}{2})^2}{\Gamma (d+1)} \left(-\tfrac{\eta ^3}{3 (N+8)^2}%
    + \tfrac{\eta ^4 (5 N+22)\alpha_{S,0}}{(N+8)^4}+\tilde{f}^{(5)} \eta^5 + O(\eta^6)\right),
\end{align}
which with our new normalization is
\begin{subequations}\label{eq:FhatLR}
\begin{align}
    \hat{F}_{\mathrm{LR}_\eta} &= N \hat{F}_b(\Delta_\phi)+\frac{N (N+2)}{16} \left(-\frac{\eta ^3}{3 (N+8)^2}%
    + \frac{\eta ^4 (5 N+22)\alpha_{S,0}}{(N+8)^4}+\tilde{f}^{(5)} \eta^5 + O(\eta^6)\right),
\end{align}
which as expected is independent of $\mu R$.
Recall that $\alpha_{S,0} \equiv 2 \psi_0\left(\frac{d}{4}\right)-\psi_0\left(\frac{d}{2}\right)+\gamma = 2 \psi_0\left(\frac{d}{4}\right)-\psi_0\left(\frac{d}{2}\right)- \psi_0(1) = 2H_{\frac{d}{4}-1} -H_{\frac{d}{2}-1}$, where $\psi_m$ denotes the polygamma function of order $m$, i.e. $\psi_m(x) \equiv \odv[m+1]{}{x}\log \Gamma(x)$, and $\gamma$ denotes the Euler-Mascheroni constant.
The next order is
\begin{align}\label{eq:FtLRnextOrd}
  \tilde{f}^{(5)} &=\frac{13 N^3-74 N^2-1088 N-2496}{4 (N+8)^6} \alpha_{S,0}^2 \notag\\
  &\quad + \frac{8 (5 N+22) \alpha_{I_4} -\left(N^3+3 N^2+114 N+368\right) \left(\psi_1\left(\frac{d}{2}\right)-\psi_1(1)\right)}{20 (N+8)^5}.
\end{align}
\end{subequations}
It is trivial to check that this matches the known $1/N$ expansion of the free energy \eqref{eq:largeNFt} \cite[\S 2.1.1]{Fraser-Taliente:2025udk} after taking $s=\frac{d+\eta}{2}$, as the complicated double sum $\alpha_{I_4}$ only appears at order $1/N^2$ (and therefore may be related to \cite[$I(\dotwo)$]{Broadhurst:1996yc}).

Differentiating \eqref{eq:FhatLR} with respect to $\eta$, and locating a perturbatively valid solution via $d=4-\epsilon$, we find that the extremum $\eta_\mathrm{SR}$ (which is a maximum) gives exactly the standard result \cite[\lstinline{DeltaE[Op[V, 0, 1]]}]{Henriksson:2022rnm} 
    \begin{align}\label{eq:DeltaSRON}
    \Delta_\phi^\mathrm{SR}\rvert_{d=4-\epsilon} &=\frac{d-\eta_\star}{4} = 1-\frac{\epsilon }{2}+(N+2) \left[\frac{\epsilon ^2}{2^2 (N+8)^2}-\frac{(N^2-56 N-272) \epsilon ^3}{2^4 (N+8)^4}\right.\\
    & -\epsilon^4 \left.\left(\frac{6 (5 N+22) }{(N+8)^5}\zeta_3+\frac{5 N^4+230 N^3-1124 N^2-17920 N-46144}{2^6(N+8)^6}\right)+O(\epsilon^5)\right],\notag
    \end{align}
where we used
\begin{equation}
  \alpha_{I_4}\rvert_{d=4-\epsilon} = 12 \zeta_3 + O(\epsilon)
\end{equation}
from \cref{app:alphaI4}. 

Incidentally, the overall $N+2$ factors in both $\delta\tilde{F}$ and $\Delta_\phi^\mathrm{SR}$ are explained by the fact that the $N=-2$ theory is trivial, because the $\mathrm{Sp}(N=-2)$ quadratic invariant of anticommuting scalars is zero, $(\Omega_{ij} \chi^i \chi^j)^2 \propto \chi_1^2 \chi_2^2=0$ \cite[\S 1]{Fei:2015kta}. %
The explosion at $N=-8$ is because the 1-loop term in the $\beta$ function vanishes at that value \cite{Fei:2015kta}. 
The $(N+8)$ in the denominator is therefore a signal that if $N=-8$ we must expand in powers of $\sqrt{\eta}$ (or $\sqrt{\epsilon}$) instead. %
 
\subsubsection{Whether the two-point function vanishes depends on field normalization}\label{sec:doesTwoPointVanish}

In \cite[(3.26)]{Paulos:2015jfa}, the two-point normalization of $\phi$ for the $N=1$ theory in the $\eta$ expansion was computed:
\begin{equation}\label{eq:rhohat}
    \expval{\phi(x) \phi(0)}_\mathrm{IR} = \frac{\hat{\rho}(g_\star)}{\abs{x}^{2\Delta_\phi}}, \quad \hat{\rho}(g) = 1+ Q g^2 + \cdots, \quad Q = \frac{\pi^d}{6} \frac{\Gamma(-\frac{d}{4})}{\Gamma(\frac{3d}{4})}.
\end{equation}
Using $\beta_g = -\eta g+ K g^2$, $K=\tfrac{3}{2} \vol S^{d-1}$, we find that $g_\star = \eta/K$. 
In our conventions, this means that the ratio of two-point functions of $\phi$ in the interacting and free theories is
\begin{equation}
\frac{C_\phi}{C_{\phi,0}} = \frac{\hat{\rho}(g_\star)}{\hat{\rho}(0)}= \hat{\rho}(g_\star).
\end{equation}
From our result \eqref{eq:FtLR}, we find
\begin{equation}
    \frac{C_\phi}{C_{\phi,0}} =\frac{\hat{F}_{\mathrm{LR}_\eta}'(\Delta_\phi) }{N \hat{F}_b'(\Delta_\phi)} = 1+\frac{\eta ^2 (N+2) \Gamma \left(-\frac{d}{4}\right) \Gamma \left(\frac{d}{2}\right)^2}{2 (N+8)^2 \Gamma \left(\frac{3 d}{4}\right)}+O(\eta ^3)
\end{equation}
which for $N=1$ matches the value of $\hat{\rho}(g_\star)$ from \eqref{eq:rhohat} to the order known.

We already have computed that $\tilde{F}_\mathrm{LR}'(\Delta^\mathrm{SR}) =0$.
Since $\Delta_\phi=\Delta^\mathrm{SR} = 1-\frac{\epsilon}{2} + \cdots$ in $d=4-\epsilon$, we see that for $\eta=\epsilon+O(\epsilon^2)$ we have that $Q g_\star^2 = -\epsilon/27 + O(\epsilon^2)$. 
Thus, with the normalization in this theory, $C_\phi/C_{\phi,0}=1-\epsilon/27+O(\epsilon^2)$ -- which does not vanish.
That is, the two-point function coefficient does \textit{not} go to zero here; it is only the combination $\tilde{F}_\mathrm{LR}'(\Delta=\frac{d-\eta}{4})$ that vanishes, where we must first truncate the expansion to a given order in $\eta$, and (\textbf{only!}) then substitute for $\eta_\mathrm{SR}(\epsilon)$ and $d=4-\epsilon$. 

This was unlike what we saw for the $\gO(N)$ model, i.e. \eqref{eq:Long_rangeONTwoPointFunc}.
Thus, the invariant statement is that at the short-range fixed point it is only the combination
\begin{equation}
   \hat{F}_\mathrm{LR}'(\Delta^\mathrm{SR}) = N\hat{F}_b'(\Delta_\phi) \frac{C_\phi}{C_{\phi,0}}, %
\end{equation}
that vanishes. In particular, as we have seen here, if $\hat{F}_b'(\Delta_\phi)$ vanishes to leading order in the perturbative parameter, $C_\phi$ does not need to vanish. 

\subsubsection{The short-range theory: numerics}\label{sec:WFnumerics}

If \eqref{eq:DeltaSRON} is plugged back into \eqref{eq:FtLRnextOrd}, we see that the extremal value of $\tilde{F}_\mathrm{LR}$ is
\begin{align}\label{eq:Ft4meps}
\tilde{F}_{\gO(N)\, \mathrm{SR},d=4-\epsilon}&= N\tilde{F}_b(\tfrac{d-2}{2}) - \frac{\pi N(N+2)}{576} \Big\{\frac{ \epsilon ^3}{(N+8)^2}+\frac{\left(13 N^2+370 N+1588\right) \epsilon ^4}{12 (N+8)^4}\notag\\
& +\frac{1}{720 (N+8)^6}\Big[647 N^4+32152 N^3+606576 N^2+3939520 N+8451008 \notag\\
&\quad -10368 \zeta_3 (5 N+22) (N+8)-30 \pi ^2 (N+8)^4\Big]\epsilon ^5+O(\epsilon ^6)\Big\}.
\end{align}
As expected, this exactly matches the computation of the free energy of the short-range $\gO(N)$ model in $d=4-\epsilon$ \cite[(1.6)]{Fei:2015oha} (see around \cite[(4.13)]{Pannell:2025ixz} for discussion of the $\epsilon^6$ term). 

As promised in \cref{sec:proposedFhat}, it is simpler when given as $\hat{F}$ instead:
\begin{align}\label{eq:Fhat4meps}
&\hat{F}_{\gO(N)\, \mathrm{SR},d=4-\epsilon}= N\hat{F}_b(\tfrac{d-2}{2})+ N(N+2) \Big\{ -\frac{\epsilon ^3}{48 (N+8)^2}-\frac{3 (3 N+14) \epsilon ^4}{32 (N+8)^4}\\
&+\frac{\epsilon ^5}{10}\left[\frac{3(5 N+22)}{(N+8)^5}\zeta _3 +\frac{(N (N (2 N+47)-1834)-17280) N-40992}{64 (N+8)^6}\right]+O(\epsilon ^6)\Big\}.\notag
\end{align}
Observe that we have eliminated both an overall prefactor and the $\pi^2\epsilon^5$ term.

Evidently, substituting in a controlled fashion $d=4-\epsilon$ and the short-range $\Delta_\phi^\star\rvert_{d=4-\epsilon}$ yields exactly the short-range free-energy. 
One might hope that putting in the known bootstrap value of $\Delta_\phi$ might somehow provide an (uncontrolled) better approximation.
This was the essence of the long-range approach of \cite[\S 4]{Giombi:2024zrt} -- but of course there is no good theoretical reason for doing so, just as there is no good reason for the $\epsilon$ expansion to work.

We can now consider the effect of adding the order $\eta^5$ terms to that computation \cite[Table 5.2, $c^{\mathrm{LRA}}$]{Giombi:2024zrt}.
Unfortunately, the agreement with the exact result in $d=2$ gets worse. 
When plugging in $s=2-2\Delta_\phi^\mathrm{SR}$ for $\Delta_\phi=\Delta_\phi^\mathrm{SR}\rvert_{N=1,2}=1/8$, we find that the changes in the value of  $c_\mathrm{LR} = 6 \hat{F}_\mathrm{LR}$ are: for $N=1$, $0.589314\to 0.59644$, and for $N=2$, $1.17135\to 1.18184$. 
In both cases we move away from the exact results, $c=\half$, $1$. %

The same is true in $d=3$, though we do not have the exact result to compare against.
Compared to \cite[Table 5.1, $\tilde{F}^{\mathrm{LRA}}$]{Giombi:2024zrt}, when we add the $\eta^5$ term we move away from the other known results -- the Pad\'e-approximated \eqref{eq:Ft4meps} and the fuzzy sphere result.
Admittedly, the results do not change by much: for $N=1$, $\tilde{F}^{\mathrm{LRA}}: \, 0.062341 \to 0.062471$; for $N=2$, $0.12454(5)\to  0.12477(5)$, and for $N=3$, $0.1868(17) \to 0.1871(17)$. 
The error indicated in the last two cases comes only from the bootstrapped $\Delta_\phi$, as for $N=2,3$ it is large enough to matter: $\Delta_{\phi,N=1} =0.518148806(24)$ \cite{Chang:2024whx}, $\Delta_{\phi,N=2} =0.519088(22)$, $\Delta_{\phi,N=3} =0.518942(51)$ \cite[Table 5]{Henriksson:2022rnm}.
We do not indicate the error from the unknown $O(\eta^6)$ term.

We might also wonder if the new normalization $\hat{F}$ improves Pad\'e approximation. As a benchmark, we might consider the two-sided Pad\'e approximation of the free field value of $\hat{F}_b(\tfrac{d-2}{2})$. 
As we show in \cref{app:FhatNumerics}, it does, principally because $\hat{F}$ varies more slowly.

\subsubsection{Higher-derivative theories}

The scaling dimensions of $\phi$ in the higher-derivative Wilson-Fisher CFTs (see \cite[(4.2)]{Gubser:2017vgc}) are also reproduced by this calculation, practically for free.
However, evidently there is no more information in this $\Delta^{\mathrm{SR}}_\phi\rvert_{\Box^k,\phi^4}$ than in $\hat{F}_\mathrm{LR}$ -- if anything, less, because it obfuscates the actual information present.
Indeed, we find that in $d=4k-\epsilon$ (of course with $k\ge 0$),
\begin{subequations}
\begin{align}
\Delta^{\mathrm{SR}}_\phi\rvert_{\Box^k, \phi^4} &= \frac{d}{2}-k+ \gamma_2 \epsilon^2 +\gamma_3\epsilon^3 + \gamma_4 \epsilon^4 + O(\epsilon^5)\\
\gamma_2 &= (-1)^{k+1} \frac{(N+2)}{2(N+8)^2 } \frac{ \Gamma (2 k)^2}{k! \Gamma (3 k)} \label{eq:gamma2MulticriticalNeq1}\\
\gamma_3 &= \gamma_2 \left(-8\gamma_2+\half (\psi_0(3k) -3\psi_0(k) + 2\psi_0(1)) + \frac{N^2-4N-24}{(N+8)^2} \alpha_{S,0,d=4k}\right),
\end{align}
\end{subequations}
where $\alpha_{S,0,d=4k}= 2 \psi_0(k)-\psi_0(2k)-\psi_0(1)$. %
$\gamma_4$ is unpleasant and straightforwardly obtainable by extremising the much simpler expression for $\hat{F}_\mathrm{LR}(\Delta_\phi)$ (except for the task of evaluating $\alpha_{I_4,d=4k-\epsilon}$ to leading order in $\epsilon$ -- we discuss that difficulty in \cref{app:alphaI4}) -- and so it is not given here.
This result for $\gamma_2$, when evaluated at $N=1$, agrees with the result for the $\Box^k$ $N=1$ $\phi^{2n}$ theory for $n=2$ \cite[(3.20)]{Safari:2017tgs}.
We believe that $\gamma_3$ and $\gamma_4$ are new results.

The second derivatives at the local point in $d=4k-\epsilon$ are
\begin{equation}
    \hat{F}_{\mathrm{LR}_\eta}''(\Delta^{\mathrm{SR}}_\phi\rvert_{\Box^k, \phi^4})|_{d=4k-\epsilon} = N\hat{F}_b''(k) +O(\epsilon) = N(-1)^k \frac{k! \Gamma (3 k)}{2 \Gamma (2 k)^2} + O(\epsilon),
\end{equation}
and so the nonunitary local CFTs alternate between being maxima and minima of $\hat{F}_\mathrm{LR}$ -- copying the behaviour of their parent theories, the local scalars.

We present the $k=2$ result explicitly to order $\epsilon^4$, since this theory has been studied in greater depth as the \textit{isotropic Lifshitz point} \cite[\S 1]{Safari:2017tgs}:
\begin{equation}
\begin{aligned}
    \frac{\gamma_{\phi,k=2}}{N+2} &= -\frac{3}{40 (N+8)^2}\epsilon^2 + \frac{23 N^2+696 N+3088}{40^2(N+8)^4}\epsilon^3+\frac{3(5 N+22)}{20(N+8)^5} \alpha_{I_4,d=8} \epsilon^4\\
    &+\frac{5101 N^4+10870 N^3-372292 N^2-1147904 N-1954880}{3\cdot 40^3(N+8)^6} \epsilon^4 + O(\epsilon^5),
\end{aligned}
\end{equation}
where $\alpha_{I_4,d=8-\epsilon} = \frac{36}{5}(6\zeta_3 -7) +O(\epsilon)\simeq 1.53$; this matches the $O(\epsilon^2)$ results of \cite[(23)]{Diehl:2002wn} and the $O(\epsilon^3)$ results of \cite[(3.6)]{Gracey:2017erc} (after using the actual beta function for their unusual squared coupling $g_1$, which is $\beta_{g_1} \equiv -\frac{\epsilon g_1^2}{2} + \beta^{(8)}(g_1)=0$ in their notation, with $d=8-\epsilon$). %

\subsection{Cubic model}

We consider the cubic theory with $\gO(N)$ symmetry encoded by the action \cite[(4.14)]{Giombi:2024zrt}
\begin{equation}
S \equiv \int_x \half \phi^i_0 \WeylOp_{s/2} \phi^i_0 +\half \sigma_0 \WeylOp_{s/2} \sigma_0 + \frac{g_{1,0}}{2} \sigma_0 \phi^i_0 \phi^i_0 + \frac{g_{2,0}}{6} \sigma_0^3
\end{equation}
To consider this theory in full generality, one should take a different fractional Laplacian for $\sigma_0$. 
However, we know that for $N=0, -2$ there exist short-range models that match this theory where $\Delta_\phi^\mathrm{SR}=\Delta_\sigma^\mathrm{SR}$ (at least formally -- of course, for $N=0$ there are no $\phi$ fields) \cite{Fei:2014yja,Fei:2014xta,Fei:2015kta,Klebanov:2021sos}. 
We therefore consider \textit{only} the long-range models with $\Delta_\phi=\Delta_\sigma=\frac{d-\eta}{3}$ (i.e. $s=\frac{d+2\eta}{3}$), making both $\sigma^3$ and $\sigma \phi_i\phi_i$ nearly marginal operators of dimension $d-\eta$.

Defining renormalized couplings 
\begin{align}
g_{i,0} \equiv \mu^\eta Z_{g,i} g_i, \quad Z_{g,i} =1+ \frac{\delta_{g_i}}{g_i}, \quad i=1,2,
\end{align}
and taking
\begin{equation}
g_i = \tilde{g}_{i} \sqrt{(4\pi)^{d/2} \Gamma(\tfrac{d}{2})}, \quad \tilde{g}_1^\star \equiv \alpha \sqrt{\eta} + O(\eta), \quad \tilde{g}_{2}^\star = \beta \sqrt{\eta} + O(\eta),
\end{equation}
the one-loop beta functions \cite[(4.16)]{Giombi:2024zrt} become
\begin{equation}\label{eq:cubicBetas}
    1+2 \alpha  (\alpha +\beta )=0, \quad 2 \beta ^3+\beta +2 \alpha ^3 N=0.
\end{equation}
For the free energy 
\begin{equation}
\delta F_\mathrm{cubic,LR} = -\frac{G_2}{2! (3!)^2} - \frac{G_4}{4!(3!)^4}
\end{equation}
we can take the diagrams from \cite[\S E.2]{Giombi:2024zrt}
\begin{subequations}
\begin{align}
G_2 &= 3! t_2 C_c^3 I_2(d-\eta), \quad G_4 = 3(3!)^3 C_c^6 (3 t_{41} G_4^{(1)} + 2 t_{42} G_4^{(2)}),\\
t_2 &= 3Ng_{1,0}^2 + g_{2,0}^2, \\
t_{41} &= (N+4)Ng_{1,0}^4 + 2Ng_{1,0}^2 g_{2,0}^2 + g_{2,0}^4,\quad t_{42} = 3Ng_{1,0}^4 + 4Ng_{1,0}^3 g_{2,0} + g_{2,0}^4,
\end{align} 
\end{subequations}
where the diagrams $G_2,G_4^{(i)}$ are defined without symmetry factors and with unit-normalized position-space propagators $1/\abs{x-y}^{2\Delta}$ (the actual propagator normalization is $C_c \equiv 1/\cF_{2\Delta,0}$ for $\Delta=\frac{d-\eta}{3}$).
To unnecessary precision, the $G_4^{(1,2)}$ are
\begin{subequations}
\begin{align}
G_4^{(1)} &=\begin{tikzpicture}[baseline=1.75ex, scale=0.8]
    \draw (0,0) -- (0,1);
  \draw[line width=0.4pt] (0,0) to[bend left =35] (1,0);
  \draw[line width=0.4pt] (0,0) to[bend right=35] (1,0);
  \draw[line width=0.4pt] (0,1) to[bend left =35] (1,1);
  \draw[line width=0.4pt] (0,1) to[bend right=35] (1,1);
    \draw (1,0) -- (1,1);
    \filldraw (0,0) circle (2pt);
    \filldraw (1,0) circle (2pt);
    \filldraw (1,1) circle (2pt);
    \filldraw (0,1) circle (2pt);
  \end{tikzpicture}=
(2R)^{4\eta }
\frac{2^{4-d} \pi ^{2 d+\tfrac{1}{2}} \cos (\tfrac{\pi  d}{6}) \Gamma (-\tfrac{d}{2}) \Gamma (-\tfrac{d}{3})}{\Gamma (\tfrac{2 d}{3}) \Gamma (\tfrac{d+1}{2})}\notag  \\
&\times\Big[\eta+ \eta^2 [2 H_{-\tfrac{d}{2}-1}+\tfrac{1}{6} (\psi_0(\tfrac{2 d}{3})-\psi_0(\tfrac{d}{3})-\psi_0(\tfrac{d}{6})+\psi_0(-\tfrac{d}{6}))] +O(\eta^3)\Big],\\
G_4^{(2)}&=\begin{tikzpicture}[baseline=1.75ex, scale=0.8]
    \draw (0,0) -- (1,0) -- (1,1) -- (0,1) -- cycle;
    \draw (0,0) -- (1,1);
    \draw (0,1) -- (1,0);
    \filldraw (0,0) circle (2pt);
    \filldraw (1,0) circle (2pt);
    \filldraw (1,1) circle (2pt);
    \filldraw (0,1) circle (2pt);
  \end{tikzpicture} =(2 R)^{4 \eta}\frac{ \pi ^{2 d+\frac{3}{2}} 2^{4-d} \Gamma (-\frac{d}{2})  }{\Gamma (d) \Gamma (\frac{d+3}{6})^3}\left[3+6 \eta H_{-\frac{d}{2}-1}+O(\eta^2)\right].
\end{align} 
\end{subequations}
The CFT free energy is 
\begin{subequations}\label{eq:FcubicLR}
\begin{align}
\tilde{F}_{\text{cubic,LR}} &= (N+1)\tilde{F}_b(\tfrac{d-\eta}{3}) - \frac{\Gamma(\tfrac{d+3}{6})^3}{24 d  \,\Gamma(\tfrac{d+1}{2})} \underbrace{\left[\beta ^2 \left(1+\beta ^2\right)+ N \alpha^2(1-\alpha ^2) \right]}_{\equiv P_{\alpha\beta}}\eta^2 +O(\eta^3),\\
    \hat{F}_{\text{cubic,LR}} &= (N+1)\hat{F}_b(\tfrac{d-\eta}{3}) -  \frac{\Gamma(\tfrac{d}{3})^3}{\Gamma(\tfrac{d}{6})^3} \frac{P_{\alpha\beta}}{\Gamma(\dotwo)} \, \frac{\eta^2}{12} +O(\eta^3),
\end{align}
\end{subequations}
where $P_{\alpha\beta}$ is independent of $d$, and was simplified from $\beta ^4+\beta ^2+\alpha ^2 N \left(3 \alpha ^2+4 \alpha  \beta +3\right)$ using the fact that $\alpha,\beta,N$ solve the beta functions \eqref{eq:cubicBetas}.

The beta functions \eqref{eq:cubicBetas} generically have six solutions under perturbative control\footnote{With $A = 4\alpha^2$, $f=\frac{N-1}{8}$, and eliminating $\beta$ via $\beta^2 = \frac{(A+2)^2}{4A}$, the beta functions become a cubic in $A$,
\begin{align}
A^3 f=(A+1)^2,\quad P_{\alpha\beta} = (A+2) \frac{A^2+5A+3}{2 A^2}.
\end{align}
The signs of $\alpha$ and $\beta$ cannot be chosen independently, so there are six fixed points, of which three are likely identical as CFTs, as they have identical data at this order.
We also see that $P_{\alpha\beta}$ can vanish only for $(N,\alpha,\beta)=(0, \pm i/\sqrt{2}, 0)$, or for four fixed points with $N=-\frac{17 \pm 4\sqrt{13}}{27}\simeq -1.16$.
}.
There are two values of $N$ for which we know the corresponding short-range model.
The long-range fixed points that become those local theories are\footnote{For $N=-2$ there are two more CFTs with $P_{\alpha\beta} = 1 \pm \frac{3\sqrt{3}}{4}i$, which we do not identify.
For the standard short-range cubic CFT where $\Delta_\phi\neq \Delta_\sigma$, this perturbative analysis of the fixed points was performed in \cite[\S 3]{Fei:2014yja}.}:
\begin{table}[ht]
\centering
\begin{tabular}{c| c c c c}
\text{Corresponding short-range theory} & $N$ & $\alpha$ & $\beta$ & $P_{\alpha\beta}$\\
\hline
\text{Yang-Lee} & $0$ & $0$ & $\pm \tfrac{i}{\sqrt{2}}$ & $-\,\tfrac{1}{4}$\\
$\mathrm{OSp}(1|2)$ & $-2$ & $\pm \tfrac{i}{\sqrt{6}}$ & $\pm i\sqrt{\tfrac{2}{3}}$ & $\tfrac{1}{6}$ %
\end{tabular}
\end{table}

\subsubsection{Short-range cubic model}

We can then perform $F$-extremisation on these two theories. 
Setting $g=g_\star(\eta)$, we look for a solution to 
\begin{equation}
    \odv{}{\eta}\hat{F}_{\text{cubic,LR},g_\star} = \pdv{}{\eta}\hat{F}_{\text{cubic,LR},g_\star} + \odv{g_{i,\star}}{\eta} \pdv{}{g_i}\hat{F}_{\text{cubic,LR},g}\Big\rvert_{g=g_\star}= \pdv{\hat{F}_{\text{cubic,LR},g_\star}}{\eta}_{g_\star}=0,
\end{equation}
where the second term vanishes because $\partial_{g_i} \hat{F} \propto \beta_i=0$.
In $d=6-\epsilon$, the perturbatively valid solution to this for \eqref{eq:FcubicLR} is
\begin{equation}
\Delta_{\phi}^\mathrm{SR} = \Delta_{\sigma}^\mathrm{SR} = \frac{d-\eta_\mathrm{SR}}{3}, \quad
\eta_\mathrm{SR} =\frac{N+1}{2}\left(\frac{\epsilon }{N+P_{\alpha\beta}+1}+O(\epsilon^2)\right), \label{eq:etaStarcubic}
\end{equation}
which matches to $O(\epsilon)$ the actual results for the Yang-Lee \cite[(2.2)]{Gracey:2025rnz} and $\mathrm{OSp}(1|2)$ \cite[(3.9)]{Fei:2015kta} theories in $d=6-\epsilon$ dimensions,
\begin{align}
 2\gamma_{\sigma,N=0}= -\frac{\epsilon }{9}-\frac{43 \epsilon ^2}{729}+O(\epsilon^3), \quad 2\gamma_{\sigma,N=-2} =-\frac{\epsilon }{15}-\frac{7\epsilon ^2}{225} +O(\epsilon^3).
\end{align}
Using that, for $\eta\propto \epsilon$,
\begin{align}
\hat{F}_b(\tfrac{d-\eta}{3})-\hat{F}_b(\tfrac{d-2}{2}) \rvert_{d=6-\epsilon}=\int_{0}^{\tfrac{\epsilon-2\eta}{6}} \odif{x}\, \hat{F}_b'(\tfrac{d-2}{2}+x) = - \frac{3}{8}(\tfrac{\epsilon-2\eta}{6})^2 + O(\epsilon^3),
\end{align}
we can plug \eqref{eq:etaStarcubic} into $\hat{F}_{\text{cubic,LR}}$ \eqref{eq:FcubicLR}, to find the short-range model's free energy
\begin{subequations}\label{eq:FcubicSR6meps}
\begin{align}
\tilde{F}_{\text{cubic,SR}}\rvert_{d=6-\epsilon}&= (N+1)\left[\tilde{F}_b(\tfrac{d-2}{2}) -\frac{\pi}{8640} \frac{P_{\alpha\beta}}{N+1+P_{\alpha\beta}} \epsilon ^2 + O(\epsilon^3)\right],\\
\hat{F}_{\text{cubic,SR}}\rvert_{d=6-\epsilon} &= (N+1)\left[\hat{F}_b(\tfrac{d-2}{2}) - \frac{P_{\alpha\beta}}{N+1+P_{\alpha\beta}} \frac{\epsilon^2}{96} + O(\epsilon^3)\right],
\end{align}
\end{subequations}
which matches $\tilde{F}_\mathrm{YL} = \tilde{F}_b(\tfrac{d-2}{2})+ \frac{\pi \epsilon^2}{25920}$ and $\tilde{F}_{\mathrm{OSp}(1|2)} = -\tilde{F}_b(\tfrac{d-2}{2})- \frac{\pi \epsilon^2}{43200}$ \cite[(5.1), (5.7)]{Giombi:2024zrt}.

The second derivative is 
\begin{equation}
    \hat{F}_{\text{cubic,LR},g_\star}''(\Delta^{\mathrm{SR}}) = -\frac{3}{4}(N+P_{\alpha\beta}+1) +O(\epsilon),
\end{equation}
Unlike the $\phi^4$ case, therefore, the type of extremum is not just governed by the free scalar contribution.
We find a maximum for Yang-Lee, and a minimum for the $\mathrm{OSp}(1|2)$ theory.

These long-range theories do not have a nice short-range identification for other values of $N$.
It is likely that they are not short-range theories at all, since we would need to independently vary $\Delta_\phi$ and $\Delta_\sigma$ to guarantee full locality. 
These additional solutions are then just the nonlocal theories satisfying
\begin{equation}
    \odv{\hat{F}}{\Delta} =   \pdv{\hat{F}}{\Delta_\phi}+  \pdv{\hat{F}}{\Delta_\sigma} \Big\rvert_{\Delta_\phi=\Delta_\sigma=\Delta} = 0.
\end{equation}
Hence, to obtain the generic-$N$ short-range model by $F$-extremisation, we need to compute $\hat{F}$ for arbitrary $\Delta_\phi=\frac{d-\eta_1}{3}$ and $\Delta_\sigma=\frac{d-\eta_2}{3}$ (rather than just the diagonal family which we consider here), and then find the extremum.
This would allow us to consider for example the $N=1$ model, which is thought to become the minimal model $M(3,8)$ in $d=2$ \cite{Klebanov:2022syt}.
We leave this for future work.

\section{Discussion}\label{sec:Discussion}

In this paper we showed that the long-range conformal field theories become local CFTs at the extrema of $\hat{F}(\{\Delta_i\})$, an observation that followed straightforwardly from the requirement that the action of a local CFT be local.
Thus, $\hat{F}$ provides a map of sorts to the space of nonlocal field theories.
For unitary theories, we then used conformal perturbation theory to show that the local CFTs actually maximise $\hat{F}(\{\Delta_i\})$ along any line of long-range CFTs, %
 something that one might expect from the fuzzy intuition given in the introduction.
We assumed throughout that the fixed points of the nonlocal actions are indeed CFTs -- rather than scale-invariant fixed points -- and that the infrared duality between the two construction routes holds.
We were able to check $F$-extremisation in perturbation theory in various examples.
This is conceptually pleasing: by taking a step back to consider more general classes of nonlocal and nonunitary CFTs, we have found an explanation for the \enquote{experimentally} observed structure in the $\gO(N)$ model in \cref{sec:motivation}.

Our observation also generalises the large-$N$ ${F}$-extremisation story in \cite{Fraser-Taliente:2024hzv}.
It clarifies why large-$N$ $F$-extremisation works: when doing it, we are actually considering the family of long-range melonic CFTs containing $n_f$ fields. $n_f-1$ of them have arbitrary scaling dimensions $\Delta_{\phi_i}$ fixed by long-range kinetic terms. 
The remaining field has its scaling dimension dynamically set to $\Delta_{\phi_{n_f}} = d-\sum_{i=1}^{n_f-1} \Delta_{\phi_i} + O(1/N)$ by the simple form of the melonic interaction. 
Extremising the free energy of that long-range theory with respect to each of the $\Delta_{\phi_i}$, we find the short-range limit, which we call the melonic CFT. %
It also resolves the question in \cite{Fraser-Taliente:2024rql} of the identity of the additional lines of theories: they are the other local (but nonunitary) limits of the melonic CFTs.
However, despite the superficial similarities, whether there is any sharp relationship between this nonlocal $F$-extremisation and supersymmetric $F$-extremisation remains unclear.

Another takeaway from this work has been that the long-range CFTs provide a useful benchmark for conformal perturbation theory calculations. 
They provide a minimal environment, with no additional near-marginal operators like $\phi(-\partial^2)^k\phi$ or the geometric couplings, which need to be accounted for in the short-range theories. 
We hope to explore this further presently \cite{Fraser-Taliente2026cpt}.

\subsection{Next steps and next questions}

A natural next step is to continue with our perturbative checks. 
Natural candidates for this would be the long-range $\phi^{2(m-1)}$ and $i\phi^{2m-1}$ models in arbitrary $d$, generalising the studies of \cite{Behan:2025ydd} and \cite{Eustachon:2026vjn} away from $d=2$.
It would be excellent if we could use our result to simplify the complex expressions \cite[(10c,11)]{Derkachov:1993uw} \cite{Vasiliev:1982dc} for the anomalous dimension corrections in the large-$N$ expansion of the vector models.
Rong's comments from \cite[\S 4]{Rong:2024vxo} also apply identically: we could also investigate the long-range versions of the multi-scalar \cite{Osborn:2020cnf,Pannell:2024sia,Benedetti:2020rrq,Benedetti:2024mqx,Harribey:2022esw}/fermion \cite{Pannell:2023tzc,Pannell:2025ajf} fixed points, Gross-Neveu-Yukawa theories \cite{Chai:2021wac,Fei:2016sgs,Gracey:2021ili,Gracey:2021yyl,Schroder:2025rka}, the non-Abelian Thirring model, QED/QCD \cite{DiPietro:2019hqe,DeCesare:2022obt} (including the various mixed-dimensional theories -- see \cite{Fraser-Taliente:2024lea} and references therein), and the nonlinear sigma model \cite{Gubser:2019uyf,Giombi:2019enr} (though we should be wary \cite{DeCesare:2025ukl,DeCesare:2026dwm}).
$F$-extremisation also likely admits an extension to long-range Lifshitz fixed points \cite{Benedetti:2024oif}; it would also be interesting to see whether CFTs with gauge fields can fit into this framework.
In \cref{sec:proposedFhat}, we suggested a new normalization $\hat{F}$ of the universal part of $F$, which removes overall $d$-dependent prefactors in a way that simplifies $\epsilon$-expanded expressions.
It would be lovely to see this simplification occur in further computations, some of which will be presented in a follow-up to this paper \cite{Fraser-Taliente2026mmls}. %

Other than at low orders, it is not immediately obvious that nonlocal $F$-extremisation will be useful in perturbative calculations of short-range data.
As we found in the case of the $\eta$ expansion in $\gO(N)$ $\phi^4$, we quickly run into sums that do not have a closed form in generic dimension.
However, it nicely explains structure in perturbative results, particularly when also considering higher-derivative theories -- and of course is a useful consistency check.

Interestingly, where the long-range CFTs coincide with solvable models, we have nonperturbative results for the locations of the extrema.
For example, consider the $d\to 2$ limit of the long-range $\gO(N)$ vector model. 
For $-2 \le N \le 2$ we know both $\Delta^\mathrm{SR}_{\hat{\phi}}$ and $\hat{F}(\Delta^\mathrm{SR}_{\hat{\phi}})=\frac{c_{2d}}{6}$ in closed form \cite[(3.4)]{Gorbenko:2020xya}, which provide two constraints on the function $\hat{F}_\mathrm{LR}(\Delta)$ \eqref{eq:FhatLR} for $d=2$.
The same holds for the $\phi^{2(m-1)}$ theories \cite{zamolodchikovConformalSymmetryMulticritical1986}, and possibly also the $i\phi^{2m-1}$ theories, though this is not clear at present \cite{Eustachon:2026vjn} \cite{Lencses:2022ira,Katsevich:2024jgq,ArguelloCruz:2025zuq,Katsevich:2025ojk,Benedetti:2026tpa}.

We close with the following broader observations and open questions about $F$-extremisation.

The fact that the derivative of the long-range free energy gives the two-point function normalization makes it clear that the sphere free energy can be computed from flat-space data alone by integrating from the known GFF value (which here is $N\tilde{F}_b(\frac{d}{4})$ exactly). 
This strongly suggests to us that the presence of the geometric couplings in the short-range computation of the conformal value of $\tilde{F}$ may be a distraction which could be removed. 
The free energy's derivative being a two-point function also appears similar to some other phenomena, to which it may or may not be related: first, gradient flow, and second, our recent observation that in conformal QED the free energy can be computed as an integral of the two-point function along a particular line of nonlocal field theories \cite{Fraser-Taliente2026qed}.

It is curious that \textit{minimisation} of the central charge $C_T$ (i.e. the coefficient $C_T$ of the two-point function of the stress tensor, which is proportional to $\hat{F}$ in $d=2$, though not generically related to $\hat{F}$ in $d\neq 2$) has been observed in the different context of the bootstrap \cite{El-Showk:2014dwa}. 
A similar phenomenon, related to the minimisation of a quantity with a schematic relation to $F$, has also been observed in simulations of CFTs on the fuzzy sphere in $d=3$ \cite{Wiese:2025zfb} and on the torus in $d=2$ \cite{Jacobsen:2024jel}.
Might these observations be somehow related?

Finally, we know the AdS duals of the conformal GFFs are massive scalars $\Phi$ on rigid AdS with Dirichlet ($\Delta>\dotwo$) or Neumann (if $\Delta <\dotwo$) boundary conditions.  %
The long-range Ising then has a dual description as a deformation of the boundary conditions of a massive scalar on rigid AdS (an AdS $\tilde{F}$-theorem has been applied to the long-range Ising model in this context in \cite[\S V.A]{Bason:2025sxb}).
The derivative with respect to $\Delta$ of the free energy is therefore related to the VEV of the mass operator $\Phi^2$.
It would be nice to understand the reasons for its vanishing in the short-range limit from the bulk perspective, both in free and interacting theories. 
Exactly in that limit, we have another bulk dual description, higher-spin gravity \cite{Aharony:2022feg}; how do the degrees of freedom rearrange themselves between the two AdS descriptions?

%% file: 09-Appendix.tex
\section{Conventions}\label{app:conventions}

We use the same conventions as defined in \cite[\S A]{Fraser-Taliente:2025udk}, and, as discussed in \cref{sec:caveats}, move smoothly between flat space and the sphere. However, for quick reference, we give a brief summary here.

We often use a simple function arising from the general Fourier transform of a traceless monomial in momentum or position space \cite[(4.18)]{Karateev:2018oml},
\begin{equation}\begin{aligned}\label{eq:FlamsDef}
&\cF_{\lambda, s} \equiv i^{-s}2^{d-\lambda}\pi^{d/2} \frac{\Gamma(\frac{d+s-\lambda}{2})}{\Gamma(\frac{\lambda+s}{2})},\\ %
\text{where } & \cF_{\lambda,s} (p^{\mu_1} \cdots p^{\mu_s} -\mathrm{traces})\, \abs{p}^{-(d-\lambda)-s}=\int_x e^{-i p\cdot x} (x^{\mu_1}\cdots x^{\mu_s}-\mathrm{traces})\, \abs{x}^{-\lambda-s}.
\end{aligned}\end{equation}

On the sphere, recall that for a propagator of the form
\begin{equation}\label{eq:desiredSphereProp}
C(x,y) \equiv \frac{C_\phi}{s(x,y)^{2\Delta}},
\end{equation}
we can schematically write
\begin{equation}\label{eq:schematicC}
S = \int_{x,y} \half \phi(x) C^{-1}(x,y) \phi(y),\quad C^{-1}(x,y) \approx \frac{1}{C_\phi} \frac{1}{\cF_{2\Delta,0}\cF_{2(d-\Delta),0}} \frac{1}{s(x,y)^{2(d-\Delta)}},
\end{equation}
dropping the local terms.
If we do not drop them, we have
\begin{equation}
    S= \int_x \half \phi(x) C^{-1} \phi(x),\quad  C^{-1}=\frac{\WeylOp_{\dotwo-\Delta}}{C_\phi \cF_{2\Delta,0} },
\end{equation}
where, for $\dotwo > \Delta > \dotwo -1$,
\begin{align} \label{eq:sphereKineticTerms}
\WeylOp_{\dotwo -\Delta} \phi(x) & = \lim_{r\to 0} \frac{1}{\cF_{2(d-\Delta),0}}\int_{y, \, s(x,y)>r} \frac{\phi(y) -\phi(x)}{s(x,y)^{2(d-\Delta)}} + \frac{1}{R^{d-2\Delta}}\frac{\Gamma(d-\Delta)}{\Gamma(\Delta)} \phi(x).
\end{align}
For lower values of $\Delta$ further subtraction terms are required, just like in flat space \cite[(3.4)]{Gubser:2019uyf}.

When we say the universal part of the free energy, we mean the part of the free energy that remains after modding out by all possible (local) contact terms for the metric $g$ (e.g. terms $\propto \Lambda^{d-2n-2m} \int \odif[d]{x} \sqrt{g}\nabla^{2m} \cR^n$ -- which yield $\delta$-functions in higher-point correlators of the stress tensor).
For free fields, we define
\begin{equation}\label{eq:FforbOrf}
    F_\phi(\Delta_\phi) = \begin{cases}
    F_b(\Delta_\phi) & \text{$\phi$ a scalar field}\\
    \Tr \spinid \, F_f(\Delta_\phi)\rvert_{r_f=2} & \text{$\phi$ a Dirac fermion}
    \end{cases}.
\end{equation}
As usual, $\tilde{F}=-\sin(\pi d/2) F$ and, as shown in \cite[\S A]{Fraser-Taliente:2025udk},
\begin{align}\label{eq:bosonF}
\tilde{F}_b(\Delta) &\equiv \int_{\frac{d}{2}}^{\Delta} \odif{\Delta} \, \tilde{F}_b'(\Delta), \quad \tilde{F}_b'(\Delta) \equiv \frac{\pi}{\Gamma(d+1)}\frac{1}{A(\Delta)A(d-\Delta)},\\ %
\label{eq:fermionF}
\tilde{F}_f(\Delta) &\equiv \int_{\frac{d}{2}}^{\Delta} \odif{\Delta} \, \tilde{F}_f'(\Delta), \quad \tilde{F}_f'(\Delta) \equiv -\frac{\pi}{\Gamma(d+1)}\frac{r_f}{A_f(\Delta)A_f(d-\Delta)}, 
\end{align}
for
\begin{equation}
A(\Delta) \equiv \frac{\Gamma(\tfrac{d}{2}-\Delta)}{\Gamma(\Delta)}, \quad A_f(\Delta)\equiv \frac{\Gamma \left(\frac{d}{2}-\Delta +\frac{1}{2}\right)}{\Gamma \left(\Delta +\frac{1}{2}\right)},
\end{equation}
and $r_f=2$ such that $\tilde{F}_f(\Delta)\rvert_{r_f=2}$ measures the degrees of freedom of one complex component of a Dirac fermion. 
Thus, we can also write $\tilde{F}_\text{Dirac fermion}(\Delta) = \tilde{F}_f(\Delta)\rvert_{r_f=2\Tr \spinid}$ instead of \eqref{eq:FforbOrf}.
Changing to our new normalization $\hat{F}$, these become the $\hat{F}$s written in terms of Euler beta functions in \eqref{eq:FhatsFB}.

\section{2PI proof of the free energy derivative} \label{app:2PI}

In this appendix we provide an all-orders perturbative proof of \eqref{eq:dFdDelEqTrGdC} in long-range $\phi^4$ theory in DREG, where we do not assume the vanishing of the one-point function of primaries.
For simplicity, we treat the one-field $\phi^4$ case, with action
\begin{equation}
S=\int_{x,y} \half \phi(x) C_{\Delta_\phi}^{-1}(x,y) \phi(y) + \frac{g_0}{4!} \int_x \phi^4,
\end{equation}
where $C_{\Delta_\phi}^{-1}(x,y) = Z_\phi \WeylOp_{\dotwo -\Delta_\phi} \delta(x-y)$ if we wish to use a fractional differential operator, but otherwise can be written as a subtracted hypersingular operator as usual \eqref{eq:sphereKineticTerms}.
We work in generic $d$ (DREG) such that there are no marginal geometric couplings to include.

We use exactly the conventions for the two-particle irreducible (2PI) formalism explained in \cite[\S 5.1]{Fraser-Taliente:2024hzv}:
\begin{equation}
\label{eq:2PI}
    \Gamma[\varphi,G]
    \;=\;
    S[\varphi]
    +\frac{1}{2}\Tr\log G^{-1}
    +\frac{1}{2}\Tr C_{\Delta_\phi}^{-1} G
    +\Gamma_2[\varphi,G],
\end{equation}
where $\varphi(x)$ is the one-point function, $G(x,y)$ is the connected two-point function and $\Gamma_2$ is (minus) the sum of 2PI vacuum diagrams built from $G$ and the interaction vertices.
At the vacuum saddle (assumed to be conformal) we have $\varphi_\star(x)=0$ and $G_\star(x,y)=\langle\phi(x)\phi(y)\rangle$, determined by
\begin{equation}
\fdv{\Gamma}{\varphi} \Big\rvert_{\varphi_\star,G_\star}=0,
    \qquad
\fdv{\Gamma}{ G}\Big\rvert_{\varphi_\star,G_\star}=0.
\end{equation}
The free energy is given by evaluating \eqref{eq:2PI} on the saddle:
\begin{equation}
    F_\mathrm{LR}(\Delta_\phi) = \Gamma[\varphi_\star(\Delta_\phi),G_\star(\Delta_\phi)].
\end{equation}
Differentiating $F$, we can drop two of the terms thanks to the saddle-point equations,
\begin{align}
  \odv{F}{\Delta_\phi} = \pdv{\Gamma}{\Delta_\phi}_{\varphi_\star,G_\star} + \Tr\left[\left(\fdv{\Gamma}{G_\star}\right)_{\Delta_\phi,\varphi_\star} \odv{G_\star}{\Delta_\phi}\right] + \int_x  \left(\fdv{\Gamma}{\varphi_\star}\right)_{\Delta_\phi,G_\star} \odv{\varphi_\star}{\Delta_\phi}.
\end{align}
Thus, explicit $\Delta_\phi$-dependence enters only through (1) the quadratic kernel $C_{\Delta_\phi}^{-1}$ and (2) the couplings to local operators, which appear only in $\Gamma_2$. 
We can now set $\varphi_\star=0$ and expand perturbatively in the sum over 2PI diagrams $D$ with symmetry factors $S_D$ and set of propagators $P_D$:
\begin{equation}
    - \Gamma_2[\varphi=0,G] \equiv \sum_D \frac{g_0^{n_D}}{S_D} \int_{x_1,\cdots, x_{n_D}}\prod_{P_D} G(\cdot, \cdot)
\end{equation}
Then the derivative for fixed $G_\star$ and $\varphi_\star=0$ is
\begin{align}
    -\pdv{\Gamma_2}{\Delta_\phi}_{G_\star} &= \odv{g_0}{\Delta_\phi}\sum_D \frac{1}{S_D} \odv{g_0^{n_D}}{g_0} \int_{x_1,\cdots, x_{n_D}}\prod_{P_D} G_\star(\cdot, \cdot) \\
    &= \odv{\log g_0}{\Delta_\phi} \sum_D  \frac{n_D}{S_D} g_0^{n_D} \int_{x_1,\cdots, x_{n_D}}\prod_{P_D} G_\star(\cdot, \cdot).
\end{align}
Since these are vacuum $\phi^4$ diagrams, the number of vertices $n_D$ is equal to half of the number of propagators $n_P$. 
Hence, we can trade $n_D$ for $\half n_P$, which can be traded for $\half \Tr G \fdv{\cdot }{G}$, since differentiating a free energy diagram with respect to $G(x,y)$ and then multiplying it by $G(x,y)$ gives exactly just a factor of $n_P$. 
Thus:
\begin{equation}
\sum_D  \frac{n_D}{S_D} g_0^{n_D} \int_{x_1,\cdots, x_{n_D}}\prod_{P_D} G_\star(\cdot, \cdot)= \half \sum_D \frac{n_P}{S_D} g_0^{n_D} \int_{x_1,\cdots, x_{n_D}}\prod_{P_D} G_\star(\cdot, \cdot)= -\half \Tr G\fdv{\Gamma_2}{G}
\end{equation}
We know the equation of motion for $G(x,y)$ is
\begin{equation}
- G^{-1}+ C_{\Delta_\phi}^{-1}  + 2\fdv{\Gamma_2}{G}=0,
\end{equation}
Since by assumption the fixed point is a long-range CFT, the field has exactly its UV scaling dimension, so $G^{-1} \propto C_{\Delta_\phi}^{-1}$.
We can then use the fact that we work in DREG, as in the bulk of the text, to see that $\Tr[G G^{-1}]\propto \Tr[G C_{\Delta_\phi}^{-1}]$ vanishes as it is $\propto I_2(d)=0$: hence $\pdv{\Gamma_2}{\Delta_\phi}_{G_\star}=0$.
Thus, as before, we wind up with the only surviving $\Delta_\phi$-dependent term in \eqref{eq:2PI} being $C_{\Delta_\phi}^{-1}$, and so
\begin{equation}
\label{eq:dFds_GCinv}
    \odv{F_{\mathrm{LR}}}{\Delta_\phi}=
    \frac{1}{2}\Tr G_\star\,\odv{C_{\Delta_\phi}^{-1}}{\Delta_\phi},
\end{equation}
recovering exactly \eqref{eq:dFdDelEqTrGdC}.

\section{The multi-operator generalisation of the \FtextOrPDF-theorem in CPT}\label{app:multiOperator}

Let us now generalise the demonstration of the $F$-theorem in CPT from \cref{sec:CPTfreeEnergy} to handle a perturbation with multiple operators. 
As a happy bonus we will also demonstrate gradient flow to second order in conformal perturbation theory.

For simplicity, to start with we take all of our operators to have the same scaling dimension $\Delta_\cO = d-\eta$; in \ref{sec:nonEqual} we will discuss the general case.
Defining a new coupling, $g^i= S_{d-1}\lambda^i/2=\pi^{d/2} \lambda^i/\Gamma(d/2)$, we find
\begin{equation}
    S=S_\text{CFT}+\sum_i \frac{\Gamma(d/2)}{\pi^{d/2}}\int_x g_0^i \cO_i, \quad
    \expval{\cO_i \cO_j}_{c,g^i=0} = \frac{C_{ij}}{\abs{x-y}^{2(d-\eta)}},
\end{equation}
where $C_{ij}$ is the generalised Zamolodchikov metric of the original CFT, which we use to raise and lower indices: $C_{ij} C^{jk} = \delta_i^k$.
We explicitly remove the $S_{d-1}/2$s in our correlators, defining:
\begin{align}
\expval{\cO_{\text{ren},i}(x_1) \cdots \cO_{\text{ren},j}(x_n)}_{g,c} \equiv \left(\tfrac{S_{d-1}}{2}\right)^{n} \mu^{-n\eta} (-1)^{n-1} \fdv{}{g^i(x_1)}\cdots \fdv{}{g^j(x_n)} F \Big\rvert_{g=\text{const}},
\end{align}
With this index convention and choice of normalization, the beta function's quadratic term is just the OPE coefficient (for $\eta=0$),
\begin{align}
\beta^i &=\odv{g^i}{\log \mu} = -\eta g^i + C\indices{^i_{jk}} g^j g^k + D\indices{^i_{jkl}} g^j g^k g^l + O(g^4) \label{eq:betaFuncMulti}.
\end{align}
The beta function is still exact in $\eta$ because we work in minimal subtraction. 
$D$ is related to our old $\beta_3$ by\footnote{
    This choice is particularly nice, because it makes the free field value of $D_{\cO\cO\cO\cO} = -1/\hat{F}_b''(\Delta)$.
} $\beta_3= \tfrac{\pi^{d}}{\Gamma(d/2)^2} D\indices{^\cO_{\cO\cO\cO}}$, and 
\begin{equation}
C\indices{^i_{jk}} = C^{ia}C_{jka},\quad D\indices{^i_{jkl}} = C^{ia} D_{ajkl}.
\end{equation}
Both $C\indices{^i_{(jk)}}$ and $ D\indices{^i_{(jkl)}}$ evidently must be totally symmetric in their \textit{lowered} indices $jk$ and $jkl$.
We also know that $C_{ijk}$, the OPE coefficient of the original CFT (at $\eta=0$), must be totally symmetric.
However, in this scheme we can also take $D_{ijkl}$ to be totally symmetric, which will be important for demonstrating gradient flow.

\subsection{The full symmetry of \texorpdfstring{$D$}{D}}

To show this, assuming that we have renormalized to leading order in $g$, let us consider the order $g^2$ term in the multi-operator generalisation of \eqref{eq:OOtoOrd4}, i.e. 
\begin{equation}
\left(\tfrac{S_{d-1}}{2}\right)^{2}  \expval{\cO_{\text{ren},i}(x)\cO_{\text{ren},j}(y)}_{g,c} = -\mu^{-2\eta}\fdv{}{g^i(x)} \fdv{F}{g^j(y)} \Big\rvert_{g =\text{const}}, \quad x \neq y.
\end{equation}
All the subdivergences are cancelled by the $C\indices{^i_{jk}}/\eta$ and $C\indices{^i_{ja}}C\indices{^a_{kl}}/\eta^2$s in $g^b$.
Thus, running through the argument from \cref{sec:countertermComputation}, the only divergences occur when the two integrated operators collapse onto either $x$ or $y$:
\begin{align}\label{eq:Iij}
    &\left(\tfrac{S_{d-1}}{2}\right)^{2} I_{ij}(x,y) \equiv \frac12\,g^k g^l \int_{w,z} \expval{\cO_i(x)\cO_j(y)\cO_k(w)\cO_l(z)}_c,\qquad x\neq y \notag\\
&  \sim \frac12\,g^k g^l \Bigg(\int_{w,z \in \mathcal{U}_x}\expval{\cO_i(x)\cO_j(y)\cO_k(w)\cO_l(z)}_c
+ \int_{w,z \in \mathcal{U}_y}\expval{\cO_i(x)\cO_j(y)\cO_k(w)\cO_l(z)}_c\\
& \qquad \qquad+\text{ subdivergences that cancel}.
\end{align}
Here, $w,z \in \mathcal{U}_a$ means an integral over $w,z \text{ near }a$, in a region of size $R$ whose precise shape does not matter, which for specificity we take to be the symmetric region  $\mathcal{U}_x\equiv \{(w,z):\ \abs{w-x}<R,\ \abs{z-x}<R,\ \abs{w-z}<R\}$.

We can now use the multi-operator generalisation of the OPE \eqref{eq:OPE},
\begin{equation}\label{eq:OPEmulti}
   \cO_i(x)\cO_k(w) \cO_l(z) \sim f\indices{_{ikl}^a}(w-x,z-x) \cO_a(x) + \text{less singular}.
\end{equation}
Considering the first term in \eqref{eq:Iij}, we can substitute the OPE to find
\begin{equation}\label{eq:firstTermInI}
    \frac12\,g^k g^l \int_{w,z \in \mathcal{U}_x}\expval{\cO_i(x)\cO_j(y)\cO_k(w)\cO_l(z)}_c \sim\frac12\,g^k g^l \left[\int_{w,z \in \mathcal{U}_0}f\indices{_{ikl}^a}(w,z)\right] \expval{\cO_a(x)\cO_j(y)}_c.  %
\end{equation}
The integral of $f$ will have a $1/\eta$ pole, and will also be fully symmetric in $(ikl)$, since it just describes an arbitrary set of three operators coming together at a point.
Thus, we define
\begin{equation}\label{eq:ffrakDef}
    \eqref{eq:firstTermInI} \sim \frac12\,g^k g^l \frac{C_{ja}}{\abs{x-y}^{2d}}\frac{\mathfrak{f}\indices{_{ikl}^a}}{\eta},
\end{equation}
and overall 
\begin{align}
    &\left(\tfrac{S_{d-1}}{2}\right)^{2}  I_{ij}(x,y) \sim  \frac{1}{\abs{x-y}^{2d}} \frac{g^k g^l}{2\eta} \left(C_{ja} \mathfrak{f}\indices{_{(ikl)}^a}  + C_{ia} \mathfrak{f}\indices{_{(jkl)}^a}\right) + \text{ subdivergences}.
\end{align}
Assuming only the symmetry $D^i_{(jkl)}$ in $g_0^i$ and so $\beta^i$ for the moment, we see that the two-point function of renormalized $\cO_i$s contains the divergence
\begin{align}
 \expval{\cO_{\text{ren},i}(x) \cO_{\text{ren},j}(y)} &\supset \frac{1}{\abs{x-y}^{2d}}\frac{g^k g^l}{2\eta} \Big[C_{ia}(3D\indices{^a_{(jkl)}} + (\tfrac{2}{S_{d-1}})^{2}\mathfrak{f}\indices{_{(jkl)}^a})\\
 &\qquad\qquad+ C_{ja} (3D\indices{^a_{(ikl)}}+(\tfrac{2}{S_{d-1}})^{2} \mathfrak{f}\indices{_{(ikl)}^a}) \Big].
\end{align}
Each pair of brackets must cancel separately  -- this must be true since the correlator should still be finite if we set $g^i(x)=0$ in a small region around $y$ (this would also be clear if we had renormalized the flat-space one-point function as in \cite{Komargodski:2016auf}).
Indeed, we find that
\begin{equation}\label{eq:Dnonsymm}
D_{i(jkl)} =\lim_{\eta \to 0} -\frac{2}{S_{d-1}} \left(\int_{\cR} \odif[d]{x} \,\expval{\cO_{(j}(0) \cO_k(x)\cO_{l)}(\hat{e}_1) \cO_{i}(\infty)}_c -\frac{ S_{d-1}}{\eta} C\indices{^m_{i(j}}C_{kl)m})\right), %
\end{equation}
for $\cR = \{x:\, \abs{x} <1, \abs{x} <\abs{\hat{e}_1 -x}\}$ (this matches the region of integration in \cite[(3.31)]{Komargodski:2016auf}).
The crucial question for gradient flow is whether this is fully symmetric.
As we show in \cref{app:Dsymmetric}, we can use conformal symmetry to show that it is, under a mild condition on the convergence of an integral and the assumption of inversion symmetry.
Hence, $D_{i(jkl)}=D_{(ijkl)}$, and so we can write
\begin{align}\label{eq:FinalDEquation}
D_{(ijkl)} &=\lim_{\eta \to 0} -\frac{2}{S_{d-1}} \left(\int_\cR \odif[d]{x} \,\expval{\cO_{(j}(0) \cO_k(x)\cO_{l)}(\hat{e}_1) \cO_{i}(\infty)}_c -\frac{ S_{d-1}}{\eta} C\indices{^m_{i(j}}C_{kl)m})\right).
\end{align}
We see that our final expressions are manifestly covariant with respect to a rescaling of $\cO$.
These symmetry properties of the coefficients in the beta function are not scheme-invariant: under a general scheme change their symmetry properties change \cite[(4.4)]{Gaberdiel:2008fn}.

\subsection{A computation of \texorpdfstring{$\hat{F}$}{\^{}F} and gradient flow}

It will be convenient to define $L=\log (2\mu R) + \frac{\psi_0(-\tfrac{d}{2})+\gamma}{2}$. This is just like in modified minimal subtraction, where we absorb certain standard factors into $\tilde{\mu} = \mu \exp(\cdots)$. 
We then have
\begin{align}
&\begin{aligned}
\delta\hat{F}_g =& -\half \eta C_{ij} g^i g^j (1+2\eta L) + \frac{1}{3} C_{ijk} g^i g^j g^k (1+6\eta L) \label{eq:FvalueMulti}\\
&+ \frac{1}{4} \left(D_{ijkl} -4 C\indices{^a_{ij}}C\indices{_{kla}} L\right)g^i g^jg^k g^l + O(g^{5-m}\eta^{m}).
\end{aligned}
\end{align}

Using this, we can confirm the Callan-Symanzik equation for $\hat{F}$. 
Recall that typically for the generating functional $W[g_{\mu\nu}, g^i] = -F$, we have \cite{Baume:2014rla}
\begin{equation}
    \left(\pdv{}{\log \mu} + \beta^i \pdv{}{g^i}\right) W = \text{local curvature terms}. %
\end{equation}
Because we are considering the universal part of both sides, with constant metric, and in general $d$ such that there are no marginal geometric anomaly terms (unlike say \cite[(2.15)]{Fei:2015kta}), we can write
\begin{align}\label{eq:CSforFhat}
    \odv{\delta \hat{F}_{g}}{\log\mu} = \beta^i \pdv{\delta \hat{F}_{g}}{g^i} + \odv{L}{\log\mu} \pdv{\delta \hat{F}_{g}}{L} = 0.
\end{align}
Thus,
\begin{equation}
    \delta\hat{F}_g = \exp(-L \beta^i \partial_i)\delta\hat{F}_g^{(0)}
\end{equation}
so the coefficient of $L$ must vanish on its own at the fixed point (as must the coefficient of $L^2$, $L^3$, etc.). 
It does so -- indeed, to this order the coefficient is just minus the beta function squared, exactly as predicted by \eqref{eq:CSforFhat}:
\begin{align}
    [L] \delta \hat{F}_{g} &=   - \eta^2 C_{ij} g^i g^j +  2\eta C_{ijk} g^i g^j g^k- C\indices{^a_{ij}}C\indices{_{kla}}  g^i g^jg^k g^l + O(g^5),\\
    & = - \beta^i \beta_i + O(g^5) = -C_{ij}\beta^i \beta^j + O(g^5).
\end{align} 

Putting all this together, we see that 
\begin{align}\label{eq:dFhatFull}
\delta\hat{F}_g &=- \mathcal{C}_{ij} \beta^i \beta^j L + \int_0^{g} \odif{g^i} \mathcal{C}_{ij}\beta^j  + O(g^{5-m} \eta^m).
\end{align}
The Callan-Symanzik equation then tells us that all of the information in \eqref{eq:dFhatFull} is contained in the coefficient of $L^0$, i.e. $\delta\hat{F}_g^{(0)} =   \delta\hat{F}_g\rvert_{L=0}$.
\begin{align}
\delta\hat{F}_g^{(0)} &=\int_0^{g} \odif{g^i} \mathcal{C}_{ij}\beta^j  + O(g^{5-m} \eta^m)
\end{align}
where the new metric is still the original CFT's Zamolodchikov metric to this order: $\mathcal{C}_{ij} = C_{ij}+O(g^{2-m}\eta^{m})$. 
This integral is path-independent, as the integrand is a total derivative (i.e. the one-form is exact):
\begin{equation}
    \pdv{}{g^i} [L^0]\delta\hat{F}_g = C_{ij} \beta^j + O(g^{4-m} \eta^m).
\end{equation}
This is gradient flow.
It is this nice form, where $\hat{F}$ is the integral of $\mathcal{C}_{ij} \beta^j$, that was the principal motivation for our definition of $\hat{F}$ in \cref{sec:proposedFhat}.
Of course, $\delta\hat{F}_g^{(0)}$ is somewhat ambiguous away from fixed points, even in this minimal subtraction-type scheme, as by changing $L$ we can modify it. However, it is evident from the Callan-Symanzik equation that modifying our modified minimal subtraction scheme is equivalent to setting $L=L' + \alpha$. 
Obviously this is equivalent to just changing the RG time by $-\alpha$, and so does not spoil the gradient flow. 

Note that we have certainly not accounted for the distinction between $\beta$ functions and $B$-function terms, as described in \cite[\S 1]{Pannell:2025ajf} and \cite{Pannell:2024sia}, \cite{Fortin:2012hn} (harkening back to \cite{Jack:1990eb}).
These lead to trivial RG cycles that do not change the physical theory. 
$B=0$ indicates a true fixed point; any nonzero value of $\beta$ then means a rotation of the couplings in flavour space along the RG flow:
\begin{align}
    B^I =0, &\quad \beta^I \neq 0, \quad \implies \odv{g^I}{\log \mu} = (S g)^I, \quad \odv{S\indices{^I_J}}{\log \mu} \overset{\text{\cite{Fortin:2012hn}}}{=} 0\\
    \implies &\quad g(\mu) =U(\mu) g(\mu_0), \quad U=\exp(-S \log (\mu/\mu_0))
\end{align}
Recalling that the theory must be invariant under simultaneous rotations of the fields and the couplings, we can just rotate the fields the other way to obtain a standard fixed point.
In the nonlocal theories of the kind that we have been considering, such local operators are not present, so we can ignore them. For example, the nonlocal kinetic term is not renormalized, so there is no $Z^{\half}$ here that can develop poles.

The gradient flow to leading order in perturbation theory becomes even simpler in the case where $g_\star \propto \sqrt{\eta}$.
This new scaling of $g$ does not modify any of the results above, only the size of the next correction:
\begin{equation}
    \pdv{\delta\hat{F}_g}{g^j} = \mathcal{C}_{ji} \beta^i + O(g^{4-2m}\eta^m), \quad \beta^i= -\eta g^i + D\indices{^i_{jkl}} g^j g^k g^l + O(g^4),
\end{equation}
for a generalised Zamolodchikov metric which to this order is just the two-point coefficient $\mathcal{C}_{ji}=C_{ji} + O(g^{2-2m}\eta^m)$.

This also confirms that the integrated one-point functions on the sphere of the renormalized primary operators $\cO_{i,\mathrm{ren}}$ vanish at the fixed point\footnote{
Since we work in $d$ dimensions in a generic CFT, there are no other nearly-marginal primaries that the operator can mix with. 
Not correctly finding the linear combination of such operators (like $\mathbb{I} \times\cR^n$) that is a primary is the usual reason for the one-point function of operators not vanishing (cf. \cite[\S D]{Giombi:2025pxx}).
}, as it should in a CFT:
\begin{equation}
   \pdv{\delta \hat{F}_g}{g^i}\Big\rvert_{g_\star} \propto \expval*{\int_x \cO_{i,\mathrm{ren}}(x)} \propto \beta^i\rvert_{g_\star} = 0
\end{equation}

Putting this all together, we see that at fixed points
\begin{align}
\delta\hat{F}_{g_\star}  = \delta\hat{F}_{g_\star}^{(0)}
= -\eta \frac{C_{ij}}{2} g_\star^i g_\star^j +\frac{ C_{ijk}}{3} g_\star^i g_\star^j g_\star^k +\frac{D_{ijkl}}{4} g_\star^i g_\star^j g_\star^k g_\star^l + O(g_\star^5)
\end{align}
If $C_{ijk}\neq 0$, we find that for $g_\star^i \propto \eta +O(\eta^2)$,
\begin{align}
\delta\hat{F}_{g_\star}  = \delta\hat{F}_{g_\star}^{(0)}
=-\frac{C_{ij}}{6}\eta g_\star^i g_\star^j +O(\eta^4),
\end{align}
agreeing with \cite[(4.38)]{Fei:2015kta}. 
Since in a unitary theory the matrix $C_{ij}$ is positive definite, and the critical couplings are real if the new fixed point is unitary, the generalised $F$-theorem is true at this order.

In our case, we must go to the next order, and use $\eta C_{ij} g_\star^i g_\star^j = D_{ijkl} g_\star^i g_\star^j g_\star^k g_\star^l+O(\eta^3)$, meaning that $g_\star \propto \pm\sqrt{\eta}(1+O(\eta))$:
\begin{align}
\delta\hat{F}_{g_\star}  = \delta\hat{F}_{g_\star}^{(0)}
= -\eta \frac{C_{ij}}{4} g_\star^i g_\star^j+ O(\eta^{3})
\end{align}
which is manifestly negative for the same reasons if both the original theory and the fixed point are unitary.
Notably, therefore, whenever the deformation $\eta$ parametrises a continuous line of CFTs (which cannot be an actual conformal manifold of CFTs, as we commented around \eqref{eq:notLocalOperator}), evidently the original CFT (assuming that it does not change with $\eta$) must be an extremum of $\hat{F}$ if the OPE coefficients $C_{ijk}$ are zero.

\subsection{Non-equal UV scaling dimensions}\label{sec:nonEqual}
A further generalisation to non-equal scaling dimensions $\Delta_i = d-\kappa^i \eta$ is possible.
After a series of nontrivial cancellations, the final result for the free energy with $L=0$ and the beta functions is straightforward:
\begin{align}
    \delta\hat{F}^{(0)}_g &= \int_0^g \odif{g^i} \mathcal{C}_{ij} \beta^j,\\
    \beta^i &= -\kappa^i \eta g^i + C\indices{^i_{jk}}g^j g^k + D\indices{^i_{jkl}} g^j g^k g^l + O(g^4),  \quad(\text{no sum on }i),\label{eq:nonEqualDimsBeta}
\end{align}
where $\mathcal{C}_{ij}=C_{ij}+O(g^2)$.
This identity relies on the UV CFT's Zamolodchikov metric $C_{ij}$ being block diagonal, i.e. $C_{ij} \propto \delta_{\kappa^i \kappa^j}$. This computation follows immediately by generalising the above, but its details will be reported elsewhere \cite{Fraser-Taliente2026cpt}.

This allows us to confirm the generalised $F$-theorem again. 
We must now consider two different cases.
If there is any vector $h^i$ such that $C\indices{^i_{jk}}h^j h^k=0$ as a vector (and $D\indices{^i_{jkl}} h^j h^k h^l \neq 0$), then we can take $g_\star^i \propto \sqrt{\eta}$ and look for fixed points.
If there are any, they satisfy
\begin{equation}\label{eq:FmaxMultiDirections}
    \delta\hat{F}_{g_\star} = \delta\hat{F}^{(0)}_{g_\star} = - \frac{C^{(\kappa)}_{ij}}{4} \eta g_\star^i g_\star^j + O(\eta^3),
\end{equation}
where $C^{(\kappa)}_{ij} = \kappa^i C_{ij} = \kappa^j C_{ij}$ (no sum). 
Otherwise, we take $g_\star^i \propto \eta$ and look for fixed points. Any such fixed points will have
\begin{equation}
    \delta\hat{F}_{g_\star} = \delta\hat{F}^{(0)}_{g_\star} = - \frac{C^{(\kappa)}_{ij}}{6} \eta g_\star^i g_\star^j + O(\eta^4).
\end{equation}
Let us assume that the UV CFT is unitary, so $C_{ij}$ is positive definite on each equal-$\kappa$ block.
Since $\kappa^i >0$ by necessity, $C^{(\kappa)}_{ij}$ is also positive definite.
Thus, if $g_\star^i$ is real, the change in $\hat{F}$ at any nontrivial IR CFT is negative in either case.

\input{095-Dsymmetric.tex}

\section{\texorpdfstring{$\phi^4$}{phi\^{}4} diagrams in the \texorpdfstring{$\eta$}{eta} expansion}\label{app:DiagramValues}

\begin{figure}[H]
\centering
\newlength{\CrossGap}
\setlength{\CrossGap}{1.15mm} %

\newcommand{\brokenedge}[4][]{%
  \draw[quartic edge,#1] (#2) -- ($(#3)!\CrossGap!(#2)$);
  \draw[quartic edge,#1] ($(#3)!\CrossGap!(#4)$) -- (#4);
}
\begin{tikzpicture}[
    x=0.95cm,
    y=0.95cm,
    quartic edge/.style={draw=black, line width=0.5pt, line cap=round, line join=round},
    quartic double/.style={quartic edge, double=white, double distance=2.8pt},
    quartic vertex/.style={circle, fill=black, inner sep=1.8pt},
    every node/.style={font=\small}
]

    \begin{scope}[shift={(0.8,3.2)}]
        \draw[quartic double] (0,0) -- (0.9,0);
        \draw[quartic edge]   (0,0) -- (0.9,0);
        \foreach \x in {0, 0.9} {
            \node[quartic vertex] at (\x,0) {};
        }
        \node at (0.45,-1.05) {$H_2$};
    \end{scope}

    \begin{scope}[shift={(2.45,2.8)}]
        \draw[quartic double] (0,0) -- (0.52,0.88) -- (1.04,0) -- cycle;
        \foreach \x/\y in {0/0, 0.52/0.88, 1.04/0} {
            \node[quartic vertex] at (\x,\y) {};
        }
        \node at (0.52,-0.65) {$H_3$};
    \end{scope}

    \begin{scope}[shift={(4.4,2.8)}]
        \draw[quartic double] (0,1) -- (1,1);
        \draw[quartic double] (0,1) -- (0,0);
        \draw[quartic double] (1,1) -- (1,0);
        \draw[quartic double] (0,0) -- (1,0);
        \foreach \x/\y in {0/0, 0/1, 1/1, 1/0} {
            \node[quartic vertex] at (\x,\y) {};
        }
        \node at (0.5,-0.58) {$H_4^{(1)}$};
    \end{scope}

    \begin{scope}[shift={(6.8,2.8)}]
        \draw[quartic edge]   (0,1) -- (1,1);
        \draw[quartic double] (0,1) -- (0,0);
        \draw[quartic edge]   (0,1) -- (0,0);
        \draw[quartic double] (1,1) -- (1,0);
        \draw[quartic edge]   (1,1) -- (1,0);
        \draw[quartic edge]   (0,0) -- (1,0);
        \foreach \x/\y in {0/0, 0/1, 1/1, 1/0} {
            \node[quartic vertex] at (\x,\y) {};
        }
        \node at (0.5,-0.58) {$H_4^{(2)}$};
    \end{scope}

    \begin{scope}[shift={(9.2,2.8)}]
        \coordinate (A) at (0,1);
        \coordinate (B) at (1,1);
        \coordinate (C) at (1,0);
        \coordinate (D) at (0,0);

        \draw[quartic double] (A) -- (B);
        \draw[quartic edge]   (A) -- (D);
        \draw[quartic edge]   (B) -- (C);
        \draw[quartic double] (D) -- (C);

        \path[name path=g43 main]  (D) -- (B);
        \path[name path=g43 break] (A) -- (C);
        \path[name intersections={of=g43 main and g43 break, by=I43}];

        \draw[quartic edge] (D) -- (B);
        \brokenedge{A}{I43}{C}

        \foreach \p in {A,B,C,D} {
            \node[quartic vertex] at (\p) {};
        }
        \node at (0.5,-0.58) {$H_4^{(3)}$};
    \end{scope}

    \begin{scope}[shift={(0,0)}]
        \draw[quartic double] (0,0.45) -- (0.75,1.2) -- (1.5,0.45);
        \draw[quartic edge]   (0,0.45) -- (1.5,0.45);
        \draw[quartic edge]   (0,0.45) -- (0.45,-0.35);
        \draw[quartic edge]   (1.5,0.45) -- (1.05,-0.35);
        \draw[quartic double] (0.45,-0.35) -- (1.05,-0.35);
        \draw[quartic edge]   (0.45,-0.35) -- (1.05,-0.35);
        \foreach \x/\y in {0/0.45, 0.75/1.2, 1.5/0.45, 0.45/-0.35, 1.05/-0.35} {
            \node[quartic vertex] at (\x,\y) {};
        }
        \node at (0.75,-0.88) {$H_5^{(1)}$};
    \end{scope}

    \begin{scope}[shift={(2.2,0)}]
        \coordinate (L)  at (0,0.45);
        \coordinate (T)  at (0.75,1.2);
        \coordinate (R)  at (1.5,0.45);
        \coordinate (BL) at (0.45,-0.35);
        \coordinate (BR) at (1.05,-0.35);

        \draw[quartic double] (L) -- (T) -- (R);
        \draw[quartic edge]   (L) -- (BL);
        \draw[quartic edge]   (R) -- (BR);
        \draw[quartic double] (BL) -- (BR);

        \path[name path=g52 main]  (BL) -- (R);
        \path[name path=g52 break] (L) -- (BR);
        \path[name intersections={of=g52 main and g52 break, by=I52}];

        \draw[quartic edge] (BL) -- (R);
        \brokenedge{L}{I52}{BR}

        \foreach \p in {L,T,R,BL,BR} {
            \node[quartic vertex] at (\p) {};
        }
        \node at (0.75,-0.88) {$H_5^{(2)}$};
    \end{scope}

    \begin{scope}[shift={(4.4,0)}]
        \draw[quartic double] (0,0.45) -- (0.75,1.2) -- (1.5,0.45);
        \draw[quartic double] (0,0.45) -- (0.45,-0.35);
        \draw[quartic double] (1.5,0.45) -- (1.05,-0.35);
        \draw[quartic double] (0.45,-0.35) -- (1.05,-0.35);
        \foreach \x/\y in {0/0.45, 0.75/1.2, 1.5/0.45, 0.45/-0.35, 1.05/-0.35} {
            \node[quartic vertex] at (\x,\y) {};
        }
        \node at (0.75,-0.88) {$H_5^{(3)}$};
    \end{scope}

    \begin{scope}[shift={(6.8,0)}]
        \draw[quartic edge]   (0,0.45) -- (0.75,1.2) -- (1.5,0.45);
        \draw[quartic double] (0,0.45) -- (0.45,-0.35);
        \draw[quartic edge]   (0,0.45) -- (0.45,-0.35);
        \draw[quartic double] (1.5,0.45) -- (1.05,-0.35);
        \draw[quartic edge]   (1.5,0.45) -- (1.05,-0.35);
        \draw[quartic edge]   (0.75,1.2) -- (0.45,-0.35);
        \draw[quartic edge]   (0.75,1.2) -- (1.05,-0.35);
        \foreach \x/\y in {0/0.45, 0.75/1.2, 1.5/0.45, 0.45/-0.35, 1.05/-0.35} {
            \node[quartic vertex] at (\x,\y) {};
        }
        \node at (0.75,-0.88) {$H_5^{(4)}$};
    \end{scope}

    \begin{scope}[shift={(9.2,0)}]
        \coordinate (L)  at (0,0.45);
        \coordinate (T)  at (0.75,1.2);
        \coordinate (R)  at (1.5,0.45);
        \coordinate (BL) at (0.45,-0.35);
        \coordinate (BR) at (1.05,-0.35);

        \draw[quartic edge]   (L) -- (T) -- (R);
        \draw[quartic edge]   (L) -- (R);
        \draw[quartic double] (L) -- (BL);
        \draw[quartic double] (R) -- (BR);
        \draw[quartic edge]   (BL) -- (BR);

        \path[name path=g55 mid]   (L) -- (R);
        \path[name path=g55 left]  (T) -- (BL);
        \path[name path=g55 right] (T) -- (BR);
        \path[name intersections={of=g55 mid and g55 left,  by=I55L}];
        \path[name intersections={of=g55 mid and g55 right, by=I55R}];

        \brokenedge{T}{I55L}{BL}
        \brokenedge{T}{I55R}{BR}

        \foreach \p in {L,T,R,BL,BR} {
            \node[quartic vertex] at (\p) {};
        }
        \node at (0.75,-0.88) {$H_5^{(5)}$};
    \end{scope}

    \begin{scope}[shift={(11.6,0)}]
        \coordinate (L)  at (0,0.45);
        \coordinate (T)  at (0.75,1.2);
        \coordinate (R)  at (1.5,0.45);
        \coordinate (BL) at (0.45,-0.35);
        \coordinate (BR) at (1.05,-0.35);

        \draw[quartic edge] (L) -- (T) -- (R);
        \draw[quartic edge] (L) -- (R);
        \draw[quartic edge] (L) -- (BL);
        \draw[quartic edge] (R) -- (BR);
        \draw[quartic edge] (BL) -- (BR);
        \draw[quartic edge] (BL) -- (R);

        \path[name path=g56 mid]   (L) -- (R);
        \path[name path=g56 left]  (T) -- (BL);
        \path[name path=g56 right] (T) -- (BR);
        \path[name path=g56 main]  (BL) -- (R);
        \path[name path=g56 break] (L) -- (BR);
        \path[name intersections={of=g56 mid and g56 left,  by=I56L}];
        \path[name intersections={of=g56 mid and g56 right, by=I56R}];
        \path[name intersections={of=g56 main and g56 break, by=I56D}];

        \brokenedge{T}{I56L}{BL}
        \brokenedge{T}{I56R}{BR}
        \brokenedge{L}{I56D}{BR}

        \foreach \p in {L,T,R,BL,BR} {
            \node[quartic vertex] at (\p) {};
        }
        \node at (0.75,-0.88) {$H_5^{(6)}$};
    \end{scope}
\end{tikzpicture}
\caption{The $\phi^4$ free energy integrals used for $H_2$, $H_3$, $H_4^{(i)}$, and $H_5^{(i)}$. $H_5^{(6)}$, which is the complete graph with $5$ vertices $K_5$, is the most difficult diagram to evaluate: it leads to the complicated double sum $\alpha_{I_4}$ which we analyse in \cref{app:alphaI4}.}
\label{fig:quarticDiagramOrder}
\end{figure}
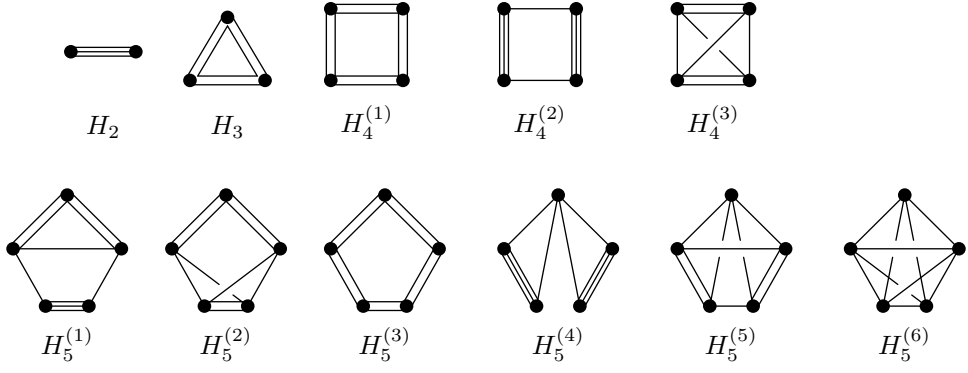

In this section, we give the diagrams required for the computation of the free energy of the long-range $\gO(N)$ $\phi^4$ theory.
The diagrams are integrated products of unit-normalized position-space propagators, without symmetry factors.
For completeness, \cref{fig:quarticDiagramOrder} also includes the lower-order topologies $H_2=I_2(4\Delta_\phi)$ and $H_3=I_3(4\Delta_\phi)$, where \cite[(B.1)]{Fei:2015kta}
\begin{align}
    I_2(\Delta) &= \int_{x,y} \frac{1}{s(x,y)^{2\Delta}} = (2R)^{2(d-\Delta)} \frac{2^{1-d} \pi^{d+\frac{1}{2}} \Gamma\!\left(\frac{d}{2} - \Delta\right)}{\Gamma\!\left(\frac{d+1}{2}\right) \Gamma(d - \Delta)} \,,\\
I_3(\Delta) &= \int_{x,y,z} \frac{1}{\left[s(x,y)\, s(y,z)\, s(z,x)\right]^{\Delta}}
= R^{3(d-\Delta)} \frac{8\pi^{\frac{3(1+d)}{2}} \Gamma\!\left(d - \frac{3\Delta}{2}\right)}{\Gamma(d)\, \Gamma\!\left(\frac{1+d-\Delta}{2}\right)^3} \,.
\end{align}
The three diagrams $H_4^{(i)}$ and the six diagrams $H_5^{(i)}$ are shown in the order used below; these are the same topologies denoted by $G_4^{(i)}$ and $G_5^{(i)}$ in \cite{Fei:2015oha}.

The values of the three diagrams $H_4^{(i)}$ are, in order,
\begin{subequations}
\begin{align}
H_4^{(i)}&= \frac{20 \pi ^{2 d} \Gamma \left(-\frac{d}{2}\right) (2R)^{4 \eta }}{\Gamma \left(\frac{d}{2}\right) \Gamma (d)} \Bigg\{\notag \\
    &\qquad\frac{2}{\eta }+4 (\psi_0(-\dotwo)+\gamma )\notag \\
    & \qquad\qquad +\eta  \left[\frac{16}{d^2}+4 (H_{-\dotwo-1}){}^2+4 \pi ^2 \csc ^2(\tfrac{\pi  d}{2})-\frac{39 \psi_1(\dotwo)}{10}-\frac{41 \pi ^2}{60}\right]+O(\eta ^2),\\
    &\qquad\frac{\eta  \Gamma (-\tfrac{d}{4}) \Gamma (\dotwo)^2}{10 \Gamma (\tfrac{3 d}{4})}+O(\eta ^2),\\
    &\qquad\frac{1}{\eta }+\frac{1}{10} (20 \psi_0(-\dotwo)-6 \psi_0(\tfrac{d}{4})+3 \psi_0(\dotwo)+17 \gamma ) \\
    &\qquad\qquad +\eta  \Big[\frac{8}{d^2}+2 (H_{-\dotwo-1}){}^2+2 \pi ^2 \csc ^2(\tfrac{\pi  d}{2})+\frac{1}{40} (2 \psi_0(\tfrac{d}{4})-\psi_0(\dotwo)+\gamma )^2  \notag\\
    &\qquad\qquad -\frac{3}{5} (\psi_0(-\dotwo)+\gamma ) (2 \psi_0(\tfrac{d}{4})-\psi_0(\dotwo)+\gamma )-\frac{81 \psi_1(\dotwo)}{40}-\frac{79 \pi ^2}{240}\Big]+O(\eta ^2)\notag
\Bigg\},
\end{align}
\end{subequations}
where we use the standard continuation of the Harmonic numbers $H_x \equiv \psi_0(x+1)+\gamma$.

The values of the six diagrams $H_{5}^{(i)}$ are, again in order,
\begin{subequations}
\begin{align}
H_5^{(i)} &= \frac{16 \pi ^{\frac{5 d}{2}} \Gamma \left(-\dotwo\right) (2R)^{5 \eta }}{3 \Gamma \left(\dotwo\right)^2 \Gamma (d)} \Bigg\{\notag \\
    &\frac{2 \Gamma \left(-\frac{d}{4}\right) \Gamma \left(\dotwo\right)^2}{\Gamma \left(\frac{3 d}{4}\right)}+O\left(\eta\right),\\
    &\frac{12}{\eta ^2}+\frac{5 \left(6 \psi_0\left(-\dotwo\right)-2 \psi_0\left(\tfrac{d}{4}\right)+\psi_0\left(\dotwo\right)+5 \gamma \right)}{\eta }\notag\\
    &\quad +\frac{1}{8} \Big[ \frac{1200}{d^2}+300 (H_{-\dotwo-1}){}^2+300 \pi ^2 \csc ^2(\tfrac{\pi  d}{2})+6 (2 \psi_0(\tfrac{d}{4})-\psi_0(\dotwo)+\gamma )^2\notag\\
    &\quad -100 (\psi_0(-\dotwo)+\gamma ) (2 \psi_0(\tfrac{d}{4})-\psi_0(\dotwo)+\gamma )-294 \psi_1(\dotwo)-51 \pi ^2\Big]+O(\eta),\\
&\frac{30}{\eta ^2}+\frac{75 (\psi_0(-\dotwo)+\gamma )}{\eta }\notag\\
& \quad +\frac{1}{8} \Big[\frac{3000}{d^2}+750 (H_{-\dotwo-1}){}^2+750 \pi ^2 \csc ^2(\tfrac{\pi  d}{2})-714 \psi_1(\dotwo)-131 \pi ^2\Big]+O(\eta),\\
& O(\eta),\\
&\frac{6}{\eta ^2}+\frac{5 (3 \psi_0(-\dotwo)-2 \psi_0(\tfrac{d}{4})+\psi_0(\dotwo)+2 \gamma )}{\eta }+\frac{1}{4} \Big[\frac{300}{d^2}\notag \\
&\quad +75 (H_{-\dotwo-1}){}^2+75 \pi ^2 \csc ^2(\tfrac{\pi  d}{2})+7 (2 \psi_0(\tfrac{d}{4})-\psi_0(\dotwo)+\gamma )^2\notag\\
&\quad -50 (\psi_0(-\dotwo)+\gamma ) (2 \psi_0(\tfrac{d}{4})-\psi_0(\dotwo)+\gamma )-78 \psi_1(\dotwo)-12 \pi ^2\Big]+O(\eta),\\
&\frac{\alpha_{I_4}}{2}+O(\eta)
\Bigg\}.
\end{align}
\end{subequations}
We define the double sum $\alpha_{I_4}$ that arises from $H_5^{(6)}$ in the next section.

The diagrams $H_4^{(i)}$ agree with the $d=3$ evaluation of \cite[(E.3)]{Giombi:2024zrt}.
It is worth noting that the limit $d\mapsto 2$ does not commute with the limit $\eta\mapsto 0$ in the intermediate diagrams, so we do not match the $d=2$ evaluation in \cite[(E.3)]{Giombi:2024zrt}.
This makes sense, as were we not working in DREG (i.e. with an explicit value of $d$ like $d=2$), we would need to pick out the coefficient of $\log R$ to obtain $\tilde{F}$ in $d=2$.
In $d=3$ the universal part of the free energy is the finite part, so the naive evaluation of the diagram works.

\subsection{\texorpdfstring{$\alpha_{I_4}$}{alpha\_\{I\_4\}} as a sum}\label{app:alphaI4}

\cite{Benedetti:2020rrq,Benedetti:2024mqx} identified a complicated $d$-dependent double sum $\alpha_{I_4}$ appearing at three loops in the beta function of long-range $\phi^4$ theory \eqref{eq:LRbetaExpansion}. 
It appears in both a free energy diagram ($H_5^{(6)}$, which is the complete graph with five vertices $K_5$) and the beta function, ending up in our result \eqref{eq:FtLRnextOrd} for $\hat{F}_{\mathrm{LR}_\eta}$ at order $\eta^5$.
In the short-range data, $\alpha_{I_4}$ is the origin of the $\zeta_3$ in the $\epsilon$ expansion, and it may also be related to the non-zeta term $I(\dotwo)$ in the $1/N$ expansion \cite{Broadhurst:1996yc}.
Here we present a slightly simplified form for it, and compute it in some particular dimensions.
Writing $S=l+m$ and $f=d/4$ for compactness, we have
\begin{equation}
    \alpha_{I_4} = \sum_{l,m=0}^\infty A_{lm},
\end{equation}
\newcommand{\ain}[2]{{a_{#1}^{(#2)}}}
\newcommand{\bin}[2]{{b_{#1}^{(#2)}}}
\begin{align}
    A_{lm}&\equiv \frac{\Gamma(2f)}{\Gamma(f)^2}\Bigg[
\frac{2\,\Pochhammer{f}{m}^2}{\Gamma(1+m)\,\Pochhammer{2f}{m}}\,
\frac{\Pochhammer{f + l + 1}{m} \Pochhammer{2 - f + l}{m}}{\Pochhammer{l + 1}{m + 1}^2} \,\left(\ain{S+1}{0}-\ain{l+1}{0}+\frac{2}{l+1}\right)
\\
&\qquad\qquad\qquad
+\frac{\Pochhammer{f}{S}^2}{\Gamma(1+S)\,\Pochhammer{2f}{S}}\,
\frac{\Pochhammer{f}{m}\,\Pochhammer{1-f}{m}}{\Gamma(1+m)^2}\,
\frac{\Pochhammer{f}{l}\,\Pochhammer{1-f}{l}}{\Gamma(1+l)^2}\,P_{lm}
\Bigg],\notag \\
P_{lm} &\equiv -\frac{1}{3}\Big(\ain{l}{0}^3+\ain{l}{2}+2\,\bin{S}{0}^3+2\,\bin{S}{2}\Big)
+\Big(\ain{m}{0}+\ain{l}{0}\Big)\Big(\bin{S}{0}^2+\bin{S}{1}\Big)\notag
\\
&\quad-2\,\bin{S}{0}\Big(2\,\psi_1(1)+\ain{m}{0}\,\ain{l}{0}+\bin{S}{1}\Big)
+\ain{m}{0}\Big(6\,\psi_1(1)+\ain{l}{0}^2+\ain{l}{1}\Big) \\
& \quad -\ain{l}{0}\Big(2\,\psi_1(1)+\ain{l}{1}\Big), \notag \\
\ain{l}{n} &\equiv 2(-1)^n \psi_n(1+l) -  \psi_n(f-l) - (-1)^n \psi_n(f+l), \\
\bin{S}{n} &\equiv 2\psi_n(f+S) - \psi_n(1+S) - \psi_n(2f+S).
\end{align}
This form was found by taking the output of \lstinline|MBsums.m|, and manually identifying cancellations and simplifications.
We were able to determine the following values of $\alpha_{I_4}$ in the limit of integer $f$ (i.e. $d\to 4k$, a limit which must be taken carefully):
\begin{equation}
\begin{array}{c|c}
d & \alpha_{I_4} \\
\hline
4  & 12 \zeta_3 \\
8  & \frac{216}{5} \left(\zeta_3-\frac{7}{6}\right) \\
12  & 360 \left(\zeta_3-\frac{173}{144}\right)\\
16  & \frac{552720}{143} \left(\zeta_3-\frac{73219}{60912}\right)\\
20  & \frac{112878360}{2431}\left(\zeta_3-\frac{984571}{819072}\right)
\end{array}.
\end{equation}
We did this by expanding and simplifying the summand in the four different sectors $(l<f,m<f)$, $(l \ge f,m<f)$, $(l<f, m \ge f)$, and $(l \ge f, m \ge f)$ in the limit of an explicit choice of $f$. 
Doing so for $f \to 1,2,3,4,5$, the sums in each sector could then be performed analytically with \lstinline|Mathematica|.
Based on this, we suspect that for integer $f=k$ it should be possible to rearrange the sum to make it manifest that $\alpha_{I_{4},d=4k} = C_k (\zeta_3-A_k)$ for rational $C_k,A_k$. 
$A_k$ should be an increasingly good rational approximation of Apéry’s constant $\zeta_3$, as the known values $A_{k=1,2,\cdots} = 0, 1.16666667, 1.20138889, 1.20204557, 1.20205672$ rapidly approach $\zeta_3=\sum_{n=1}^\infty n^{-3} = 1.20205690$, with a rough exponential fit $\zeta_3-A_k \sim 10^2 e^{-4 k}$ for $k>1$.

\begin{figure}
\centering
\includegraphics[width=0.72\linewidth]{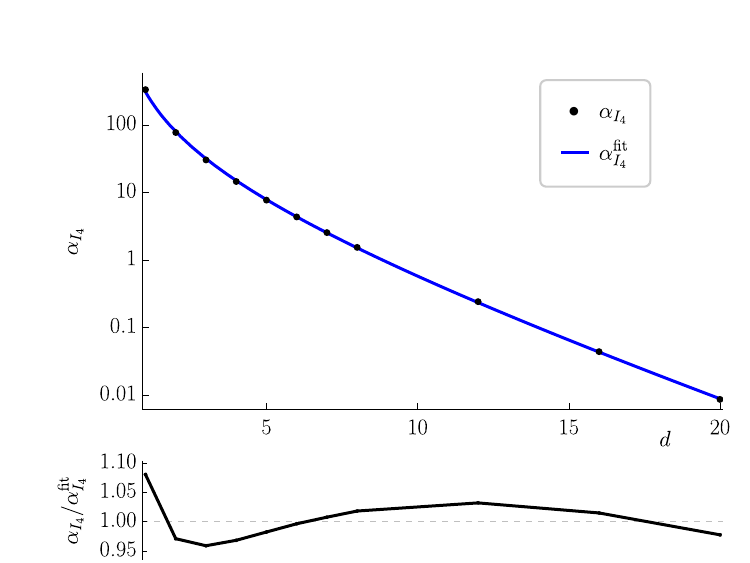}
\caption{The known values of the $K_5$ integral $\alpha_{I_4}$, compared to the exponential fit \eqref{eq:alphaFit}.}
\label{fig:alphaI4plot}
\end{figure}

For other values of $d$, our compact sum form efficiently reproduces the numerical results of \cite[(3.19)]{Benedetti:2024mqx}, i.e. $\alpha_{I_4,d=1}= 330.008919215(21)$, $\alpha_{I_4,d=2}= 76.62828703(7)$, and $\alpha_{I_4,d=3}= 30.1026152(6)$.
We also find $\alpha_{I_4,d=5}=7.6521970(6)$, $\alpha_{I_4,d=6} = 4.3142853(6)$, and $\alpha_{I_4,d=7}=2.532102(1)$.
This decreases exponentially, and so in \cref{fig:alphaI4plot} we show a rough fit to $\alpha_{I_4}$ of
\begin{equation}\label{eq:alphaFit}
\alpha_{I_4,d=4f}^{\text{fit}} \sim 52 f^{-\frac{3}{2}} e^{-1.25 f},
\end{equation}
which is accurate to within $\sim 10\%$. 

\section{\FhatextOrPDF improves approximants}\label{app:FhatNumerics}

In this appendix we show that when we only know the data to a fixed order in $\epsilon$, using the normalization $\hat{F}$ instead of $\tilde{F}$ appears to improve the extrapolation of the free energy from $d=4$ to $d=3$ (and even to $d=2$) in the $\epsilon$ expansion, whether naive or Pad\'e-improved.
We consider first the case of the free scalar, which we can solve exactly, as a benchmark; then we consider the critical Ising model (the $N=1$ $\phi^4$ CFT).

\subsection{The free scalar}

We can compute its $\epsilon$ expansion to $O(\epsilon^2)$ analytically\footnote{The coefficient $f_2=\frac{47}{120}
-\frac{\pi^2}{360}
-\frac{\log(4\pi)}{8}
-\frac{3\zeta_3}{8\pi^2}
+\half \sum_{\substack{m,n\ge 3\\ m,n\ \mathrm{odd}}}
\frac{(\zeta_m-1)(\zeta_n-1)}{(m+n+1)(m+n+3)}.$}, and to any desired order numerically.
Since our perturbative data for the Ising CFT below is to order $\epsilon^5$, we do the same here:
\begin{align}
\hat{F}_s &\equiv \hat{F}_b(\dotwo -1)|_{d=4-\epsilon} = \frac{1}{15}+\epsilon  \left(\log A-2 \zeta '(-3)-\tfrac{\gamma }{15}-\tfrac{139}{720}\right)+ f_2 \epsilon^2+ O(\epsilon ^3),\\
\hat{F}_s^{\text{Naive}^{\epsilon^5}}&=0.0666667+0.00646072 \epsilon +0.00262106 \epsilon ^2+0.00107121 \epsilon ^3 \label{eq:FhatFreeExpandedO5}\\
&\quad+0.000441669 \epsilon ^4+0.000183847 \epsilon ^5+O(\epsilon ^6),\notag
\end{align}
where $A$ is \href{https://en.wikipedia.org/wiki/Glaisher-Kinkelin_constant}{Glaisher's constant}.
We observe that all of the coefficients in the expansion are positive, something which appears to be true to all orders.

The exact values in $d=2$ and $d=3$ are 
\begin{equation}\label{eq:FhatsD23}
    \hat{F}_s\rvert_{d=2} = \frac{1}{6}, \quad \hat{F}_s\rvert_{d=3} =\frac{3}{2\pi^2} \left(\log (2)-\frac{3}{2 \pi ^2}\zeta_3\right)=0.0775801.
\end{equation}
These can be compared to (1) just evaluating \eqref{eq:FhatFreeExpandedO5} naively; (2) computing the four one-sided Pad\'e$_{[m,n]}$ approximants with total order $m+n=5$.
In $d=3$, we can also use the known $\hat{F}_s|_{d=2}=\tfrac{1}{6}$ to (3) do the same for the five two-sided Pad\'e$_{[m,n]}$ with total order $m+n=6$. 
Comparing the results from application of the same procedure to $\tilde{F}_s$, we find that $\hat{F}_s$ gives results that are better by an order of magnitude in $d=3$:
\begin{table}[H]
\centering
\begin{tabular}{c|c|c|c}
$d=2$  & Exact $F^\mathrm{x}$ & Naive $\epsilon^5$ & One-sided Pad\'e$_{[n,5-n],n=1,2,3,4}$ \\
\hline
$\hat{F}_s$
& $\frac{1}{6}$
& $0.669551\hat{F}_s^\mathrm{x}$
& $\{0.845433,1.00792,1.00789,0.845006\}\,\hat{F}_s^\mathrm{x}$
\\
$\tilde{F}_s$
& $\frac{\pi}{6}$
& $0.493004\tilde{F}_s^\mathrm{x}$
& $\{0.695695,0.880851,0.771736,0.877848\}\,\tilde{F}_s^\mathrm{x}$
\end{tabular}
\begin{tabular}{c|c|c|c|c}
$d=3$ & Exact $F^\mathrm{x}$ & Naive $\epsilon^5$ & One-sided Pad\'e$_{5}$ & Two-sided Pad\'e$_6$ \\
\hline
$\hat{F}_s$
& $0.0775801$
& $0.995891\,\hat{F}_s^\mathrm{x}$ %
& $0.9999(78\pm15)\,\hat{F}_s^\mathrm{x}$ %
& $1.0000(8\pm5)\,\hat{F}_s^\mathrm{x}$ %
\\
$\tilde{F}_s^\mathrm{x}$
& $0.0638071$
& $0.966386\,\tilde{F}_s^\mathrm{x}$
& $0.999(88\pm28)\,\tilde{F}_s^\mathrm{x}$
& $1.001(05\pm32)\,\tilde{F}_s^\mathrm{x}$
\end{tabular}
\caption{Comparison of naive truncation and one-sided Pad\'e approximants for the free scalar in $d=2$ and $d=3$. In $d=3$ we can also compare a two-sided Pad\'e approximant using the known $d=2$ data.
The naive truncation is already noticeably better for $\hat{F}_s$ than for $\tilde{F}_s$, and the Pad\'e estimates are as well. In $d=3$ the Pad\'e approximants are close to each other, so we just give the mean and standard error ($=$ standard deviation$/\sqrt{n}$).}
\end{table}

\begin{figure}[H]
\centering
\includegraphics[width=0.72\linewidth]{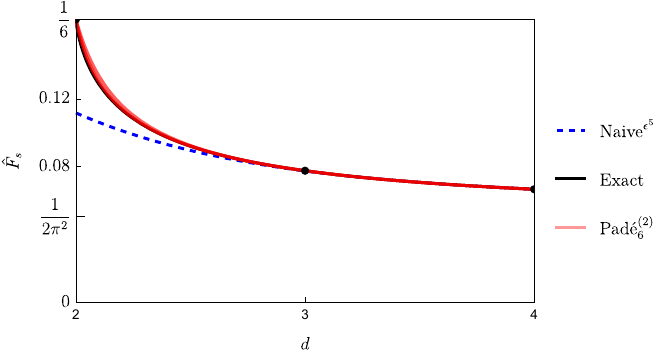}
\caption{Comparison of the naive $\epsilon$ expansion to order $\epsilon^5$, the exact result (see \cref{fig:FfreePlot}), and the two-sided Pad\'e approximants for $\hat{F}_s$ between $d=2$ and $d=4$. The one-sided approximants are almost identical to the two-sided ones, and so they are not shown. We also mark the asymptotic limit $\lim_{d\to\infty} \hat{F}_s=1/(2\pi^2)$.
}
\label{fig:FhatSPadeComparison}
\end{figure}

One plausible reason for this order of magnitude is that $\hat{F}$ varies more slowly across dimensions. 
There is only a factor of $5/2$ between $\hat{F}_{s,d=4}=1/15$ and $\hat{F}_{s,d=2}=1/6$, compared to the factor of $30$ between $\tilde{F}_{s,d=2}=\tfrac{\pi}{180}$ and $\tilde{F}_{s,d=4}=\frac{\pi}{6}$.

\subsubsection{Applied to the Ising model}

We can therefore combine \eqref{eq:FhatFreeExpandedO5} with \eqref{eq:Fhat4meps} to find
\begin{equation}
    \begin{aligned}
    \hat{F}_{SR}^{\text{Naive}^{\epsilon^5}} = \frac{1}{15} +0.006461 \epsilon +0.002621 \epsilon ^2+0.000300 \epsilon ^3-0.000287 \epsilon ^4+0.000149 \epsilon ^5+O(\epsilon ^6).
    \end{aligned}
\end{equation}

The naive evaluation of this in $d=2$ is now not bad, as we show in \cref{tab:NumericsIsing}.
We no longer know the answer in $d=3$, but for comparison we give $\tilde{F}_{\mathrm{SR}} = 0.061(2\pm 5)$ (and so $\hat{F}_{\mathrm{SR}} = 0.074(4\pm 6)$), which were determined using the fuzzy sphere \cite{Hu:2024pen}.
Once again, using $\hat{F}$ dramatically improves the one-sided match in $d=2$, and reduces the standard error of Pad\'e approximants in $d=3$ (but only slightly -- and we exclude the one approximant  Pad\'e$_{[5,1]}^{(2)}$ that gives a pole).
Once again, the likely reason is the slow variation of $\hat{F}_\mathrm{SR}$ across dimensions, changing only from $\tfrac{1}{15}$ to $\tfrac{1}{12}$ as $d$ changes from $4$ to $2$.
We find reasonable agreement with the fuzzy sphere value, although it remains consistently below the $\epsilon$ expansion results.

\begin{table}[H]
\centering
\begin{tabular}{c|c|c|c}
$d=2$ & Exact $F^\mathrm{x}$  & Naive $\epsilon^5$ & One-sided Pad\'e$_{[n,5-n],n=1,2,3,4}$ \\
\hline
$\hat{F}_{\mathrm{SR}}$
& $\frac{1}{12}$
& $1.11165\hat{F}_{\mathrm{SR}}^{\text{x}}$
& $\{1.07018,1.09469,1.08353,1.08257\}\,\hat{F}_{\mathrm{SR}}^{\text{x}}$
\\
$\tilde{F}_{\mathrm{SR}}$
& $\frac{\pi}{12}$
& $0.912679\tilde{F}_{\mathrm{SR}}^{\text{x}}$
& $\{8.52381,-0.557729,1.06954,1.17771\}\,\tilde{F}_{\mathrm{SR}}^{\text{x}}$
\end{tabular}
\begin{tabular}{c|c|c|c|c}
$d=3$ & Fuzzy sphere & Naive $\epsilon^5$ & One-sided Pad\'e$_{5}$ & Two-sided Pad\'e$_6$ \\
\hline
$\hat{F}_{\mathrm{SR},d=3}/\hat{F}_{s,d=3}$
& \multirow{2}{*}{$0.95(9\pm 8)$}
& $0.97847$ 
& $0.977(86\pm 14)$ 
& $0.975(8\pm 7)^{\star}$ 
\\
$\tilde{F}_{\mathrm{SR},d=3}/\tilde{F}_{s,d=3}$
&
& $0.97134$
& $0.98(23\pm25)$
& $0.976(5\pm 9)$
\end{tabular}
\caption{Comparison of naive truncation and Pad\'e approximants for the critical Ising model in $d=2$ and $d=3$. 
The $y$ axis is narrow, since $\hat{F}_\mathrm{SR}$ does not change much between $d=2,4$.
Note that the one-sided Pad\'e for $\tilde{F}$ fails disastrously in $d=2$.
In $d=3$, we do not have an exact result. To make comparison easy, we normalize by the exact value for a single free scalar from \eqref{eq:FhatsD23};
we find a tight grouping of Pad\'es, so we just give the mean and standard error (the $\star$ indicates that we drop Pad\'e$_{[5,1]}^{(2)}$ for $\hat{F}$, which has a pole visible in \cref{fig:FhatSRPadeComparison}).
Our two-sided Pad\'e for $\tilde{F}$ reproduces \cite[(3.20)]{Fei:2015oha}.
}
\label{tab:NumericsIsing}
\end{table}

\begin{figure}[H]
\centering
\includegraphics[width=0.72\linewidth]{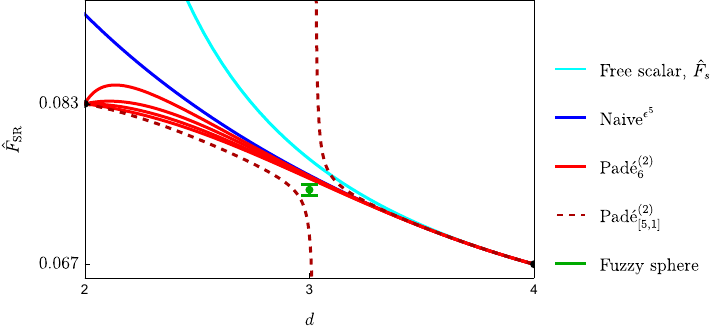}
\caption{Comparison of the naive $\epsilon$ expansion to order $\epsilon^5$ and the one-sided and two-sided Pad\'e approximants for $\hat{F}_{\mathrm{SR}}$ between $d=2$ and $d=4$. We mark the exact values $\hat{F}_{\mathrm{SR},d=4}=1/15$ and $\hat{F}_{\mathrm{SR},d=2}=1/12$, together with the fuzzy-sphere estimate $\hat{F}_{\mathrm{SR},d=3}=0.074(4\pm 6)$ \cite{Hu:2024pen}. The exact results in $d=2,4$ are marked with black dots.}
\label{fig:FhatSRPadeComparison}
\end{figure}

\section{Spectral continuity: the short-range limit of the LR CFT is the SR CFT plus a GFF}\label{sec:subtlety}

One must be slightly careful in matching up the CFTs constructed by the GFF$+$local and CFT$+\hat{\phi}\chi$ routes. %
However, the details of this matching and the flows between the various CFTs are not particularly relevant to the extremisation discussion, and are therefore relegated to this appendix.

The constructions in \cref{sec:creatingLR} and the duality described there state that the local limit of a long-range CFT is a short-range CFT plus a set of decoupled generalised free fields (i.e. the $\lambda \to 0$ limit of the CFT$+\hat{\phi}\chi$ route)\footnote{We discuss the flows between these CFTs in \cref{sec:flows}.}.
This applies both for the Ising model and for the free scalar (see \eqref{eq:FreeScalarWithChi}).

Let us now review the evidence for these extra fields: essentially, they serve to ensure continuity of the spectrum of the LR theory in the SR limit.
Considering our two different routes for the Ising CFT, where we have only a single extra field $\chi$, we see that:
\begin{itemize}
\item Naively counting primaries in the long-range theory, %
we have $\phi$ and $\phi^3$, which are evidently distinct primaries -- although they are related by the nonlocal equation of motion $(-\partial^2)^{\dotwo-\Delta_\phi} \phi \propto \phi^3$, and so have scaling dimensions $\Delta_\phi$ and $d-\Delta_\phi$ (one can also think about $\phi^3$ as being a physical shadow operator \cite[\S 7]{Paulos:2015jfa}).
\item It is a standard result for the \textit{local} Ising CFT that $\hat{\phi}^3$ is a level-2 conformal descendant of the fundamental field $\hat{\phi}$ thanks to the equation of motion\footnote{See \cite{Rychkov:2025vet} for a clarification of the word \enquote{is} here: essentially, $\hat{\phi}^3$ and $\partial^2 \hat{\phi}$ are the same operator up to contact terms, which conformal symmetry does not care about.
A slogan for this might be that they differ only in the QFT, not in the CFT.}. 
There is no primary of dimension $\Delta^\mathrm{SR}_{\hat{\phi}}+2$.
\end{itemize}
So naively, if we did not account for the field $\chi$, we would appear to lose a primary in the short-range limit.
The extra scalar field resolves this confusion\footnote{
$\chi$ is also one reason the long-range Ising model is not seen in the usual bootstrap. 
If we add in a third relevant scalar primary, the \enquote{island} of the Ising CFT reconnects to the \enquote{mainland} \cite[\S 4.2]{Behan:2018hfx} (see also \cite{vli25} \cite{Behan:2023ile,Behan:2020nsf,Behan:2021tcn,behanBootstrappingContinuousFamilies2019}).
Of course, if the presence of a local conserved stress tensor is also assumed, the long-range theory will not appear.}; its presence keeps the number of primaries the same in both theories.
That is, the short-range limit of the long-range CFT is a short-range CFT with a decoupled GFF$_\chi$.
Matching up the low-lying scalar primaries on each side, we see that this requires us to make the nontrivial identifications:
\begin{center}
\begin{tabular}{ c|cc } 
Operator & $\lim_{\Delta \to \Delta^\mathrm{SR}_{\hat{\phi}}}$ GFF$+$local  & CFT + GFF$_\chi$ \\ 
 \hline
$\Delta=\Delta^\mathrm{SR}_{\hat{\phi}}$ & $\phi$ & $\hat{\phi}$ \\ 
$\Delta=d-\Delta^\mathrm{SR}_{\hat{\phi}}$ & $(-\partial^2)^{\dotwo-\Delta^\mathrm{SR}_{\hat{\phi}}}\phi \overset{\text{eom}}{\propto} \phi^3$ & $\chi$\\
Level two descendant & $\partial^2 \phi$& $\partial^2 \hat{\phi} \overset{\text{eom}}{\propto} \hat{\phi}^3$ 
\end{tabular}
\end{center}
Notably, the conformal operator that we call $\phi^3$ in the first route is not the same as the operator $\hat{\phi}^3$ in the second route (which is not even a primary).

Another problem that $\chi$ solves is the following; it was first presented as an additional argument for the infrared duality in \cite[(2.14)]{Behan:2018hfx}. 
We know that the local CFT has a local stress tensor (by definition): $\hat{\phi} \times \hat{\phi} \supset T_{\mu\nu}$. 
However, in the nonlocal theory the spin-two operator $\tilde{T}^{\mu\nu}$ appearing in the OPE of $\phi \times \phi$ is not conserved for generic $\Delta$. 

Thus, in the CFT$+\hat{\phi}\chi$ construction there should be some vector operator $V_\mu$ of dimension $d+1$ that can appear on the right-hand side of 
\begin{equation}
\partial_\mu \tilde{T}^{\mu\nu} \sim (\Delta_\phi -\Delta^\mathrm{SR}_{\hat{\phi}}) V^\nu + O((\Delta_\phi -\Delta^\mathrm{SR}_{\hat{\phi}})^2),
\end{equation}
an equation encoding the non-conservation of $\tilde{T}^{\mu\nu}$ away from $\Delta_\phi =\Delta^\mathrm{SR}_{\hat{\phi}}$.
Once again, in the pure Ising CFT without the GFF, there is no such operator of dimension $d+1$.
However, if there is a decoupled GFF $\chi$ of dimension $d-\Delta^\mathrm{SR}_{\hat{\phi}}$ hanging around, then
\begin{equation}
V_\mu \equiv \Delta^\mathrm{SR}_{\hat{\phi}} \hat{\phi} \partial_\mu \chi - (d-\Delta^\mathrm{SR}_{\hat{\phi}})\chi \partial_\mu \hat{\phi}
\end{equation}
straightforwardly has $\Delta_V=(d+1)$. 
This means that when we perturb with $\hat{\phi}\chi$, $V_\mu$ can combine with the conserved $T^{\mu\nu}$ to give the lowest spin-2 local primary $\tilde{T}^{\mu\nu}$.

\subsection{Extra \enquote{stuff} in the continuous-data theories}

This story fits into a broader narrative in CFT: \textbf{to uplift the data of a family of CFTs from discrete to continuous, we need to add extra data} \cite[\S 2.5]{Henriksson:2025kws}.
For example, the $d\to 2$ limit of the $\phi^{2m}$ theories contains not only a consistent subsector that is the A-series minimal model $\cM_{m+1,m+2}$, but also other operators \cite{Zan:2026oyb}.
The evanescent operators decouple in the limit of integer dimension \cite{Zan:2026oyb,Hogervorst:2015akt,DiPietro:2017vsp}; turning this round, this means that we need to add them back in if we want to promote an integer-$d$ CFT to real-$d$.
Similar comments apply to the $\gO(N)$ model \cite[\S 5.5]{Maldacena:2011jn} \cite{Gorbenko:2020xya}.
In the long-range context, the extra data that must be added to the long-range theory is exactly a decoupled GFF, which is remarkably simple\footnote{This is easy to explain when the long-range theory is unitary, since the positivity of operator degeneracies means that the extra \enquote{stuff} must just decouple.
 This is not so when upgrading from $d\in \mathbb{N}$ to $d \in \mathbb{R}$ \cite{Zan:2026oyb}.}.

\begin{figure}
\centering
\includegraphics[width=\linewidth]{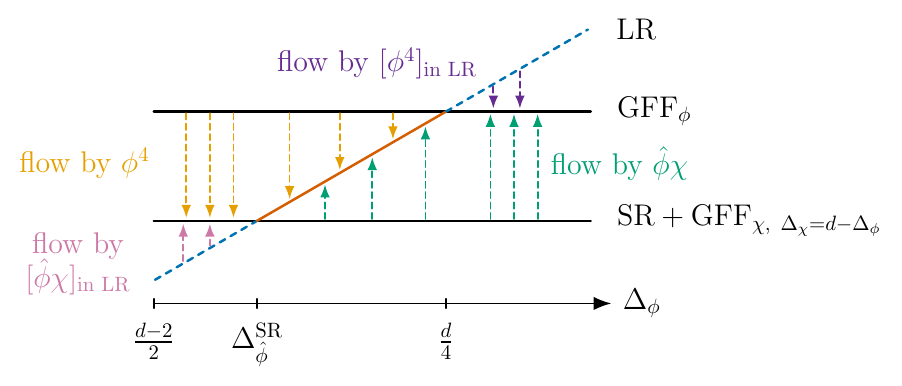}
\caption{A portrait of the allowed IR flows in theory space for the LR Ising model. 
The $x$-axis is an unusual direction, which does not correspond to a deformation by a local operator; it is the long-range parameter.
The $y$ axis corresponds to deformation by various local operators.
 We have three conformal field theories parametrised by $\Delta_\phi$: the \textbf{generalised free field GFF$_\phi$}, the \textbf{decoupled product SR$+$GFF$_\chi$ with $\Delta_\chi=d-\Delta_\phi$}, and the long-range CFT LR. In the window $\Delta^\mathrm{SR}_{\hat{\phi}}<\Delta_\phi<\tfrac{d}{4}$, indicated in \textcolor{red}{red}, the long-range CFT can be reached in the \textcolor{red}{IR} from either of the other two by the relevant deformations $\phi^4$ and $\hat{\phi}\chi$, whose flows are shown; the former agrees with \cref{fig:Fextremisation}. 
 Outside this window, LR is unstable (a \textcolor{blue}{UV} fixed point), which we indicate by \textcolor{blue}{dashing it}.
 In the window $\tfrac{d-2}{2}<\Delta_\phi < \Delta^\mathrm{SR}_{\hat{\phi}}$, we know that we can perturb by whatever operator becomes $\hat{\phi}\chi$ to flow from LR to SR+GFF${}_\chi$.
 We also know that when deforming GFF$_\phi$ by $\phi^4$ we end up at SR+GFF${}_\chi$, unless the relevant operator $\phi(-\partial^2)\phi$ generated by the OPE is tuned away.
 For $\Delta_\phi > \tfrac{d}{4}$, the LR fixed point is unstable to $[\phi^4]_{\text{in LR}}$, which takes it back to GFF$_\phi$ (see \eqref{eq:LRbetaExpansion}, which describes the perturbative flows around the top-right crossing). 
 Of course, all of these CFTs are unstable to $\phi^2$, which simply produces a flow to the empty CFT.
This figure is an adaptation of \cite[Figure 2]{Behan:2017emf}, where we have added in the unstable regimes of the long-range theory. }
\label{fig:CFTlines}
\end{figure}

\subsection{Flows between long-range CFTs}\label{sec:flows}

Throughout this paper we have neglected the RG flows between our various CFTs, always working at the formal fixed points of the beta functions: we were agnostic about whether they were real, complex, unitary, or non-unitary.
However, to make contact with the picture of \cite{Behan:2017emf}, we will now review the RG flows related to the long-range Ising CFT.
Unlike \cite{Behan:2017emf}, we will consider the long-range theory \textit{outside} the window $\Delta^\mathrm{SR}_{\hat{\phi}}<\Delta_\phi<\tfrac{d}{4}$.

We have three families of CFTs parametrised by $\Delta_\phi$: 
\begin{enumerate}
\item GFF$_{\phi}$: a generalised free field of dimension $\Delta_\phi$.
\item SR+GFF$_{\chi,\Delta_\chi=d-\Delta_\phi}$, which is a pair of decoupled CFTs.
\item LR: the long-range Ising CFT.
\end{enumerate}
We show the flows between these CFTs in \cref{fig:CFTlines}, where the arrows indicate flows towards the IR.
Conjecturally, $\tilde{F}\propto \hat{F}$ (as shown in \cref{fig:Fextremisation}) serves as a height function on this theory space (i.e. it would be the $z$-axis), and we are only allowed to flow downhill.
We make two observations:
\begin{itemize}
    \item 
For this picture of $\hat{F}$ as a height function to be consistent, we must also be consistent about whether or not we include the contribution from $\chi$ to the free energy of each of these CFTs. 
That is, we give the decoupled pair SR+GFF${}_\chi$ a value of $\hat{F}=\hat{F}_\mathrm{SR}$ (not $\hat{F}_\mathrm{SR}+\hat{F}_b(\Delta_\chi)$); however, this agrees perfectly with $\hat{F}$ if we use our normalization of $Z_\mathrm{LR}$ in \eqref{eq:ZLRwithAndWithoutChi} and set $\lambda=0$.
As discussed in \cref{sec:notincludechi}, we could also add back $\hat{F}_b(d-\Delta_\phi)$ everywhere. Along each fixed-$\Delta_\phi$ slice of \cref{fig:CFTlines}, this is just a constant shift of $\hat{F}$, and so the allowed flows are not modified.
\item In the window $\Delta_\phi>\tfrac{d}{4}$, we have the schematic operator $\phi^4$, which is precisely $:\phi^4:$ around the GFF and something more complicated, which we call $[\phi^4]_{\text{in LR}}$, around LR. It is the operator that allows us to flow from the LR to the GFF (again, we can use perturbation theory to check this). 
In the same way, in the window $\Delta_\phi < \Delta^\mathrm{SR}_{\hat{\phi}}$, we know that formally, flowing \enquote{the wrong way up the RG flow}, we can move from SR+GFF${}_\chi$ to LR using the perturbation $\hat{\phi}\chi$ (we can even compute the beta function in PT \cite{Behan:2017emf}). 
However, as suggested by \cite{Sak:1973oqx}, in this window perturbing the LR Ising by the relevant operator $:\phi(-\partial^2)\phi:$ should take it to the SR theory (possibly with a decoupled sector).
This suggests that these two operators may be related in some way -- something which we leave for future work.
\end{itemize}

%% file: 095-Dsymmetric.tex
\newcommand{\eone}{\hat e_1}

\section{\texorpdfstring{$D_{ijkl}$}{D\_{}ijkl} is totally symmetric}\label{app:Dsymmetric}

In this appendix we show that in our minimal subtraction scheme the cubic coefficient in the beta function \eqref{eq:betaFuncMulti} is totally symmetric when all its indices are lowered, under a mild assumption on the convergence of an integral.

We consider scalar primaries $\cO_i$ of common dimension $\Delta=d-\eta$ for $\eta$ small. It is clearer to first treat the case $C_{ijk}=0$.
Define the lowered triple-OPE kernel from \eqref{eq:OPEmulti} by
\begin{equation} \label{eq:flowered}
    f_{jkl|i}(w,z)\equiv C_{ia} f\indices{_{jkl}^a}(w,z)
    =
    \langle \cO_j(0)\cO_k(w)\cO_l(z)\cO_i(\infty)\rangle_c.
\end{equation}
If we write
\begin{equation}
    z=r\,\eone,
    \qquad
    w=r\,x,
\end{equation}
then scaling tells us that
\begin{equation}\label{eq:equalDeltas}
    f_{jkl|i}(w,z)=r^{-2\Delta}G^{(\eta)}_{jkli}(x),
    \qquad
    G^{(\eta)}_{jkli}(x)\equiv
    \expval{\cO_j(0)\cO_k(x)\cO_l(\eone)\cO_i(\infty)}_c.
\end{equation}
We repeat the definition \eqref{eq:ffrakDef} of the local triple-collision residue
\begin{equation}\label{eq:fPoleDef}
    \int_{(w,z)\in\mathcal{U}_0}\odif[d]{w}\,\odif[d]{z}\,f_{jkl|i}(w,z) = \frac{\mathfrak{f}_{jkl|i}}{\eta}+O(\eta^0),
\end{equation}
with the symmetric region of integration
\begin{equation}\label{eq:U0def}
    \mathcal{U}_0\equiv
    \{(w,z):\ |w|<R,\ |z|<R,\ |w-z|<R\},
\end{equation}
for some arbitrary finite value $R$.
Our aim is to prove that $\mathfrak{f}_{jkl|i}$ is totally
symmetric using the inversion symmetry of the four-point correlator: i.e. $\mathfrak{f}_{jkl|i}=\mathfrak{f}_{(jkl|i)}=\mathfrak{f}_{(ijkl)}$.
That is, it does not depend on which of the four operators is the \enquote{spectator} at infinity $\cO_i(\infty)$ that does not participate in the divergence.
The finiteness condition is $3D_{i(jkl)}+(\tfrac{2}{S_{d-1}})^{2} \mathfrak{f}_{(jkl)|i}=0$;
if $\mathfrak{f}_{ijkl}$ is totally symmetric, we can then cancel the divergence with a totally symmetric $D_{ijkl}=-\frac13\,(\tfrac{2}{S_{d-1}})^{2} \mathfrak{f}_{ijkl}$.

To do this, we decompose the $\mathfrak{f}_{jkl|i}$ integral into three regions, and show by a change of coordinates that it is manifestly symmetric in $(jkl)$. 
Then we show that it is also equal to $\mathfrak{f}_{jki|l}$ by crossing symmetry: there exists a conformal transformation that fixes $0$, swaps $\eone$ and $\infty$, and just permutes the 3 regions of integration. 
This suffices to make $\mathfrak{f}$ fully symmetric.

We will need to assume that inversion symmetry holds (which is not required in general for a CFT \cite[(2.47)]{Osborn2019cftlectures}) and that:
\begin{equation}\label{eq:ASS1}
\begin{gathered}
\text{The integrals $J^{(\eta)}_{jkl|i}$ defined in
\eqref{eq:Jdef} are finite}\\
\text{for sufficiently small $\eta$ and regular at $\eta=0$.}
\end{gathered}
\end{equation}
As usual, this is an assumption that they are finite in DREG, where we ignore the contributions from any relevant operators that arise.

\subsection{Symmetry in the lowered indices}

We now define a set of maps that generate the permutation group of the three collision points $(0,w,z)$, have Jacobian $|\text{J}|=1$, and preserve $\mathcal{U}_0$.
\begin{equation}\label{eq:maps}
    \sigma_{jk}(w,z)=(-w,z-w),
    \qquad
    \sigma_{jl}(w,z)=(w-z,-z),
    \qquad
    \sigma_{kl}(w,z)=(z,w).
\end{equation}
Separately, bosonic symmetry and translation
invariance give that \eqref{eq:flowered} satisfies
\begin{equation}
    f_{jkl|i}(w,z)
    =
    f_{kjl|i}(-w,z-w)
    =
    f_{lkj|i}(w-z,-z)
    =
    f_{jlk|i}(z,w).
\end{equation}
Using each of these in \eqref{eq:fPoleDef} and changing variables via the maps \eqref{eq:maps} then yields that $\mathfrak{f}$ is symmetric in the triple indices.
\begin{equation}\label{eq:fjklSym}
    \mathfrak{f}_{jkl|i}=\mathfrak{f}_{(jkl)|i}.
\end{equation}

\subsection{Factoring out the pole}

Let us now use spherical symmetry to move to the variables $z=r\,\eone$, $w=r\,x$, with Jacobian $\odif[d]{z}\,\odif[d]{w}=S_{d-1}\,r^{2d-1}\,\odif{r}\,\odif[d]{x}$, where $r$ runs from $0$ to the region edge, which is
\begin{equation}\label{eq:regionEdgeDef}
    0<r<R\rho(x),
    \qquad
    \rho(x)\equiv
    \min\!\left\{1,\frac{1}{\abs{x}},\frac{1}{\abs{\eone-x}}\right\}.
\end{equation} 
The symmetric region $\mathcal{U}_0$ then decomposes according to which of $|w|$, $|w-z|$, or $|z|$ is smallest. 
This gives the three regions
\begin{subequations}
\begin{align}
    \cR_{01}&\equiv\{x:\ \abs{x}<1,\ \abs{x}<\abs{\eone-x}\},\\
    \cR_{12}&\equiv\{x:\ \abs{\eone-x}<1,\ \abs{\eone-x}<\abs{x}\},\\
    \cR_{02}&\equiv\{x:\ 1<\abs{x},\ 1<\abs{\eone-x}\},
\end{align}
\end{subequations}
which partition $\mathbb{R}^d$ up to boundaries of measure zero.
Hence, \eqref{eq:fPoleDef} becomes
\begin{align}
   &\int_{(w,z)\in\mathcal{U}_0}\odif[d]{w}\,\odif[d]{z}\,f_{jkl|i}(w,z)\\
    &\qquad=
    S_{d-1}\sum_{\alpha\in\{01,12,02\}}
    \int_{\cR_{\alpha}}\odif[d]{x}
    \int_0^{R\rho(x)}\frac{\odif{r}}{r}\,r^{2\eta}G^{(\eta)}_{jkli}(x) \equiv
    \frac{S_{d-1}}{2\eta}\,J^{(\eta)}_{jkl|i},\notag
\end{align}
where we used
\begin{equation}
    \int_0^{R\rho(x)}\frac{\odif{r}}{r}\,r^{2\eta}
    =
    \frac{(R\rho(x))^{2\eta}}{2\eta},
\end{equation}
and defined
\begin{equation}\label{eq:Jdef}
    J^{(\eta)}_{jkl|i}\equiv
    \sum_{\alpha\in\{01,12,02\}}
    \int_{\cR_{\alpha}}\odif[d]{x}\,
    (R\rho(x))^{2\eta}G^{(\eta)}_{jkli}(x).
\end{equation}
We know that this integral is convergent for $\eta=0$, so it is not unreasonable that the deformation by $\eta$ is also convergent, though we have not proved it. 
Our assumption \eqref{eq:ASS1} then just means that $J^{(\eta)}_{jkl|i}$ is regular at $\eta=0$, and so
\eqref{eq:fPoleDef} gives
\begin{equation}\label{eq:fFromJ}
    \mathfrak{f}_{jkl|i} = \frac{S_{d-1}}{2}\,J^{(0)}_{jkl|i}.
\end{equation}

\subsection{Transforming the integrand}

Define the point reflection $M$ and the inversion $I$,
\begin{equation}
    M(x)\equiv \eone-x,
    \qquad
    I(x)\equiv \frac{x}{x^2},
    \qquad
    S\equiv M\circ I\circ M.
\end{equation}
Then
\begin{align}
    S(x)&=\eone+\frac{x-\eone}{\abs{x-\eone}^2},
    \qquad
    &S^2&=\mathrm{id},\\
\label{eq:SdistanceRelations}
    \abs{S(x)-\eone}&=\frac{1}{\abs{\eone-x}},
    \qquad
    & \abs{S(x)} &=\frac{\abs{x}}{\abs{\eone-x}}.
\end{align}
The regions are transformed as
\begin{equation}\label{eq:regionImages}
    S(\cR_{01})=\cR_{01},
    \qquad
    S(\cR_{12})=\cR_{02},
    \qquad
    S(\cR_{02})=\cR_{12},
\end{equation}
where the fact that, for example,  $S(x\in \cR_{01})\in \cR_{01}$ is easy to check by direct computation, and the equality of the two sets follows from $S^2=\mathrm{id}$.
The cutoff profile transforms with an overall factor,
\begin{equation}\label{eq:rhoTransform}
    \rho(S(x))
    =
    \min\!\left\{1,\frac{\abs{\eone-x}}{\abs{x}},\abs{\eone-x}\right\}
    =
    \abs{\eone-x}\,\rho(x).
\end{equation}

We now use permutation symmetry of the Euclidean four-point
function together with conformal covariance. The point reflection $M(x)=\eone-x$ exchanges $(\cO_a,0)$ and $(\cO_c,\eone)$, so
\begin{equation}
    G^{(\eta)}_{abcd}(x)=G^{(\eta)}_{cbad}(M(x)).
\end{equation}
The inversion $I(x)=x/x^2$ exchanges $(\cO_a,0)$ and $(\cO_d,\infty)$, and so %
\begin{equation}
    G^{(\eta)}_{abcd}(x)=\abs{x}^{-2\Delta}G^{(\eta)}_{dbca}(I(x)).
\end{equation}
Hence, when applied to $G_{jkil}$, $S=M\circ I\circ M$ fixes $\cO_j(0)$ at $0$, moves $\cO_i(\eone)$ to $\infty$ and $\cO_l(\infty)$ to $\eone$.
Thus,
\begin{equation}\label{eq:CanonicalCrossing}
    G^{(\eta)}_{jkli}(x)=\abs{\eone-x}^{-2\Delta}G^{(\eta)}_{jkil}(S(x)).
\end{equation}

Now set $y=S(x)$, $x=S(y)$. Since
\begin{equation}
    \odif[d]{x}=\frac{\odif[d]{y}}{\abs{\eone-y}^{2d}},
    \quad
    \abs{\eone-x}=\abs{\eone-y}^{-1},
    \quad
    \rho(x)=\rho(S(y))=\abs{\eone-y}\,\rho(y),
\end{equation}
\eqref{eq:CanonicalCrossing} gives
\begin{equation}
    G^{(\eta)}_{jkli}(x)=\abs{\eone-y}^{2\Delta}G^{(\eta)}_{jkil}(y),
    \qquad
    (R\rho(x))^{2\eta}=\abs{\eone-y}^{2\eta}(R\rho(y))^{2\eta}.
\end{equation}
Therefore,
\begin{equation}
\begin{aligned}
    (R\rho(x))^{2\eta}G^{(\eta)}_{jkli}(x)\,\odif[d]{x}
    &=
    \abs{\eone-y}^{2(\eta+\Delta-d)}
    (R\rho(y))^{2\eta}G^{(\eta)}_{jkil}(y)\,\odif[d]{y}\\
    &=
    (R\rho(y))^{2\eta}G^{(\eta)}_{jkil}(y)\,\odif[d]{y},
\end{aligned}
\end{equation}
since $\Delta=d-\eta$. Hence,
\begin{align}
    \int_{\cR_{\alpha}}\odif[d]{x}\,
    (R\rho(x))^{2\eta}G^{(\eta)}_{jkli}(x)
    =
    \int_{S(\cR_{\alpha})}\odif[d]{y}\,
    (R\rho(y))^{2\eta}G^{(\eta)}_{jkil}(y).
\end{align}
Summing over $\alpha\in\{01,12,02\}$ and using \eqref{eq:regionImages}, we obtain
\begin{equation}\label{eq:Jexchange}
    J^{(\eta)}_{jkl|i}
    =
    J^{(\eta)}_{jki|l}
\end{equation}
for every $\eta$ for which the integrals in \eqref{eq:Jdef} converge. 
By \eqref{eq:fFromJ}, then,
\begin{equation}\label{eq:spectatorExchange}
    \mathfrak{f}_{jkl|i}=\mathfrak{f}_{jki|l}.
\end{equation}

Equation \eqref{eq:fjklSym} gives the full $S_3$ action on $(jkl)$, while
\eqref{eq:spectatorExchange} gives the transposition $(il)$. Conjugating $(il)$ by
permutations of $(j,k,l)$ gives $(ij)$ and $(ik)$ as well, so these symmetries
generate all of $S_4$. As desired, therefore, $\mathfrak{f}$ is fully symmetric:
\begin{equation}
    \mathfrak{f}_{jkl|i}=\mathfrak{f}_{(ijkl)}.
\end{equation}

\subsection{The case \texorpdfstring{$C_{ijk}\neq 0$}{C\_ijk =/= 0}}

In the generic case, we must replace the bare local residue by the subtracted one, because when $C_{ijk}\neq 0$, the three OPE regions contain pairwise subdivergences.
These occur in the channels
\begin{equation}
    (jk), \qquad (kl), \qquad (lj),
\end{equation}
with coefficients
\begin{equation}\label{eq:Adef}
    A_{jkl|i}\equiv
    C^m{}_{jk}C_{mli}
    +
    C^m{}_{kl}C_{mji}
    +
    C^m{}_{lj}C_{mki}
    =
    3\,C_{mi(j}C^m{}_{kl)},
\end{equation}
which must be subtracted off, as shown in \eqref{eq:FinalDEquation}.
This is already totally symmetric in $(ijkl)$ because of the symmetry of $C_{(ijk)}$, so we have nothing more to do: $A_{jkl|i} = A_{(i|jkl)} = A_{(ijkl)}$.
The fact that the new expression for $D_{ijkl}$ is finite is easily confirmed by applying the OPE $\cO_{i}(0) \cO_j(x)\sim \frac{C\indices{^m_{ij}}}{x^{\Delta}} \cO_m(0)$ inside to the correlator. 
Because we are in the region $\cR$ we need only do this once,
\begin{equation}
\begin{aligned}
&\int_{\cR} \odif[d]{x}\expval{\cO_{i}(0) \cO_j(x)\cO_k(\eone)\cO_{l}(\infty)}_c \sim \int_{\cR} \odif[d]{x}\frac{C\indices{^m_{ij}}}{x^{\Delta}} \expval{\cO_m(0)\cO_k(\eone)\cO_{l}(\infty)} \\
&= \int_{\cR} \odif[d]{x}\frac{C\indices{^m_{ij}}}{x^{\Delta}}C_{mkl} = S_{d-1} C\indices{^m_{ij}} C_{mkl}  \int_0^{\sim 1} \frac{\odif{r}}{r} r^{d-\Delta} =S_{d-1} C\indices{^m_{ij}} C_{mkl}  \frac{1}{\eta} + O(\eta),
\end{aligned}
\end{equation}
where the precise value of \enquote{$\sim 1$} does not matter, just as in \eqref{eq:Rnotcontribute}.
We note as an aside that at the next order it is probable that the quartic term in the beta function will not generically be fully symmetric when its index is lowered with $C_{ij}$; this is just a signal that the metric $\mathcal{C}_{ij}$ becomes nontrivial. 

Because each of the three sectors just permutes the three non-spectator operators, we can just explicitly symmetrize over them and restrict to the region $\cR \equiv \cR_{01}$, yielding the result \eqref{eq:Dnonsymm}. 
The argument above then tells us that this is fully symmetric, as desired.

\subsection{Arbitrary \texorpdfstring{$\Delta_i=d-\kappa^i\eta$}{Delta\_i=d-kappa\^{}i eta}}\label{sec:DsymmArbKappa}

It is straightforward to repeat the $D$-symmetry argument for non-equal near-marginal dimensions, and so we do so here, anticipating the work in \cite{Fraser-Taliente2026cpt}. 
Define
\begin{equation}
    \Delta_i=d-\kappa^i\eta,
    \qquad
    \widehat{\kappa}^{jkl}{}_i\equiv
    \kappa^j+\kappa^k+\kappa^l-\kappa^i,
\end{equation}
in 
\begin{equation}
        G^{(\eta)}_{jkli}(x)\equiv \expval{\cO_j(0)\cO_k(x)\cO_l(\eone)\cO_i(\infty)}_c.
\end{equation}
leading to
\begin{equation}\label{eq:ffrakDefForKappa}
\int_{(w,z)\in\mathcal{U}_0}\odif[d]{w}\,\odif[d]{z}\,f_{jkl|i}(w,z) = \frac{S_{d-1}}{\widehat{\kappa}^{jkl}{}_i\eta}\, J^{(\eta,\kappa)}_{jkl|i}.
\end{equation}
The geometry of course does not change: the maps \eqref{eq:maps}, the three regions
\eqref{eq:regionImages}, and the region edge \eqref{eq:regionEdgeDef} do not depend on the values of the dimensions. 
Likewise, the proof of the triple symmetry \eqref{eq:fjklSym} is unchanged, since
$\widehat{\kappa}^{jkl}{}_i$ in \eqref{eq:ffrakDefForKappa} is symmetric in $(jkl)$.

The first change is to the overall scaling of the correlator. 
With $z=r\,\eone$ and $w=r\,x$, \eqref{eq:equalDeltas} becomes 
\begin{equation}
    f_{jkl|i}(w,z)
    =
    r^{-\Delta_j-\Delta_k-\Delta_l+\Delta_i}\,
    G^{(\eta)}_{jkli}(x)
    =
    r^{-2d+\widehat{\kappa}^{jkl}{}_i\eta}\,
    G^{(\eta)}_{jkli}(x).
\end{equation}
Thus, \eqref{eq:Jdef} becomes
\begin{equation}\label{eq:JdefKappa}
    J^{(\eta,\kappa)}_{jkl|i}\equiv
    \sum_{\alpha\in\{01,12,02\}}
    \int_{\cR_{\alpha}}\odif[d]{x}\,
    (R\rho(x))^{\widehat{\kappa}^{jkl}{}_i\eta}G^{(\eta)}_{jkli}(x),
\end{equation}
and the analogue of \eqref{eq:fFromJ} is
\begin{equation}\label{eq:fFromJKappa}
    \mathfrak{f}_{jkl|i}
    =
    \frac{S_{d-1}}{\widehat{\kappa}^{jkl}{}_i}\,
    J^{(0,\kappa)}_{jkl|i},
\end{equation}
assuming the same regularity at $\eta=0$ as in \eqref{eq:ASS1}.

The crossing relation is also modified.
Since the operator at the integrated point $x$ has dimension $\Delta_k$, \eqref{eq:CanonicalCrossing} becomes
\begin{equation}\label{eq:CanonicalCrossingKappa}
    G^{(\eta)}_{jkli}(x)= \abs{\eone-x}^{-2\Delta_k}G^{(\eta)}_{jkil}(S(x)).
\end{equation}
Setting $y=S(x)$ as before, and using \eqref{eq:rhoTransform}, gives
\begin{equation}\label{eq:WeightedCrossingKappa}
    (R\rho(x))^{\widehat{\kappa}^{jkl}{}_i\eta}G^{(\eta)}_{jkli}(x)\,\odif[d]{x}
    =
    \abs{\eone-y}^{(\widehat{\kappa}^{jkl}{}_i-2\kappa^k)\eta}
    (R\rho(y))^{\widehat{\kappa}^{jkl}{}_i\eta}G^{(\eta)}_{jkil}(y)\,\odif[d]{y}.
\end{equation}
Unlike the equal-dimension case, the extra factor on the right-hand side is not identically $1$, so the exact-$\eta$ \eqref{eq:Jexchange} no longer holds in general.  However, 
\begin{equation}
    \abs{\eone-y}^{(\widehat{\kappa}^{jkl}{}_i-2\kappa^k)\eta}=1+O(\eta),
    \qquad
    (R\rho(y))^{\widehat{\kappa}^{jkl}{}_i\eta}
    =
    (R\rho(y))^{\widehat{\kappa}^l{}_{jki}\eta}(1+O(\eta)),
\end{equation}
so by the regularity assumption \eqref{eq:ASS1} we still obtain
\begin{equation}\label{eq:JexchangeKappa}
    J^{(0,\kappa)}_{jkl|i}=J^{(0,\kappa)}_{jki|l}.
\end{equation}
Together with \eqref{eq:fjklSym}, we find that the $\eta^0$ shape integral is still totally symmetric:
\begin{equation}\label{eq:JKappaFullySymmetric}
    J^{(0,\kappa)}_{jkl|i}=J^{(0,\kappa)}_{(ijkl)}.
\end{equation}

Observe that it is $J^{(0,\kappa)}_{jkl|i}$, not $\mathfrak{f}_{jkl|i}$, which is fully symmetric: \eqref{eq:fFromJKappa} shows that $\mathfrak{f}$ now has the extra factor $(\widehat{\kappa}^{jkl}{}_i)^{-1}$. Happily, the quartic counterterm $D\indices{^i_{jkl}}$ also has the same factor $1/(\widehat{\kappa}^{jkl}{}_i\eta)$, so after lowering the first index we are still permitted to take
\begin{equation}
    D_{i(jkl)}=D_{(ijkl)}.
\end{equation}